\documentclass[conference]{IEEEtran}
\IEEEoverridecommandlockouts
\usepackage{hyperref}
\usepackage{cite}
\usepackage{amsmath,amssymb,amsfonts,mathtools}
\usepackage{algorithmic}
\usepackage{graphicx}
\usepackage{textcomp}
\usepackage{xcolor}
\usepackage{enumitem}
\usepackage[utf8]{inputenc} 

\usepackage{makecell}
\usepackage{comment}
\def\BibTeX{{\rm B\kern-.05em{\sc i\kern-.025em b}\kern-.08em
    T\kern-.1667em\lower.7ex\hbox{E}\kern-.125emX}}

\newtheorem{theorem}{Theorem}
\newtheorem{corollary}{Corollary}
\newtheorem{proposition}{Proposition}
\newtheorem{definition}{Definition}

\newtheorem{lemma}{Lemma}
\newtheorem{Claim}{Claim}

\newenvironment{Proof}[1]{\medskip\par\noindent{\bf Proof:\,}\,#1}{{\mbox{\,$\blacksquare$}\par}}

\usepackage{array}
\newcolumntype{L}{>{\centering\arraybackslash}m{2.5cm}}
\newcolumntype{M}{>{\centering\arraybackslash}m{3.3cm}}

\usepackage{booktabs}

\begin{document}

\title{Correlated Privacy Mechanisms for Differentially Private Distributed Mean Estimation
}

\author{
\IEEEauthorblockN{
\begin{minipage}{0.45\textwidth}
    \centering
    Sajani Vithana\\
    \textit{School of Engineering and Applied Sciences} \\
    \textit{Harvard University} \\
    sajani@seas.harvard.edu
\end{minipage}
\begin{minipage}{0.45\textwidth}
    \centering
    Viveck R. Cadambe\\
    \textit{School of Electrical and Computer Engineering} \\
    \textit{Georgia Institute of Technology} \\
    viveck@gatech.edu
\end{minipage}
}\\
\IEEEauthorblockN{
\begin{minipage}{0.45\textwidth}
    \centering
    Flavio P. Calmon\\
    \textit{School of Engineering and Applied Sciences} \\
    \textit{Harvard University} \\
    flavio@seas.harvard.edu
\end{minipage}
\begin{minipage}{0.45\textwidth}
    \centering
    Haewon Jeong\\
    \textit{Department of Electrical and Computer Engineering} \\
    \textit{University of California, Santa Barbara} \\
    haewon@ece.ucsb.edu
\end{minipage}
}
}

\maketitle

\begin{abstract}
  Differentially private distributed mean estimation (DP-DME) is a fundamental building block in privacy-preserving federated learning, where a central server estimates the mean of $d$-dimensional vectors held by $n$ users while ensuring $(\epsilon,\delta)$-DP. Local differential privacy (LDP) and distributed DP with secure aggregation (SA) are the most common notions of DP used in DP-DME settings with an untrusted server. LDP provides strong resilience to dropouts, colluding users, and adversarial attacks, but suffers from poor utility. In contrast, SA-based DP-DME achieves an $O(n)$ utility gain over LDP in DME, but requires increased communication and computation overheads and complex multi-round protocols to handle dropouts and attacks. In this work, we present a generalized framework for DP-DME, that captures LDP and SA-based mechanisms as extreme cases. Our framework provides a foundation for developing and analyzing a variety of DP-DME protocols that leverage correlated privacy mechanisms across users. To this end, we propose CorDP-DME, a novel DP-DME mechanism based on the correlated Gaussian mechanism, that spans the gap between DME with LDP and distributed DP. We prove that CorDP-DME offers a favorable balance between utility and resilience to dropout and collusion. We provide an information-theoretic analysis of CorDP-DME, and derive theoretical guarantees for utility under any given privacy parameters and dropout/colluding user thresholds. Our results demonstrate that (anti) correlated Gaussian DP mechanisms can significantly improve utility in mean estimation tasks compared to LDP -- even in adversarial settings -- while maintaining better resilience to dropouts and attacks compared to distributed DP.

\end{abstract}


\section{Introduction}\label{intro}

Distributed mean estimation (DME) is a fundamental building block  in a number of applications ranging from federated learning  \cite{fl1,fl2}, distributed stochastic gradient descent  \cite{dsgd,cpsgd,sgd2,vqsgd} to distributed sensor network computations \cite{sensor}. Differentially private DME (DP-DME) refers to the setting where a central server aims to estimate the mean of $d$ dimensional  vectors held by $n$ distributed users while ensuring differential privacy (DP) \cite{DP1,dp} of the users' vectors. The utility of DP-DME is measured by the mean squared error (MSE) between the estimate at the server and the true mean. DP-DME has been studied under multiple notions of DP: 
\begin{enumerate}[label=(\roman*)]
    \item  \emph{Central DP (CDP)} \cite{cdp1}, where a trusted server collects user vectors, computes the mean, and adds noise to ensure DP;
    \item  \emph{Local DP (LDP)} \cite{PrivUnit,optimal_algorithms}, where each user independently perturbs their vector before sending to a potentially adversarial server that performs aggregation;
    \item \emph{Distributed DP with secure aggregation (SA)} \cite{distributeddp,ddg,skellam,dpsecagg,dpsecagg2}, where users perturb vectors locally and use a SA protocol \cite{secagg,secaggplus,lightsecagg} to ensure the server only views the sum of the perturbed vectors, and
    \item \emph{Shuffle DP \cite{encode,shuffle1,shuffle2}}, where the locally perturbed vectors of the users are shuffled by a trusted shuffler before sending to the server for aggregation. 
\end{enumerate}

LDP and distributed DP with SA are the most common approaches to DP-DME that eliminate the need for a trusted third party.\footnote{Other DP-DME approaches exist, employing distributed DP in decentralized (graph-based) settings \cite{dp_corr2,decentralized}. A comparison of our work with these approaches is provided in Section~\ref{related}.}  LDP  and SA offer distinct advantages for DME, such as reduced overhead in LDP and enhanced utility in SA. In this work, we address a fundamental question:
\begin{center}
\emph{How should we design a DP-DME mechanism that achieves a higher utility than LDP while significantly reducing the overhead relative to SA?} 
\end{center}
In order to answer this question, we first provide more details on the advantages and limitations of each method.

LDP-based DME protocols \cite{PrivUnit,fast_private_mean} are simple, single-round methods that require nearly no coordination among users. It assumes the strongest threat model in DP-DME with an adversarial server and no constraints on the number of users colluding with the server or dropouts. However, the robust privacy guarantees of LDP-based DME come at a significant utility cost. For the same level of privacy, the MSE resulted by LDP-based DME is higher than the MSE of CDP by a factor of $O(n)$ \cite{private_estimation_lb}.



In contrast, distributed DP via SA achieves the same privacy-utility trade-off as CDP-based DME by relying on perfect coordination and synchronization among users, along with a complex multi-party cryptographic protocol executed over multiple rounds. The key feature of SA is that the server only learns the sum of the users' uploads. In other words, SA ensures that the data received by the server is independent of the user's inputs, conditioned on the sum. This allows users to apply less noise to their private vectors than LDP, thus improving utility. 

The utility gain in distributed DP with SA comes at a high coordination cost. Practical implementations of SA (e.g., the SecAgg\footnote{A detailed description of the SecAgg protocol is provided in Appendix~\ref{secaggdet}, along with a simple four-user example illustrating the key ideas.} protocol in \cite{secagg}) operate in two main phases: an offline phase for pre-processing (shared randomness generation) and an online phase where user-vectors are uploaded and aggregated by the server. SecAgg relies on perfectly (anti) correlated pair-wise random noise, uniformly sampled from a finite field for each pair of users. For this, SecAgg employs the Diffie-Hellman key exchange among all participating users during the offline phase. The use of finite field noise requires a costly multi-round dropout recovery mechanism to handle even a single dropout in the online phase. To facilitate this, each user distributes components of all pairwise random seeds as secret shares with every other user during the offline phase. Such complex dropout recovery mechanisms result in increased communication, computational overhead, synchronization, and the need for extensive shared randomness among users.
As discussed in \cite{dpsecagg}, the increased cost of multi-round interactions between users and the server can thwart the efficient training of large machine learning models, particularly in federated learning settings with unreliable communication links. Furthermore, SecAgg enforces strict constraints on the maximum number of dropouts and colluding users, and fails if these limits are exceeded. SecAgg's multiple communication rounds between the server and users also make the protocol vulnerable to malicious server attacks \cite{secaggattack1}. 

In this paper, we develop a DP-DME framework that addresses the shortcomings of both LDP-based and SA-based DME. In particular, we identify the following underlying factors contributing to inefficiencies.
\begin{itemize}
    \item\emph{Increased MSE with LDP:} LDP-based mechanisms are independent across users, leading to increased MSE.
    \item \emph{Complexity and overheads of SA:} The complexity and overheads of SA protocols largely stems from its multi-round dropout handling mechanism,
    \item \emph{Strict dropout thresholds in SA:} The lack of flexibility in SA-based DP-DME, due to strict limits on dropouts and colluding users, as well as the reliance on complex multi-round dropout recovery mechanisms, arise from the use of random noise from a finite field.
    
\end{itemize}

Our key insight is that, ultimately, the privacy guarantee of SA-based DP-DME is $(\epsilon,\delta)$-DP, which obviates overly conservative cryptographic measures such as sampling random noise from a finite field. Building on this observation, we define a DP-DME framework that is end-to-end DP and simplifies dropout handling, resulting in protocols that are more flexible and resilient to attacks and dropouts, compared to DP-DME with SA. Specifically, we target techniques that: (i) like LDP, can be performed with one round without requiring complex cryptographic approaches to handle dropouts and collusions, (ii) like SA, provide MSE comparable to CDP, (iii) like LDP, are resilient to privacy attacks, and do not cease to operate beyond dropout/colluding user thresholds (see Table~\ref{res}\footnote{Table~\ref{res} provides a comparison of the two main approaches to DP-DME and CorDP-DME. For SA-based DP-DME, SecAgg \cite{secagg} is selected as the underlying SA protocol for fair comparison based on the common assumptions made. Further comparisons between CorDP-DME and other SA protocols are discussed in Section~\ref{SA}.}).

\begin{figure}
  \centering
  \includegraphics[scale=0.65]{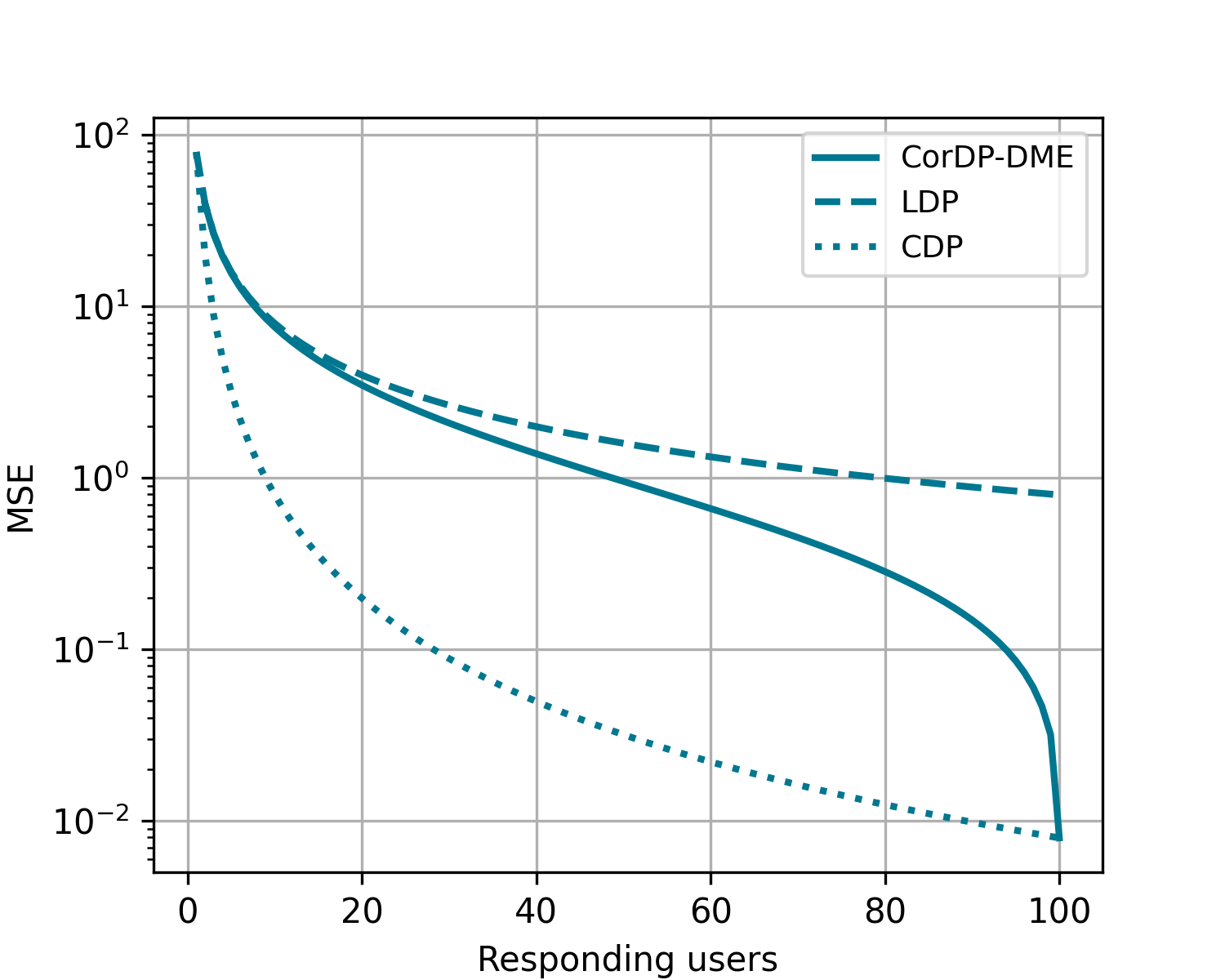}
    \caption{MSE of LDP, CDP and CorDP-DME with different numbers of responding users for a setting with $n=100$, $\epsilon=2$, $\delta=10^{-5}$. CorDP-DME coincides with CDP when all users respond. All three mechanisms coincide when only one user responds. CorDP-DME always outperforms LDP. In general, CorDP-DME spans the gap between DME with LDP and CDP. }
    \label{theory}
\end{figure}

Our starting point is a novel geometric interpretation (see Sec.~\ref{why}) of the privacy-utility trade-off in DP-DME, which reflects how (anti) correlated mechanisms can outperform independent (LDP) mechanisms. Our analysis demonstrates that SA, in essence, represents an extreme case within the broader spectrum of DP-DME mechanisms described by this geometric framework. Unlike SA, which requires correlated noise over a finite field, we consider adding correlated Gaussian noise directly to real-valued vectors held by users. The use of correlated Gaussian noise leads to single-round DP-DME protocols\footnote{Single-round protocols refer to those with only one round in the (online) DME phase, excluding the (offline) noise generation phase.} that are robust to dropouts and colluding users. We couple this starting point with a DP framework for quantifying the impact of arbitrary noise correlations on privacy, particularly in the face of user dropouts and collusion. We apply this framework to  develop a simple, one-round DP-DME protocol called \textbf{CorDP-DME}.  After an initial offline phase to establish shared randomness, CorDP-DME uses an optimized correlated Gaussian mechanism among users and provides the utility of CDP without resorting to SA in absence of dropouts (See Table~\ref{res}). In the presence of dropouts and collusions, CorDP-DME achieves significantly reduced MSE relative to LDP yet retains its flexibility, resilience, and simplicity compared to SA. Notably, CorDP-DME spans the gap between the two extremes of LDP and distributed DP with SA (Fig.~\ref{theory}).

From a broader theoretical perspective, we present a generalized framework for DP-DME that serves as a foundation for developing a range of DP-DME protocols with varying properties. LDP-based and SA-based DME represent extreme points within this spectrum, each with distinct characteristics. Our framework enables the design of protocols that combine the key advantages of both extremes -- such as enhanced utility and reduced complexity and overheads -- while also supporting theoretical insights into fundamental limits and trade-offs.

\subsection{Our Contributions}


\begin{itemize}
    \item We introduce a generalized distributed DP framework for DME that accounts for users that execute privacy mechanisms that are arbitrarily correlated with each other. The proposed model considers both dropouts and colluding users. DME with LDP and distributed DP with SA are extreme cases of the proposed generalized DP-DME framework (Section~\ref{problem}). 
    \item We provide a novel geometric interpretation of the privacy-utility trade-off in DP-DME, which demonstrates how (anti) correlated privacy mechanisms can outperform independent privacy mechanisms used by LDP (Section~\ref{why}).
    \item We perform an information-theoretic analysis on the correlated Gaussian mechanism for DP-DME. We derive the optimum noise parameters and the decoding strategy at the server that minimizes the MSE for a target $(\epsilon,\delta)$ privacy parameters and number of dropout/colluding user thresholds. We also provide a converse result that establishes the optimum noise covariance structure that minimizes the MSE of unbiased mean estimates (Section~\ref{results}).
    \item Building on the information-theoretic analysis, we propose CorDP-DME, a single-round DP-DME mechanism that spans the gap between DME with LDP and distributed DP with SA in terms of privacy-utility trade-offs and resilience to dropouts and attacks (Section~\ref{proposed}).
\end{itemize}

\begin{table*}[!tb]
    \centering
    \footnotesize
    \begin{tabular}{M M M M}
    \toprule
      & \textbf{LDP} \cite{optimal_algorithms} & \textbf{Distributed DP w/ SecAgg} \cite{secagg,ddg} & \textbf{CorDP-DME (ours)}  \\ \toprule
      Rounds in protocol  & 
       Single round  & Multiple rounds & Single round  \\

    \midrule 
    MSE: no dropouts  & $O\left(\frac{d}{n\min\{\epsilon,\epsilon^2\}}\right)$  & $O\left(\frac{d}{n^2\min\{\epsilon,\epsilon^2\}}\right)$  & $O\left(\frac{d}{n^2\min\{\epsilon,\epsilon^2\}}\right)$\\
\midrule 
MSE: dropouts $\leq\ p$  & $O\left(\frac{d}{(n-p)\min\{\epsilon,\epsilon^2\}}\right)$  & $O\left(\frac{d}{(n-p)^2\min\{\epsilon,\epsilon^2\}}\right)$  & $O\left(\frac{dp}{n(n-p)\min\{\epsilon,\epsilon^2\}}\right)$\\
\midrule 
    Computation & \makecell{User: $O(d)$ \\ Server: $O(d)$} & \makecell{User: $O(n^2+dn)$\\ Server: $O(dn^2)$} & \makecell{User: $O(dn)$\\  Server: $O(d)$}\\
    \midrule 
    Communication: pre-processing  & 
       None required  & \makecell{User: $O(n)$ key exchanges\\ $O(n)$ secret shares\\ Server: $O(n^2)$ key exchanges\\ $O(n^2)$ secret shares} & \makecell{User: $O(n)$ key exchanges\\ Server: $O(n^2)$ key exchanges}  \\
       \midrule 
\makecell{Communication:\\ DME phase}  & 
       \makecell{User: $O(d)$\\ Server: $O(dn)$}  & \makecell{User: $O(d+n)$\\ Server: $O(dn+n^2)$}  & \makecell{User: $O(d)$\\ Server: $O(dn)$}  \\
       \midrule 
       Storage & \makecell{User: $O(d)$\\ Server: $O(d)$}  & \makecell{User: $O(d+n)$\\ Server: $O(d+n^2)$} & \makecell{User: $O(d+n)$\\ Server: $O(d)$}\\
\midrule 
       Dropouts & Gradual rise in MSE with dropouts  & Gradual rise in MSE with dropouts up to the threshold, MSE surge afterwards & Gradual rise in MSE with dropouts\\
       \midrule 
     Collusion & $(\epsilon,\delta)$-DP with any number of colluding users  & $(\epsilon,\delta)$-DP up to $c<n/3$ colluding users, sudden drop in privacy afterwards  & $(\epsilon,\delta)$-DP up to \emph{any} $c$ colluding users, graceful privacy decay afterwards\\
       
       \bottomrule
       \\
    \end{tabular}
    \caption{Comparison of CorDP-DME with existing approaches for DP-DME.}
    \label{res}
\end{table*}

\subsection{Related Work}\label{related}

\subsubsection{DME with LDP} In DME with LDP, each user perturbs their vectors independently to satisfy $(\epsilon,\delta)$-DP. As the user has the complete control over the perturbations made to their private vectors, DME with LDP assumes the strongest threat model in DP-DME. The strong privacy and security guarantees in LDP comes at a large utility cost. The MSE of DME with LDP is $O(n)$ times higher than the MSE of CDP. Optimality results on DME with LDP have been provided in \cite{minimax_procedures,optimal_algorithms,PrivUnit}, based on the lower bounds derived in \cite{private_estimation_lb}. A number of order optimal LDP-DME algorithms and fundamental results on communication constraints have been provided in \cite{minimax_rates,local_randomizers,comm-privacy-accuracy2020,fast_private_mean,instance,comm-privacy-utility,optimal_compression,comm2023}. CorDP-DME fundamentally differs from LDP by allowing for the privacy mechanisms among different users to be arbitrarily correlated. In fact, the system model introduced in this work is a generalization of additive DME mechanisms with LDP. The lower bounds established in \cite{private_estimation_lb} for LDP are not applicable to CorDP-DME, as they do not account for potential correlations between the privacy mechanisms of different users. These bounds are derived under more restrictive assumptions.


\begin{figure}[!htb]
    \centering
  \includegraphics[scale=0.4]{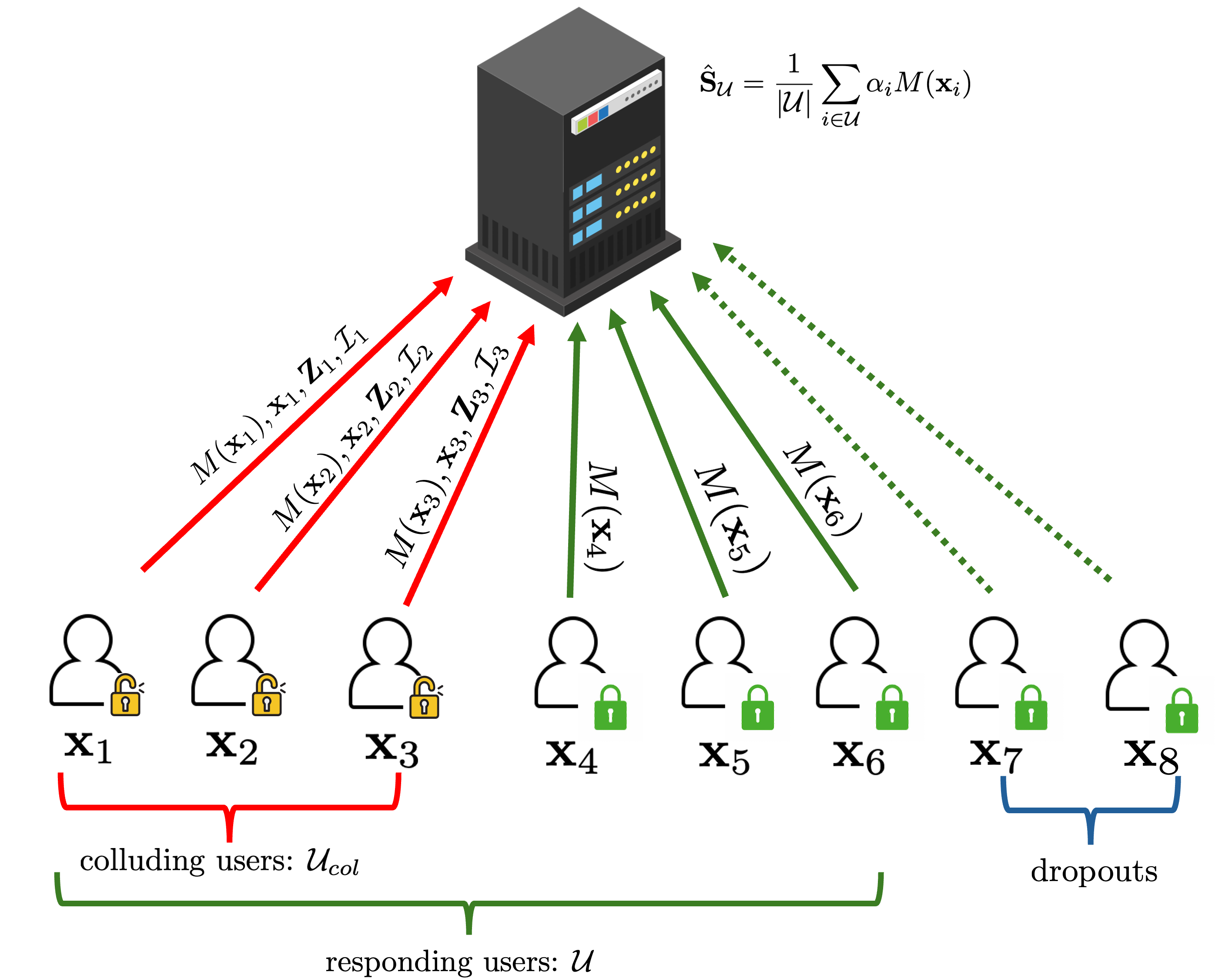}
    \caption{System model: Each user sends a perturbed vector $M(\mathbf{x}_i)$ and the central server decodes the mean through linear decoding, using the uploads of the responding users. We assume that there can be upto $c$ colluding users and $n-t$ dropouts. The server learns all the random variables observed by the colluding users.}
    \label{model}
\end{figure}

\subsubsection{Correlated noise in DP-DME} The application of correlated noise in DP-DME has been studied in \cite{dpsecagg,dpsecagg2,ddg,dp_corr1,dp_corr2,decentralized}. Our work is closely related to DECOR \cite{decentralized}, CAPE \cite{dp_corr1}, and GOPA \cite{dp_corr2}  as they all incorporate the use of correlated Gaussian noise, and achieve MSEs of the same order as CDP in the absence of dropouts. CAPE and DECOR fundamentally differ from ours by not considering the effects of dropouts. GOPA operates similarly to SecAgg-based DP-DME, but it is designed for general graph structures, and functions over the reals while incorporating correlated noise. The dropout handling mechanisms of both GOPA \cite{dp_corr2} and SecAgg \cite{secagg} require multiple rounds of communication, incurring substantial overheads. In fact, because it handles general graph structures, GOPA, unlike SecAgg, does not necessarily achieve CDP performance even after multiple rounds of the protocol. CorDP-DME primarily stands out from GOPA \cite{dp_corr2}, SecAgg-based DP-DME \cite{secagg}, and all comparable correlated privacy mechanisms \cite{dpsecagg,dpsecagg2,ddg,dp_corr1,dp_corr2,decentralized}, in that it provides a systematic \emph{single-round} approach to handle dropouts with no additional communications with the remaining users, no additional shared information among users beyond what is required to generate correlated randomness (unlike \cite{dp_corr2,secagg} which require additional information to be shared among users for dropout recovery), no additional computations at the server, and with no significant loss of MSE. To achieve this, CorDP-DME optimizes the joint distribution of the noise variables across all users to minimize the MSE for the worst-case dropouts and collusions, for any given dropout and collusion thresholds. We provide fundamental privacy-utility trade-offs for arbitrary settings with dropouts and colluding users, considering single-round protocols.
Additionally, we present converse results that establish the optimum covariance structure, and derive the optimum noise parameters that minimize the MSE (exact optimal) under different dropout/collusion settings, as opposed to aiming for order-optimality.

\subsubsection{Secure Aggregation (SA)}\label{SA}

SA is a multi-party computation (MPC) protocol that enables computation of the sum of data from multiple participants while ensuring that only the aggregate result is revealed, and individual inputs remain private. The first practical protocol, SecAgg \cite{secagg}, relies on pairwise random noise generation between all user pairs, as described in Section~\ref{intro} and Appendix~\ref{secaggdet}. Subsequent protocols, such as SecAgg+ \cite{secaggplus} and LightSecAgg \cite{lightsecagg}, aim to reduce communication and computational overhead.

SecAgg+ modifies SecAgg by connecting each user with only a randomly selected subset of $O(\log n)$ users for pairwise random noise generation. This reduces the communication and computational overheads of SecAgg+ compared to SecAgg at the expense of a $O\left(\frac{1}{n}\right)$ probability of the protocol failing.\footnote{This failure probability is separate from the protocol's termination due to the number of dropouts exceeding a given threshold.}\footnote{This is explained in Lemmas 3.8-3.9 in \cite{secaggplus}. When $\eta,\sigma_1\sim O(\log n)$, the probability of termination is $2^{-\eta}=O\left(\frac{1}{n}\right)$.} 
LightSecAgg removes the need for pairwise noise and eliminates the secret-share reconstruction step used in SecAgg and SecAgg+ during dropout recovery, thereby reducing the overheads. However, it requires the transmission of entire $d$-length vectors during initialization and dropout recovery, as opposed to sharing only the random seeds in SecAgg and SecAgg+, which can be challenging when $d$ is large.


A direct comparison between CorDP-DME (zero probability of failure) and DP-DME with SecAgg+ ($O\left(\frac{1}{n}\right)$ probability of failure) highlights that CorDP-DME achieves lower computational overhead at the server and reduced communication costs during the online phase, requiring $O(d)$ communication per user and $O(dn)$ at the server, compared to SecAgg+'s $O(d+\log n)$ per user and $O(dn+n\log n)$ at the server. However, in the offline phase for noise initialization, CorDP-DME incurs higher communication overhead ($O(n^2)$ at the server and $O(n)$ per user) than SecAgg+ ($O(n\log n)$ at the server and $O(\log n)$ per user). The key advantage of CorDP-DME over all existing secure aggregation protocols is its single-round design, which eliminates user synchronization issues, avoids additional communication and computation from extra rounds, prevents privacy leaks caused by delayed responses, and mitigates the risk of multi-round privacy attacks.

It is also worth noting that a new version of CorDP-DME can be developed by solving a new problem where a certain probability of protocol failure is allowed, to reduce the initialization costs. This version of CorDP-DME can simply (randomly) select only $O(\log n)$ pairwise random seeds per user to achieve lower initialization (offline) costs as in SecAgg+ while also enjoying its lower online costs. This approach requires optimizing the noise correlations (covariance matrix) among users under a sparsity constraint. 

\textbf{Malicious users and multiple servers:} Addressing malicious users is a critical challenge in secure aggregation, as they can compromise the aggregate by providing incorrect inputs. Single-server approaches, such as ACORN \cite{acorn} and \cite{mal_user}, utilize cryptographic techniques for input validation, while multi-server methods like PRIO \cite{prio} and \cite{talwar} employ secret sharing in a non-colluding server setup to validate inputs while maintaining user privacy. In this work, we assume users are honest-but-curious and leave input validation as a topic for future research. Multi-server aggregation methods like PRIO \cite{prio} and \cite{talwar} differ from CorDP-DME in following ways.  1) They use multiple servers and require at least one of them to be non-colluding, while CorDP-DME uses a single server with no assumptions on collusion. 2) In \cite{prio} and \cite{talwar}, users act independently, with no correlation required among them, ensuring that user dropouts have no impact on the mechanism. In contrast, CorDP-DME utilizes a privacy mechanism based on correlated noise among participating users, making it essential to account for dropouts. 3) The use of secret sharing in \cite{prio} and \cite{talwar} is fundamentally different from that of CorDP-DME: In PRIO \cite{prio} and \cite{talwar}, each user's input is divided into secret shares and distributed among the non-colluding servers. In CorDP-DME, (partially) additive secret shares of random noise terms are distributed among multiple users.

Other studies on multi-server SA, such as \cite{multiserver1, multiserver2}, investigate optimal communication rates under perfect user-privacy (conditioned on the aggregate). However, these approaches rely on the strong assumption of non-colluding servers, which can be challenging in adversarial scenarios. CorDP-DME avoids this by operating in a single-server.

\section{Problem Formulation}\label{problem}


We consider a distributed mean estimation (DME) setting with $n$ honest-but-curious users, each holding an independent $d$-dimensional vector, and a central server that estimates the mean of these vectors while ensuring $(\epsilon,\delta)$-DP. We assume a centralized communication model where each user is only connected to the server (see Fig.~\ref{model}). Let $\mathcal{U}_{all}\coloneq\{1,\dotsc,n\}$ denote the set of all users. Each user $i\in\mathcal{U}_{all}$ generates a private vector $\mathbf{x}_i\in\mathbb{B}^d$, where $\mathbb{B}^d\subset \mathbb{R}^d$ is the unit ball,\footnote{Even though we present the DP-DME analysis for vectors in the unit ball, this is not a necessary condition. The necessary condition is for the vectors to be bounded. The MSE for a case where the vectors lie in a ball of radius $c$ scales as $c^2$ times the MSE of the case corresponding to the unit ball.} and sends an encoded (distorted) version of $\mathbf{x}_i$ given by, 
\begin{align}\label{encode}
M(\mathbf{x}_i)\coloneq \mathbf{x}_i+\mathbf{Z}_i
\end{align}
to the server, where $\mathbf{Z}_i\in\mathbb{R}^d$  are (not necessarily independent) random noise variables. We assume an honest-but-curious server that attempts to compromise user-privacy, and allow for up to $c$ colluding users and $n-t$ dropouts. Let $\mathcal{U}_{col}\subset\mathcal{U}_{all}$, $|\mathcal{U}_{col}|\leq c$ denote the set of users colluding with the server, i.e., the server has access to all the random variables observed by the users in $\mathcal{U}_{col}$, including $\mathbf{x}_i$, $\mathbf{Z}_i$ and $M(\mathbf{x}_i)$ for $i\in\mathcal{U}_{col}$. Let $\mathcal{U}\subseteq\mathcal{U}_{all}$, $|\mathcal{U}|\geq t$ be the set of responding users at a given time. We assume $c<t$. The goal is to obtain the mean of the vectors of the responding users given by $\mathbf{S}_{\mathcal{U}}=\frac{1}{|\mathcal{U}|}\sum_{i\in\mathcal{U}}\mathbf{x}_i$. Based on the encoded vectors received from $\mathcal{U}$, the server estimates the mean using a linear function $d_{\mathcal{U}}:\mathbb{R}^{d|\mathcal{U}|}\rightarrow\mathbb{R}^d$, defined as:
\begin{align}\label{decoder}
\hat{\mathbf{S}}_{\mathcal{U}}&\coloneq d_{\mathcal{U}}([M(\mathbf{x}_{j_1}),\dotsc,M(\mathbf{x}_{j_{|\mathcal{U}|}})])=\frac{1}{|\mathcal{U}|}\sum_{i\in\mathcal{U}} \alpha_iM(\mathbf{x}_i),
\end{align}
where $\mathcal{U}=\{j_1,\dotsc,j_{|\mathcal{U}|}\}$, $j_1<j_2<\dotsc<j_{|\mathcal{U}|}$, and $\alpha_i\in\mathbb{R}$ for $i\in\mathcal{U}$ are constants that define the decoding function for a given $\mathcal{U}$. We assume that $\alpha_i$ for $i\in\mathcal{U}_{all}$ are independent of $\{M(\mathbf{x}_j)\}_{j\in\mathcal{U}}$. In this work, we analyze single-round DP-DME settings defined by the encoding and decoding steps in \eqref{encode}-\eqref{decoder}.

In the data encoding step in \eqref{encode}, each user $i$ has to first generate the noise terms $\mathbf{Z}_i$, and then encode the private vector as $M(\mathbf{x}_i)$. We refer to the data-independent noise generation step as the offline phase. The data-dependent vector encoding and decoding steps in \eqref{encode}-\eqref{decoder} belong to the online phase. Depending on the required covariance structure of $(\mathbf{Z}_1,\dotsc,\mathbf{Z}_n)$, each user $j\in\mathcal{U}_{all}$ exchanges a set of random variables denoted by $\mathcal{I}_j$ with the other users\footnote{For example, $\mathcal{I}_j$ can be the set of random seeds shared by user $j$ with other users when generating $\mathbf{Z}_j$, such that $(\mathbf{Z}_1,\dotsc,\mathbf{Z}_n)$ satisfies some required covariance structure.} in the offline phase to facilitate the generation of $\mathbf{Z}_j$.\footnote{This is a common practice in any correlated privacy mechanism such as SecAgg \cite{secagg,secagg2} to generate the correlated noise prior to data communication, with no trusted party present. In successive DP-DME scenarios, such as federated learning with iterative user updates, the offline phase may be executed a single time to establish all necessary shared randomness for the entire process.} 
In the online phase, user $i$ sends $M(\mathbf{x}_i)$ to the server as per \eqref{encode}, and the server outputs $\hat{\mathbf{S}}_{\mathcal{U}}$. 
In the DP-DME protocol, after both the offline and online phases, the server has access to the encoded vectors of all users $\{M(\mathbf{x}_j)\}_{j=1}^n$ and all the random variables observed by the colluding users $\{\mathcal{I}_j,\mathbf{x}_j,\mathbf{Z}_j\}_{j\in\mathcal{U}_{col}}$. 

Next, we define the privacy constraint considered in this work, which generalizes the privacy definitions used in LDP \cite{minimax_procedures,optimal_algorithms} and distributed DP with SecAgg \cite{ddg,dpsecagg}.


\vspace{0.1cm}

\begin{definition}[Generalized $(\epsilon,\delta)$-DP for DME]\label{def1}
Let $\mathcal{D}_i=\{\mathbf{x}\}_{j\neq i}\cup\mathbf{x}_i$ and $\mathcal{D}'_i=\{\mathbf{x}\}_{j\neq i}\cup\mathbf{x}_i'$  be two neighboring datasets that only differ in the vector of user $i$ for any $i\in\mathcal{U}_{all}\setminus\mathcal{U}_{col}$. Let $\mathcal{G}_i=\{\{M(\mathbf{x}_j)\}_{j\in\mathcal{U}_{all}\setminus\{i\}}, \{\mathbf{x}_j,\mathbf{Z}_j,\mathcal{I}_j\}_{j\in\mathcal{U}_{col}}\}$ denote all the random variables observed by the server from all users except user $i$, for any $i\in\mathcal{U}_{all}\setminus\mathcal{U}_{col}$. For a given $\epsilon\geq0$ and $\delta\in(0,1)$, a DP-DME scheme ensures $(\epsilon,\delta)$-DP if the following is satisfied.
\begin{align}\label{dp_eq}
    \mathbb{P}(M(\mathbf{x}_i)\in\mathcal{A}|\mathcal{G}_i)\leq e^\epsilon\mathbb{P}(M(\mathbf{x}_i')\in\mathcal{A}|\mathcal{G}_i)+\delta,
\end{align}
$\forall \mathcal{D}_i,\mathcal{D}'_i$, $\forall i\in\mathcal{U}_{all}\setminus\mathcal{U}_{col}$ and $\forall\mathcal{A}\subset\mathbb{R}^d$ in the Borel $\sigma$-field.
\end{definition}
To motivate the privacy definition in \eqref{dp_eq}, we discuss an alternative DP definition that compares the observed distributions from two neighboring datasets (comparing the joint distributions as opposed to the conditional ones, following the original DP framework), and show that the privacy constraint in Definition~\ref{def1} is stronger. A detailed explanation is included in Appendix~\ref{privacydef}.
Definition~\ref{def1} generalizes the privacy constraints used in DP-DME with LDP \cite{PrivUnit,optimal_algorithms,minimax_procedures} and distributed DP with SA \cite{ddg,dpsecagg}, and offers a privacy framework to analyze and compare DP-DME mechanisms with arbitrary correlations among users. The proofs of LDP and distributed DP with SA being special cases of Definition~\ref{def1} are given in Appendix~\ref{privacydef}. In this work, we use the privacy constraint in Definition~\ref{def1} to perform fair comparisons of arbitrarily correlated DP mechanisms with the two extreme cases of LDP and distributed DP with SA. In addition, the privacy constraint in Definition~\ref{def1} gives rise to useful geometric interpretations of the privacy-utility trade-offs in DP-DME, as explained in Section~\ref{why}. 


\begin{definition}[$(n,t,c,\epsilon,\delta)$-DP-DME scheme] For a system of $n$ users with up to $c$ colluding users and at least $t$ responding users, and given privacy parameters $(\epsilon, \delta)$, a DP-DME scheme is defined by:
\begin{itemize}
    \item the encoding mechanism:  the joint distribution $\mathcal{D}_Z$ of $(\mathbf{Z}_1,\dotsc,\mathbf{Z}_n)$ that satisfies the gnearlized $(\epsilon, \delta)$-DP for DME as given in Definition~\ref{def1}, and
    \item the decoding mechanism: linear decoding functions $d_{\mathcal{U}}:\mathbb{R}^{d|\mathcal{U}|}\rightarrow\mathbb{R}^d$, for each subset $\mathcal{U}\subseteq \mathcal{U}_{all}$.
\end{itemize}
\end{definition}
The accuracy of an $(n,t,c,\epsilon,\delta)$-DP-DME scheme is measured by the MSE between the mean estimate at the server $\hat{\mathbf{S}}_{\mathcal{U}}$ and the true mean $\mathbf{S}_{\mathcal{U}}=\frac{1}{|\mathcal{U}|}\sum_{i\in{\mathcal{U}}}\mathbf{x}_i$. 

\begin{definition}[MSE of an $(n,t,c,\epsilon,\delta)$-DP-DME scheme]\label{def3} Consider an $(n,t,c,\epsilon,\delta)$ DP-DME  scheme with a joint distribution $\mathcal{D}_Z$ of $(\mathbf{Z}_1,\dotsc,\mathbf{Z}_n)$ and a decoder $d_{\mathcal{U}}$ characterized by $\boldsymbol{\alpha}_{\mathcal{U}}=[\alpha_{j_1},\dotsc,\alpha_{j_{|\mathcal{U}|}}]^T$ for each $\mathcal{U}\subseteq\mathcal{U}_{all}$. For a given $\mathcal{U}\subseteq\mathcal{U}_{all}$, the MSE is defined as,
\begin{align}\label{mse_def}
    &\mathrm{MSE}(\mathcal{D}_Z,\mathcal{U},\boldsymbol{\alpha}_{\mathcal{U}})\nonumber\\
    &\triangleq\sup_{\mathbf{x}_{j_1},\dotsc,\mathbf{x}_{j_{|\mathcal{U}|}}\in\mathbb{B}^d}\mathbb{E}\left[\left\|\hat{\mathbf{S}}_{\mathcal{U}}-\mathbf{S}_{\mathcal{U}}\right\|^2\right]\\
    &=\sup_{\mathbf{x}_{j_1},\dotsc,\mathbf{x}_{j_{|\mathcal{U}|}}\in\mathbb{B}^d}\mathbb{E}\left[\left\|\frac{1}{|\mathcal{U}|}\sum_{i\in\mathcal{U}} \alpha_iM(\mathbf{x}_i)-\frac{1}{|\mathcal{U}|}\sum_{i\in{\mathcal{U}}}\mathbf{x}_i\right\|^2\right]
\end{align}
The expectation is over $\mathcal{D}_Z$, and $\|\cdot\|$ denotes the $L_2$ norm.
\end{definition}
The goal of this work is to find the optimum $(n,t,c,\epsilon,\delta)$-DP-DME scheme that minimizes the MSE in Definition~\ref{def3} for the worst case dropouts and colluding users. Specifically, we characterize the minimum MSE given by,
\begin{align}
    &\mathrm{MMSE}\nonumber\\
    &\triangleq \inf_{\mathcal{D}_Z}\quad
    \max_{\substack{\mathcal{U}\subseteq\mathcal{U}_{all}\\ t\leq|\mathcal{U}|\leq n}} \quad\inf_{\boldsymbol{\alpha}_{\mathcal{U}}}\quad\mathrm{MSE}(\mathcal{D}_Z,\mathcal{U},\boldsymbol{\alpha}_{\mathcal{U}})\\
    &=\inf_{\mathcal{D}_Z}\quad
    \max_{\substack{\mathcal{U}\subseteq\mathcal{U}_{all}\\ t\leq|\mathcal{U}|\leq n}} \quad\inf_{\boldsymbol{\alpha}_{\mathcal{U}}}\nonumber\\
    &\quad\sup_{\mathbf{x}_{j_1},\dotsc,\mathbf{x}_{j_{|\mathcal{U}|}}\in\mathbb{B}^d}\mathbb{E}\left[\left\|\frac{1}{|\mathcal{U}|}\sum_{i\in\mathcal{U}} \alpha_iM(\mathbf{x}_i)-\frac{1}{|\mathcal{U}|}\sum_{i\in{\mathcal{U}}}\mathbf{x}_i\right\|^2\right]\label{main_opt}
\end{align} 
while satisfying the privacy constraint in Definition~\ref{def1} for any $\mathcal{U}_{col}\subset\mathcal{U}_{all}$ with $|\mathcal{U}_{col}|\leq c$. The parameters to be optimized in \eqref{main_opt} are the joint distribution $\mathcal{D}_Z$ of $(\mathbf{Z}_1,\dotsc,\mathbf{Z}_n)$ considering the worst case dropouts and colluding users, and the decoder $\boldsymbol{\alpha}_{\mathcal{U}}$ for each $\mathcal{U}\subseteq\mathcal{U}_{all}$.\footnote{We show that the optimum $\alpha_{\mathcal{U}}$ only depends on $|\mathcal{U}|$ and not directly on each $\mathcal{U}$ (See Proposition~\ref{vec_opt_dec} for details).} The $\max$ and $\sup$ terms in \eqref{main_opt} characterize the worst case dropouts and the private vectors of the users, that give the worst case MSE. In this paper, we establish a minimax result by minimizing the maximum MSE. The order of optimizations in \eqref{main_opt} is explained as follows. The noise distribution $\mathcal{D}_Z$ is determined prior to the mean estimation step in \eqref{decoder}, and is fixed irrespective of the number of responding users at a given time. Therefore, the problem is formulated to optimize $\mathcal{D}_Z$ for the worst case dropouts and colluding users. The decoder on the other hand is used at the time of estimation as shown in \eqref{decoder}, and is optimized based on the number of responding users for any fixed $\mathcal{D}_Z$. This allows the decoder to make use of the information on the set of responding users at any given time. 

\vspace{0.2cm}

\section{Correlated Gaussian Mechanism for DP-DME: A Geometric Interpretation}\label{why}

In this section, we illustrate the privacy-utility trade-off of DP-DME using a geometric approach. We show how carefully tuned (anti) correlated noise can significantly outperform DME with independent noise (LDP) while ensuring the same level of privacy, with no dropouts. This also presents a re-interpretation of the SecAgg scheme --- but  over $\mathbb{R}^d$ in the DP framework rather than over finite fields. Building up on this idea, in the subsequent sections of this paper, we analyze the privacy-utility trade-offs even with the presence of dropouts and/or collusions, and show that carefully tuned correlated noise continues to provide superior performance to independent noise (see Section~\ref{results} for the results).


\begin{figure*}[t]
    \centering
    \includegraphics[scale=0.62]{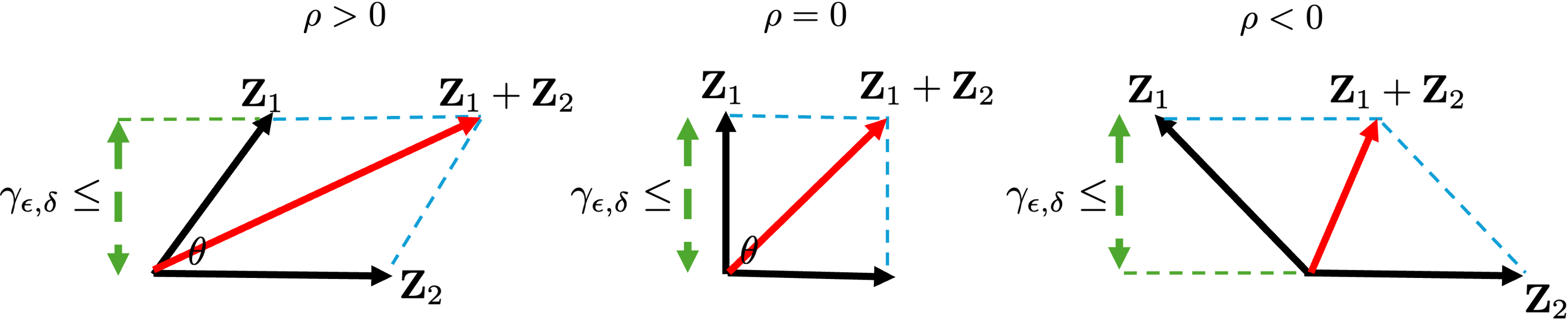}
    \caption{Geometric interpretation of the privacy-utility trade-off in DP-DME for different correlation coefficients among the users' noise distributions: Noise vectors $\mathbf{Z}_1$ and $\mathbf{Z}_2$ are represented as vectors in $\mathcal{H}$ with magnitude $\sigma \sqrt{d}$ and angle $\theta=\cos^{-1}\rho$ between them. The privacy constraint on $\mathbf{x}_i$ enforces that the orthogonal component of $\mathbf{Z}_i$ relative to $\mathbf{Z}j$ (for $i \neq j$) has a magnitude bounded below by a constant $\gamma_{\epsilon, \delta}$. The MSE is proportional to the magnitude of $\mathbf{Z}_1+\mathbf{Z}_2$.}
    \label{privacy_fig}
\end{figure*}

Consider a simple two-user setting with no dropouts and no colluding users, where the private vectors of the two users are given by $\mathbf{x}_1$ and $\mathbf{x}_2$, with $\mathbf{x}_1,\mathbf{x}_2\in\mathbb{B}^{d}$. User $i$ sends $M(\mathbf{x}_i)=\mathbf{x}_i+\mathbf{Z}_i$ to the server, where $\mathbf{Z}_i\sim\mathcal{N}(\mathbf{0}_d,\sigma^2\mathsf{I}_d)$ for $i=1,2$, with $\mathbf{0}_d$, $\mathsf{I}_d$, and  $\sigma^2$ representing the all zeros vector of size $d\times1$, identity matrix of size $d\times d$, and a constant, respectively. The elements of $\mathbf{Z}_1$ and $\mathbf{Z}_2$ are correlated as follows:
\begin{align}
    \mathbb{E}[Z_{1,j}Z_{2,k}]=\begin{cases}
        \rho\sigma^2, & j=k,\quad j,k\in\{1,\dotsc,d\}\\
        0, & j\neq k,\quad j,k\in\{1,\dotsc,d\},
    \end{cases}
\end{align}
where $Z_{i,j}$ is the $j$-th element of $\mathbf{Z}_i$ for $i\in\{1,2\}$, $j\in\{1,\dotsc,d\}$ and $\rho$ is the correlation coefficient between $Z_{1,k}$ and $Z_{2,k}$. For simplicity, let the decoder be $\boldsymbol{\alpha}$=[1,1]. Therefore, the server's estimation of the mean of the vectors of the two users is given by $\hat{S}=\frac{1}{2}(\mathbf{x}_1+\mathbf{x}_2+\mathbf{Z}_1+\mathbf{Z}_2)$. The estimation error is $\frac{1}{2}(\mathbf{Z}_1+\mathbf{Z}_2)$, which is quantified in terms of the MSE given by $\frac{1}{4}\mathbb{E}[\|\mathbf{Z}_1+\mathbf{Z}_2\|^2]$. The goal is to minimize $\mathbb{E}[\|\mathbf{Z}_1+\mathbf{Z}_2\|^2]$ while satisfying the privacy constraint in \eqref{dp_eq}.

To define the problem geometrically, consider the following vector representation. Let $\mathcal{H}$ be an inner product space consisting of all Gaussian random variables of dimension $d$. For any $A,B\in\mathcal{H}$, let the inner product be defined as $\langle A,B\rangle_{\mathcal{H}}=\mathbb{E}[A^TB]$. With this definition, the noise vectors $\mathbf{Z}_1,\mathbf{Z}_2$ in this example are represented as vectors with magnitude:
\begin{align}
    \|\mathbf{Z}_i\|_{\mathcal{H}}=\sqrt{\langle \mathbf{Z}_i,\mathbf{Z}_i\rangle_{\mathcal{H}}}=\sqrt{\mathbb{E}[\mathbf{Z}_i^T\mathbf{Z}_i]}=\sqrt{d\sigma^2}.
\end{align}
The angle between $\mathbf{Z}_1$ and $\mathbf{Z}_2$ in $\mathcal{H}$ (denoted by $\theta$) corresponds to the correlation coefficient $\rho$ as:
\begin{align}
    \langle \mathbf{Z}_1,\mathbf{Z}_2\rangle_{\mathcal{H}}=d\sigma^2\cos\theta=\mathbb{E}[\mathbf{Z}_1^T\mathbf{Z}_2]=\rho\sigma^2d
\end{align}
Now, the MSE given by $\frac{1}{4}\mathbb{E}[\|\mathbf{Z}_1+\mathbf{Z}_2\|^2]$ is represented as $\frac{1}{4}\|\mathbf{Z}_1+\mathbf{Z}_2\|_{\mathcal{H}}^2$, and minimizing MSE is equivalent to minimizing $\|\mathbf{Z}_1+\mathbf{Z}_2\|_{\mathcal{H}}^2$.\footnote{Recall that $\|\cdot\|$ and $\|\cdot\|_{\mathcal{H}}$ denote the $L_2$ norm and vector norm in $\mathcal{H}$, respectively.} 


Next, we illustrate the privacy constraint. For this example, \eqref{dp_eq} simplifies to,
\begin{align}\label{egpri}
    \mathbb{P}(M(\mathbf{x}_i)\in\mathcal{A}|M(\mathbf{x}_j))\leq e^\epsilon\mathbb{P}(M(\mathbf{x}_i')\in\mathcal{A}|M(\mathbf{x}_j))+\delta,
\end{align} 
for $i,j\in\{1,2\}$, $i\neq j$. To geometrically illustrate the privacy constraint, define,
\begin{align}
   \boldsymbol{Z}_{i}^{\parallel}: = \frac{\langle \boldsymbol{Z}_{i},\boldsymbol{Z}_{j}\rangle_{\mathcal{H}} \boldsymbol{Z}_{j}}{\|\boldsymbol{Z}_{j}\|_{\mathcal{H}}^2} = (\cos\theta) \boldsymbol{Z}_{j} 
\end{align}
and $\boldsymbol{Z}_{i}^{\bot} := \boldsymbol{Z}_{i}-\boldsymbol{Z}_{j}^{\parallel}$
to be the components of $\boldsymbol{Z}_i$ in $\mathcal{H}$, that are parallel and orthogonal to $\boldsymbol{Z}_j$, respectively, for $j\neq i$.
The privacy constraint on $\boldsymbol{x}_i$ in \eqref{egpri} simplifies to:
\begin{align}\label{geopriv}
    \|\boldsymbol{Z}_i^\bot\|_{\mathcal{H}} \geq\gamma_{\epsilon,\delta},\quad i\in\{1,2\}
\end{align}
where $\gamma_{\epsilon,\delta}$ is a constant that depends on the given $\epsilon$ and $\delta$. The intuition behind this constraint is as follows. $\mathbf{Z}_i^\bot$ denotes the component of $\mathbf{Z}_i$ that remains independent of $\mathbf{Z}_j$. The uncertainty in $\mathbf{Z}_i^\bot$ is critical for preserving the privacy of $\mathbf{x}_i$, thus requiring a lower bound on its variance (See Appendix~\ref{addwhy} for a rigorous proof).

Our goal of minimizing MSE while satisfying the privacy constraint is equivalent to minimizing $\|\mathbf{Z}_1+\mathbf{Z}_2\|_{\mathcal{H}}^2$ (red vectors in Fig.~\ref{privacy_fig}) while preserving $\|\boldsymbol{Z}_i^\bot\|_{\mathcal{H}}\geq\gamma_{\epsilon,\delta}$ (green dotted line in Fig.~\ref{privacy_fig}). The core insight of CorDP-DME is to find the best $\sigma$ (the norm of $\mathbf{Z}_1$ and $\mathbf{Z}_2$) and $\rho$ (the angle between $\mathbf{Z}_1$ and $\mathbf{Z}_2$). 

To gain insight into how $\rho$ and $\sigma$ affect the MSE for a given level of privacy, let us start with the simplest case where $\rho = 0$. Since we assume Gaussian noise, $\rho = 0$ corresponds to independent noise. Hence, this is simply LDP: adding sufficient noise at each user by setting $\| \boldsymbol{Z}_i\|_{\mathcal{H}} \geq \gamma_{\epsilon,\delta}$, without any correlation in the noise. Due to this lack of correlation, the angle between the noise vectors is always a right angle. As a result, $\| \boldsymbol{Z}_1 + \boldsymbol{Z}_2 \|_{\mathcal{H}}^2 = 2 \cdot d \sigma^2$. When there are more than two users ($n > 2$), it is straightforward to see that the total noise will be $n \cdot d\sigma^2$.

Now, let us consider the case when $\rho \neq 0$. When $\rho > 0$ (the leftmost plot in Fig.\ref{privacy_fig}), this results in $\theta < \pi/2$. At the same time, to maintain $\| \boldsymbol{Z}_i \|_{\mathcal{H}} \geq \gamma_{\epsilon, \delta}$, the norms of $\mathbf{Z}_1$ and $\mathbf{Z}_2$ must increase. These two factors together increase $\|\mathbf{Z}_1 + \mathbf{Z}_2 \|_{\mathcal{H}}^2$, which is undesirable. However, when $\rho < 0$ (the rightmost plot in Fig.\ref{privacy_fig}), which corresponds to $\theta > \pi/2$, the anti-correlated components of $\mathbf{Z}_1$ and $\mathbf{Z}_2$ cancel out, making $\| \mathbf{Z}_1 + \mathbf{Z}_2 \|_{\mathcal{H}}^2$ smaller than when $\rho = 0$. Note that $\sigma$ still has to grow in this case to meet the privacy requirement.

\begin{figure*}[t]
    \centering
    \includegraphics[scale=0.65]{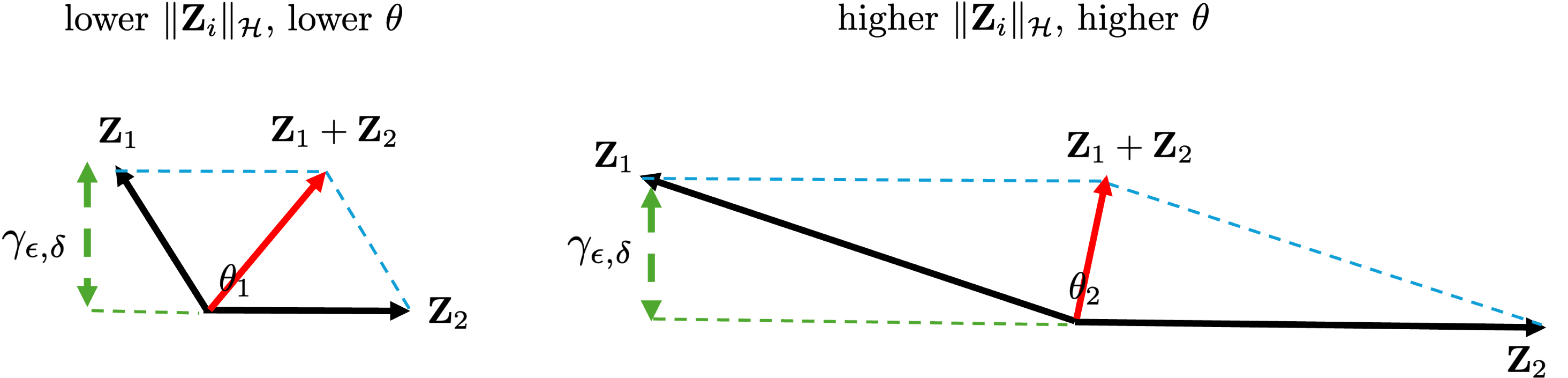}
    \caption{Variation of the MSE with changing noise parameters $\sigma^2$ and $\rho$: Increasing $\|\mathbf{Z}_i\|_{\mathcal{H}}=\sigma\sqrt{d}$ and $\theta=\cos^{-1}\rho$ while maintaining the orthogonal distance between $\mathbf{Z}_1$ and $\mathbf{Z}_2$ at $\gamma_{\epsilon,\delta}$ for privacy, decreases $\|\mathbf{Z}_1+\mathbf{Z}_2\|_{\mathcal{H}}$ (the MSE).}
    \label{mag}
\end{figure*}

As we understand how anti-correlated noise can reduce MSE while satisfying the same privacy constraint, let us consider how best we can design the anti-correlation. As you can see in Fig.~\ref{mag}, as  $\theta$ increases, $\sigma$ must also grow, but 
$\| \mathbf{Z}_1 + \mathbf{Z}_2 \|_{\mathcal{H}}^2$ becomes progressively smaller. In fact, when $\theta\to\pi$ and $\sigma\to\infty$, $\mathbf{Z}_1 + \mathbf{Z}_2$ becomes almost perpendicular to the individual noise vectors $\mathbf{Z}_1$ and $\mathbf{Z}_2$. In this limit, $\| \mathbf{Z}_1 + \mathbf{Z}_2 \|_{\mathcal{H}}^2$ will approach $\gamma_{\epsilon,\delta}^2$. Note that in this case, the MSE of the sum is the same as if there were only one user; the noise from two users does not add up as it does in the LDP case. It is easy to generalize that even with $n$ users, the total MSE given by $\|\sum_{i=1}^n\mathbf{Z}_i\|^2_{\mathcal{H}}$ still approaches $\gamma_{\epsilon,\delta}^2$, and remains independent of $n$. In fact, this MSE is the same as in CDP! (See Appendix~\ref{opt_para} for rigorous proofs).

In essence, this example shows that for any given privacy parameters $\epsilon,\delta$, the optimized correlated Gaussian mechanism can achieve the same utility as CDP even without a trusted server. However, the correlated Gaussian scheme described above faces a significant shortcoming. If a user drops out, the residual noise of the remaining user/s cannot be canceled resulting in significantly high MSEs; this is because it uses noise with arbitrarily large variance (recall that the lowest possible MSE is achieved when $\|\mathbf{Z}_i\|_{\mathcal{H}}^2 = E[||\mathbf{Z}_{i}||^2] \to\infty$). SecAgg---one of the main correlated privacy mechanisms used in practice---faces the same shortcoming. In SecAgg, users' data is quantized to a finite field, and then a noise random variable that is uniformly distributed over the field elements is added to the quantized data and sent to the server. If a user drops out, this added noise cannot be canceled and the server input from the remaining users is statistically independent of the users' data, which in effect results in a large MSE. SecAgg circumvents dropouts through additional rounds where remaining users' share these non-canceled noise variables with the server enabling the server to cancel it. In fact, SecAgg (and its variants like GOPA \cite{dp_corr2}) is a complex protocol that requires more additional rounds to guard against further dropouts, data leaks and malicious behaviour in the previous rounds. Alternatively, this geometric interpretation offers a more straightforward method for addressing user dropouts, removing the necessity for additional communication rounds by appropriately adjusting the noise distribution. This gives rise to the CorDP-DME protocol introduced in this paper. In CorDP-DME we optimize the noise parameters---$\rho$ and $\sigma^2$---directly accounting for dropouts and colluding users (up to a threshold). As our analysis demonstrates, this approach significantly improves upon LDP in terms of MSE, while avoiding additional rounds of communication. In fact, in Section~\ref{results}, we show that for any dropout threshold, CorDP-DME outperforms LDP with independent Gaussian noise.  

\section{Proposed Approach: CorDP-DME Protocol}\label{proposed}

In this section, we provide a comprehensive overview of CorDP-DME, along with a numerical example demonstrating the optimal noise parameters under different scenarios involving user dropouts and collusion. CorDP-DME operates in two phases: a data-independent offline phase, where correlated noise is generated (details in Section~\ref{offline}), and a data-dependent online phase. During the online phase, each user sends the encoded vectors from \eqref{encode} to the server, which then computes the mean estimate using \eqref{decoder}, completing the protocol. Similar to LDP-based DME, this approach remains simple and straightforward. Despite using correlated noise, CorDP-DME requires no additional communication rounds or computations to manage dropouts. The encoding and decoding processes in \eqref{encode}-\eqref{decoder} are designed to handle dropouts and collusion seamlessly within the same round, while still ensuring privacy guarantees. Consequently, the primary focus of CorDP-DME is the design of the encoding and decoding functions, which is obtained by solving the optimization problem in \eqref{main_opt}.

In this paper, we study the case where $\mathcal{D}_Z$ in \eqref{main_opt} is multivariate Gaussian. Specifically, we choose,
\begin{align}
    \mathbf{Z}_i\sim\mathcal{N}(\mathbf{0}_d,\sigma^2\mathsf{I}_d),\quad i\in[1:n]
\end{align}
with the all zeros vector of size $d\times1$ denoted by $\mathbf{0}_d$ and the identity matrix of size $d\times d$ denoted by $\mathsf{I}_d$. The $k$th element of $\mathbf{Z}_i$ is denoted by $Z_{i,k}$ for $i\in[1:n]$ and $k\in[1:d]$. The $Z_{i,k}$s are distributed as: 
\begin{align}\label{cov_str}
    \begin{pmatrix}
        Z_{1,k}\\Z_{2,k}\\\vdots\\Z_{n,k}
    \end{pmatrix}\sim\mathcal{N}\left(\mathbf{0}_n,\sigma^2\begin{pmatrix}
        1 & \rho  & \dotsc & \rho\\
        \rho & 1  & \dotsc & \rho\\
        \vdots & \vdots & \vdots & \vdots\\
        \rho & \rho  & \dotsc & 1
    \end{pmatrix}_{n\times n}\right)
\end{align}
for all $k\in[1:d]$. Moreover, we let $\mathbb{E}[Z_{i,k}Z_{j,k'}]=0$, $\forall i,j$, $\forall k\neq k'$.
We denote this class of distributions as $\mathcal{D}_Z^G$. In Section~\ref{results}, we present the solution to $\eqref{main_opt}$ for $D_Z=D_Z^G$. Furthermore, for the case of unbiased mean estimates, we show that the covariance structure in $D_Z^G$ is the optimal among all covariance matrices corresponding to all Gaussian distributions (see Theorem~\ref{converse}). 

Next, we provide the main results of this paper, that characterize the optimal encoding decoding functions and the corresponding minimum MSEs.


\subsection{Main Results}\label{results}

In this section, we provide the solution to \eqref{main_opt} while satisfying the privacy constraint in Definition~\ref{def1}, for the class of multivariate Gaussian distributions $\mathcal{D}_Z^G$ specified in \eqref{cov_str}. We present the optimum decoder and the optimum noise distribution, along with the corresponding minimum MSE for the general case with arbitrary $n,t,c,\epsilon,\delta$ in Theorem~\ref{thm1}. Subsequently, in Corollaries~\ref{cor1}-\ref{cor3}, we derive the MSEs of the special cases, with specific $t$ and $c$. Furthermore, Theorems~\ref{unbiased} and~\ref{converse} provide further achievability and converse results for unbiased estimates. The following notation is used throughout the paper.
\begin{align}\label{main_notation}
    &\mathrm{MSE}(\sigma^2,\rho,\mathcal{U},\boldsymbol{\alpha}_{\mathcal{U}})\nonumber\\
    &=\sup_{\mathbf{x}_{j_1},\dotsc,\mathbf{x}_{j_{|\mathcal{U}|}}\in\mathbb{B}^d}\mathbb{E}\left[\left\|\frac{1}{|\mathcal{U}|}\sum_{i\in\mathcal{U}}\alpha_i(\mathbf{x}_i+\mathbf{Z}_i)-\frac{1}{|\mathcal{U}|}\sum_{i\in\mathcal{U}}\mathbf{x}_i\right\|^2\right]
\end{align}
In Proposition~\ref{vec_opt_dec}, we state the optimum decoder, $\boldsymbol{\alpha}^*_{\mathcal{U}}$, (with $\alpha_i^*$ as its $i$th component) that minimizes the MSE for any set of responding users $\mathcal{U}$, and any given noise distribution characterized by $\sigma^2$ and $\rho$. 

\begin{proposition}[Optimum decoder]\label{vec_opt_dec} For any fixed $\mathcal{U}\subseteq\mathcal{U}_{all}$ satisfying $|\mathcal{U}|\geq t$, and for any $\sigma^2$, $\rho$, the optimum decoder is:
 \begin{align}\label{opt_decoder}
        \boldsymbol{\alpha}^*_{\mathcal{U}}&=\arg\min_{\boldsymbol{\alpha}_{\mathcal{U}}}\mathrm{MSE}(\sigma^2,\rho,\mathcal{U},\boldsymbol{\alpha}_{\mathcal{U}})\\
        &=\frac{1}{1+\frac{d\sigma^2}{|\mathcal{U}|}(1+\rho(|\mathcal{U}|-1))}\mathbf{1}_{|\mathcal{U}|}.
\end{align}
The corresponding MSE is given by,
\begin{align}
    \min_{\boldsymbol{\alpha}_{\mathcal{U}}}\mathrm{MSE}(\sigma^2,\rho,\mathcal{U},\boldsymbol{\alpha}_{\mathcal{U}})\!=\!\left(1\!+\!\frac{|\mathcal{U}|/d}{\sigma^2(1+\rho(|\mathcal{U}|\!-\!1))}\right)^{-1}
\end{align}
\end{proposition}
The proof of Proposition~\ref{vec_opt_dec} is given in Appendix~\ref{pfprop1}. Next, we provide Theorem~\ref{thm1}, which characterizes the MMSE in \eqref{main_opt} for $\mathcal{D}_Z=\mathcal{D}_Z^G$ while ensuring the privacy constraint in Definition~\ref{def1}, for any $t$ and $c$. Note that finding the optimum $\mathcal{D}_Z^G$ in \eqref{main_opt} is equivalent to optimizing $\rho$ and $\sigma^2$ that minimizes the MSE while satisfying the privacy constraint in Definition~\ref{def1}, based on the structure of $\mathcal{D}_Z^G$ specified in \eqref{cov_str}.

\vspace{0.1cm}
\begin{theorem}[Optimum noise distribution]\label{thm1} For any given $\epsilon>0$, $\delta\in(0,1)$, $t\leq n$ and $c<t$, the optimum $\sigma^2$ and $\rho$ that solves \eqref{main_opt} for $\mathcal{D}_Z=\mathcal{D}_Z^G$  while satisfying the generalized privacy constraint in Definition~\ref{def1} for any $\mathcal{U}_{col}\subset\mathcal{U}_{all}$ with $|\mathcal{U}_{col}|\leq c$ is characterized by,
\begin{align}
(\sigma^2_*,\rho_*)=\arg\min_{\sigma^2,\rho}\max_{\substack{\mathcal{U}\subseteq\mathcal{U}_{all}\\ t\leq|\mathcal{U}|\leq n}} \min_{\boldsymbol{\alpha}_{\mathcal{U}}}\mathrm{MSE}(\sigma^2,\rho,\mathcal{U},\boldsymbol{\alpha}_{\mathcal{U}})\nonumber    
\end{align}
\begin{align}\label{opt_sig}
    \sigma^2_*&\!=\!\begin{cases}
        \infty, & t=n\\
        \frac{\sigma^2_{\epsilon,\delta}(n^2-2n-cn+2)}{(n-c)^2}\\
    +\frac{\sigma^2_{\epsilon,\delta}(n-c-1)(n+c-2nc+t(n+c-2))}{(n-c)^2\sqrt{(t-c)(n-t)(n-c-1)}}, & c\!<\!t\!<\!n.
    \end{cases}\\
    \rho_*&\!=\!\begin{cases}
    -\frac{\sigma^2_*-\sigma^2_{\epsilon,\delta}}{\sigma^2_*(n-1)-\sigma^2_{\epsilon,\delta}(n-2)}, & c=1\\
        \frac{-(n-2)\left(1-\frac{\sigma^2_{\epsilon,\delta}}{\sigma^2_*}\right)-c}{2(n-1)(c-1)}\\
    \quad+\frac{\!\sqrt{\left(\!(n-2)\left(\!1-\frac{\sigma^2_{\epsilon,\delta}}{\sigma^2_*}\right)-c\!\right)^2\!\!+4(n-c-1)\left(1-\frac{\sigma^2_{\epsilon,\delta}}{\sigma^2_*}\!\right)}}{2(n-1)(c-1)}, & c\neq1
    \end{cases}\label{opt_r}
    \end{align}
with $\sigma_{\epsilon,\delta}\!=\!\inf_{\hat{\sigma}>0}\{\hat{\sigma};\Phi\!\left(\frac{1}{\hat{\sigma}}\!-\!\frac{\epsilon\hat{\sigma}}{2}\right)-e^\epsilon\Phi\!\left(-\frac{1}{\hat{\sigma}}\!-\!\frac{\epsilon\hat{\sigma}}{2}\right)\leq\delta\}$, where $\Phi$ is the standard Gaussian CDF. The resulting minimum MSE is given by,
\begin{align}
    \min_{\sigma^2,\rho}\max_{\substack{\mathcal{U}\subseteq\mathcal{U}_{all}\\ t\leq|\mathcal{U}|\leq n}} &\min_{\boldsymbol{\alpha}_{\mathcal{U}}}\mathrm{MSE}(\sigma^2,\rho,\mathcal{U},\boldsymbol{\alpha}_{\mathcal{U}})\nonumber\\
    &=\left(1+\frac{t/d}{\sigma^2_*(1+\rho_*(t-1))}\right)^{-1}\label{dropcollmmse}
\end{align}
\end{theorem}
The proof of Theorem~\ref{thm1} is given in Appendix~\ref{pfthm1}. For a fixed $\epsilon$ and $\delta$, the value of $\sigma^2_{\epsilon,\delta}$ in Theorem~\ref{thm1} is the minimum variance of the Gaussian noise added to achieve $(\epsilon,\delta)$-DP in the standard Gaussian mechanism with an $L_2$ sensitivity of $2$. For any $\epsilon,\delta,t$ and $c$, Theorem~\ref{thm1} shows that $\rho_*\leq0$ always holds. This implies that the (anti) correlated Gaussian mechanism outperforms (or performs equally when $c=t-1$) the independent Gaussian mechanism for any $\epsilon,\delta,t,c$. Next, we upper bound $\sigma^2_{\epsilon,\delta}$ using the results from \cite{gauss,reviewGauss,DP1} to better interpret the dependency of the MSE on $\epsilon$ and $\delta$. Generally, we assume that $\delta=10^{-5}$.
\begin{proposition}[Bounds on $\sigma^2_{\epsilon,\delta}$]\label{ubsnodrop}
    The following upper bounds hold for $\sigma^2_{\epsilon,\delta}$, where $\eta=1+2\sqrt{\ln\frac{1}{2\delta}}$ for $\delta\in(0,0.05]$ and $\eta=1+2\sqrt{\ln10}$ for $\delta\in(0.05,1)$.
    \begin{align}\label{ubs}
        \sigma^2_{\epsilon,\delta}&\leq\begin{cases}
            \frac{8\ln(1.25/\delta)}{\epsilon^2}, & \epsilon,\delta\in(0,1)\\
            \frac{2\eta^2}{\epsilon}, & \epsilon\geq1, \delta\in(0,1)
        \end{cases}
    \end{align}
       
\end{proposition}
The proof of Proposition~\ref{ubsnodrop} is given in Appendix~\ref{pfcoro1}. In Corollaries~\ref{cor1}-~\ref{cor3}, we consider special cases of Theorem~\ref{thm1}, and provide simplified MSE results using the bounds in Proposition~\ref{ubsnodrop}. 
\begin{corollary}[Without collusion, Without dropouts]\label{cor1}
    For $t=n$, $c=0$, the minimum MSE while satisfying $(\epsilon,\delta)$-DP in Definition~\ref{def1} with $\mathcal{U}_{col}=\emptyset$ is given by,
    \begin{align}
    \min_{\sigma^2,\rho,\boldsymbol{\alpha}_{\mathcal{U}_{all}}}\mathrm{MSE}&(\sigma^2,\rho,\mathcal{U}_{all},\boldsymbol{\alpha}_{\mathcal{U}_{all}})\nonumber\\
    &=\begin{cases}
        O\left(\frac{d}{n^2}\frac{\ln(1/\delta)}{\min\{\epsilon,\epsilon^2\}}\right), & \text{if } n^2>>d\\
        O(1), & otherwise.
    \end{cases} \nonumber
\end{align}
\end{corollary}

\begin{corollary}[Without collusion, with dropouts]\label{cor2}
    For $t<n$ and $c=0$, the minimum MSE while satisfying $(\epsilon,\delta)$-DP in Definition~\ref{def1} with $\mathcal{U}_{col}=\emptyset$ is given by,
    \begin{align}
    \min_{\sigma^2,\rho}&\max_{\substack{\mathcal{U}\subset[1:n]\\ t\leq|\mathcal{U}|< n}} \min_{\boldsymbol{\alpha}_{\mathcal{U}}}\mathrm{MSE}(\sigma^2,\rho,\mathcal{U},\boldsymbol{\alpha}_{\mathcal{U}})\nonumber\\
    &\quad=\begin{cases}
        O\left(\frac{d(n-t)}{tn}\frac{\ln(1/\delta)}{\min\{\epsilon,\epsilon^2\}}\right), & \text{if }\frac{nt}{n-t}>>d\\
        O(1), & otherwise.
    \end{cases}\label{dropoutmmse}
\end{align}
\end{corollary}
\begin{corollary}[With collusion, without dropouts]\label{cor3}
    For $t=n$ and $c>0$ the minimum MSE while satisfying $(\epsilon,\delta)$-DP in Definition~\ref{def1} for any $\mathcal{U}_{col}\subset\mathcal{U}_{all}$ with $|\mathcal{U}_{col}|\leq c$ is given by,
    \begin{align}
    &\min_{\sigma^2,\rho,\boldsymbol{\alpha}_{\mathcal{U}_{all}}}\mathrm{MSE}(\sigma^2,\rho,\mathcal{U}_{all},\boldsymbol{\alpha}_{\mathcal{U}_{all}})\nonumber\\
    &\quad=\begin{cases}
        O\left(\frac{d}{n(n-c)}\frac{\ln(1/\delta)}{\min\{\epsilon,\epsilon^2\}}\right), & \text{if }n(n-c)>>d\\
        O(1), & otherwise.
    \end{cases}
\end{align}
\end{corollary}
For the case of no dropouts and no collusions, CorDP-DME achieves the same order of MSE as CDP (and DP-DME with SecAgg) for any $\epsilon$ and $\delta$, as shown in Corollary~\ref{cor1}. In the absence of dropouts, CorDP-DME reduces the MSE compared to LDP by a factor of $O(n)$ without collusion, and by $O(n-c)$ with collusion, as shown in Corollaries~\ref{cor1} and ~\ref{cor3}.

For the case of dropouts, CorDP-DME achieves MSEs that lie in between LDP and CDP (Corollary~\ref{cor2} and Section~\ref{experiments}).
If the maximum number of dropouts is $O(n^p)$ for any $p<1$, Corollary~\ref{cor2} implies that the minimum MSE is $O\left(\frac{d\sigma^2_{\epsilon,\delta}}{tn^{1-p}}\right)$, which has a scaling advantage over the case with independent noise, i.e., $\rho=0$, that has a minimum MSE of $O\left(\frac{d\sigma^2_{\epsilon,\delta}}{t}\right)$.

The optimum decoder in Proposition~\ref{vec_opt_dec} results in a biased estimate of the mean. As we assume linear decoders $\boldsymbol{\alpha}_{\mathcal{U}}$ that are independent of $M(\mathbf{x}_i)$, $\forall i$ and zero mean Gaussian noise for the privacy mechanism, an unbiased estimate is obtained if and only if $\boldsymbol{\alpha}_{\mathcal{U}}=\mathbf{1}_{\mathcal{U}}$ for any $\mathcal{U}\subseteq\mathcal{U}_{all}$, in \eqref{decoder}. Theorem~\ref{unbiased} characterizes the optimum noise distribution and the resulting MSE for the case of unbiased estimates.

\begin{theorem}[Unbiased mean estimate]\label{unbiased} Let $\tilde{\mathbf{S}}_{\mathcal{U}}=\frac{1}{|\mathcal{U}|}\sum_{i\in\mathcal{U}}(\mathbf{x}+\mathbf{Z}_i)$ be an unbiased estimate of the true mean $\mathbf{S}_\mathcal{U}=\frac{1}{|\mathcal{U}|}\sum_{i\in\mathcal{U}}\mathbf{x}_i$, where $(\mathbf{Z}_1,\dotsc,\mathbf{Z}_n)$ follows distributions from $\mathcal{D}_Z^G$. For any given $\epsilon>0$, $\delta\in(0,1)$, $t\leq n$ and $c<t$, the $\sigma^2_*$ and $\rho_*$ given in \eqref{opt_sig} and \eqref{opt_r} satisfy,
    \begin{align}\label{unbiased_eq}
(\sigma^2_*,\rho_*)&=\arg\min_{\sigma^2,\rho}\quad\max_{\substack{\mathcal{U}\subseteq\mathcal{U}_{all}\\ t\leq|\mathcal{U}|\leq n}}\quad\sup_{\mathbf{x}_{j_1},\dotsc,\mathbf{x}_{j_{|\mathcal{U}|}}\in\mathbb{B}^d}\nonumber\\
&\quad\mathbb{E}\left[\left\|\frac{1}{|\mathcal{U}|}\sum_{i\in\mathcal{U}}(\mathbf{x}_i+\mathbf{Z}_i)-\frac{1}{|\mathcal{U}|}\sum_{i\in\mathcal{U}}\mathbf{x}_i\right\|^2\right]
\end{align}
while satisfying $(\epsilon,\delta)$-DP in Definition~\ref{def1} for any $\mathcal{U}_{col}\subset\mathcal{U}_{all}$ with $|\mathcal{U}_{col}|\leq c$. The resulting MSE is given by:
\begin{align}
    \min_{\sigma^2,\rho}\quad\max_{\substack{\mathcal{U}\subseteq\mathcal{U}_{all}\\ t\leq|\mathcal{U}|\leq n}} &\sup_{\mathbf{x}_{j_1},\dotsc,\mathbf{x}_{j_{|\mathcal{U}|}}\in\mathbb{B}^d}\nonumber\\
&\mathbb{E}\left[\left\|\frac{1}{|\mathcal{U}|}\sum_{i\in\mathcal{U}}(\mathbf{x}_i+\mathbf{Z}_i)-\frac{1}{|\mathcal{U}|}\sum_{i\in\mathcal{U}}\mathbf{x}_i\right\|^2\right]\nonumber\\
    &=\frac{d\sigma^2_*(1+\rho_*(t-1))}{t}
\end{align}
\end{theorem}

The proof of Theorem~\ref{unbiased} is given in Appendix~\ref{pfunbiased}. The simplified MSEs of the unbiased case corresponding to the settings in Corollaries~\ref{cor1}-\ref{cor3} are given by $O\left(\frac{d}{n^2}\frac{\ln(1/\delta)}{\min\{\epsilon,\epsilon^2\}}\right)$, $O\left(\frac{d(n-t)}{tn}\frac{\ln(1/\delta)}{\min\{\epsilon,\epsilon^2\}}\right)$
and $O\left(\frac{d}{n(n-c)}\frac{\ln(1/\delta)}{\min\{\epsilon,\epsilon^2\}}\right)$, respectively. 

\vspace{0.2cm}

The results in Theorems~\ref{thm1} and~\ref{unbiased} are based on the specific covariance structure employed by the class of distributions in $\mathcal{D}^G_Z$, i.e., the structure in \eqref{cov_str}. We will now demonstrate that, for the case of unbiased estimates, this covariance structure is optimal among all possible covariance matrix configurations. It is important to note that the influence of colluding users on the privacy constraint is dependent on the protocol, as the random variables accessible to the server are determined by the protocol’s specific procedures. In Theorem~\ref{converse}, we establish the general converse result, which characterizes the optimal covariance structure in the absence of collusion.
\vspace{0.2cm}

\begin{theorem}[Converse - optimal covariance structure]\label{converse} Let $\tilde{\mathbf{S}}_{\mathcal{U}}=\frac{1}{|\mathcal{U}|}\sum_{i\in\mathcal{U}}(\mathbf{x}+\tilde{\mathbf{Z}}_i)$ be an unbiased estimate of $\mathbf{S}_\mathcal{U}$, where $\tilde{\mathbf{Z}}_i\sim\mathcal{N}(\mathbf{0}_d,\sigma^2_i\mathsf{I}_d)$ for $i\in[1:n]$. Let $[\tilde{Z}_{1,k},\dotsc,\tilde{Z}_{n,k}]^T\sim\mathcal{N}(\mathbf{0}_n,\Sigma)$ for $k\in[1:d]$, where $\tilde{Z}_{i,k}$ is the $k$th coordinate of $\tilde{\mathbf{Z}}_i$ and $\Sigma$ is symmetric positive definite. Define the corresponding MSE of $\tilde{\mathbf{S}}_{\mathcal{U}}$ as, 
\begin{align}\label{unbiased_def}
\mathrm{MSE}(\Sigma)&=\max_{\substack{\mathcal{U}\subseteq\mathcal{U}_{all}\\ t\leq|\mathcal{U}|\leq n}}\quad\sup_{\mathbf{x}_{j_1},\dotsc,\mathbf{x}_{j_{|\mathcal{U}|}}\in\mathbb{B}^d}\nonumber\\
&\qquad\mathbb{E}\left[\left\|\frac{1}{|\mathcal{U}|}\sum_{i\in\mathcal{U}}(\mathbf{x}_i+\tilde{\mathbf{Z}}_i)-\frac{1}{|\mathcal{U}|}\sum_{i\in\mathcal{U}}\mathbf{x}_i\right\|^2\right]
\end{align} 
$\forall\Sigma=\Sigma^T,\quad\Sigma\succ0$ satisfying $(\epsilon,\delta)$-DP in Definition~\ref{def1} with $\mathcal{U}_{col}=\emptyset$, 
\begin{align}\label{conv_eq}
    \mathrm{MSE}\left(\frac{1}{n!}\sum_{i=1}^{n!}\Pi_i(\Sigma)\right)\leq\mathrm{MSE}(\Sigma)
\end{align}
where $\Pi_i(\Sigma)=P_i\Sigma P_i^T$ is the $i$-th permutation of $\Sigma$ defined by the permutation matrix $P_i$. Moreover, $\tilde{\Sigma}=\frac{1}{n!}\sum_{i=1}^{n!}\Pi_i(\Sigma)$ satisfies $(\epsilon,\delta)$-DP in Definition~\ref{def1} with $\mathcal{U}_{col}=\emptyset$.

\end{theorem}

The proof of Theorem~\ref{converse} is given in Appendix~\ref{pfconverse}. As $\tilde{\Sigma}$ has all equal diagonal and all equal off diagonal entries, optimizing over the class of distributions $\mathcal{D}^G_Z$ in \eqref{cov_str} yields the optimal MSE among all Gaussian mechanisms (with i.i.d. coordinates that are arbitrarily correlated among users) for unbiased estimates.

\begin{figure}
    \centering
    \includegraphics[scale=0.4]{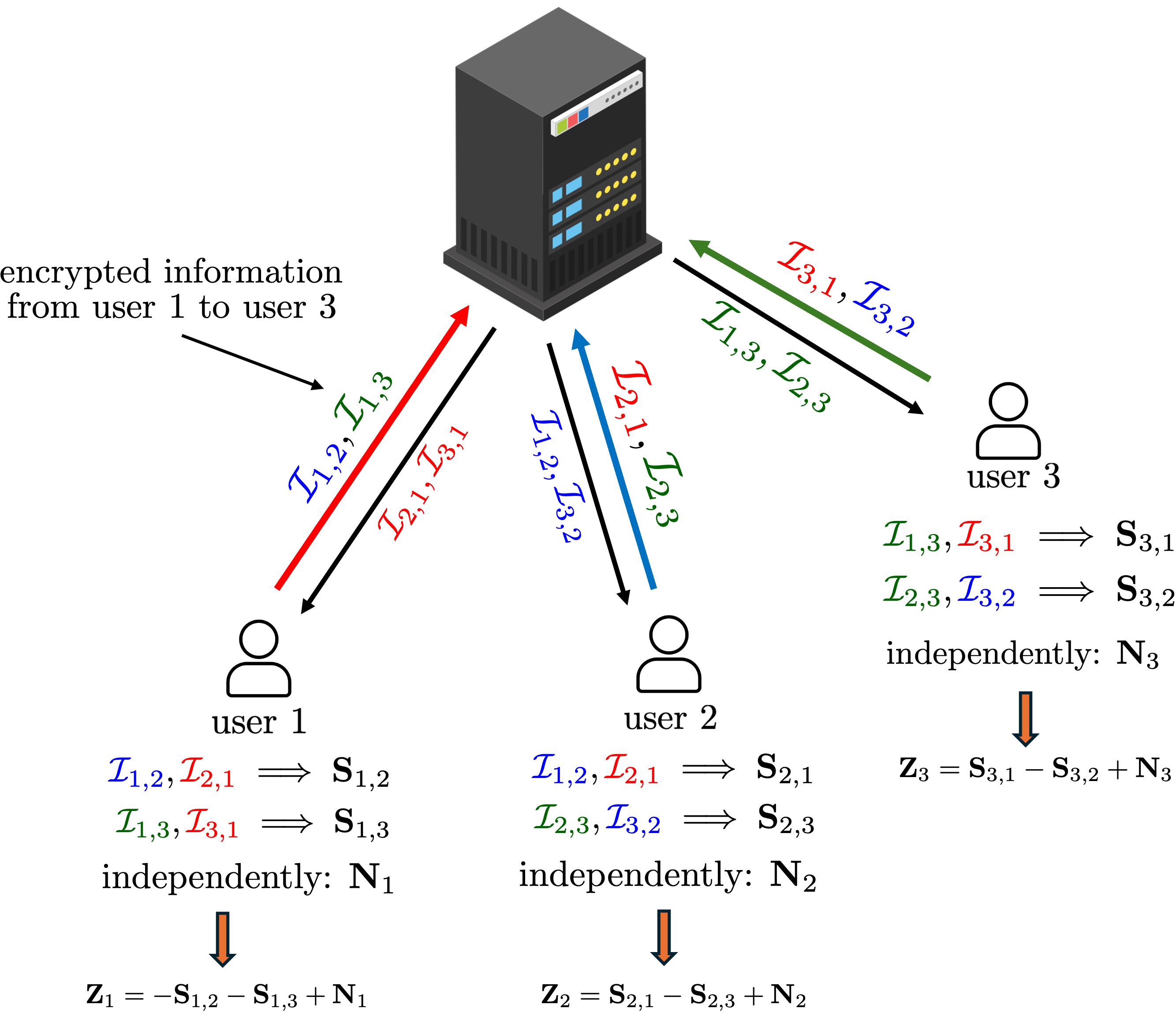}
    \caption{Overview of the offline phase of CorDP-DME in a three-user example: Each user $i$ sends secure information to other users $j$ (denoted by $\mathcal{I}_{i,j}$) to determine the common random seeds required for both users $i$ and $j$ to generate the same shared random variable $\mathbf{S}_{i,j}$.}
    \label{offline_fig}
\end{figure}

\begin{table*}[!tb]
    \centering
    \footnotesize
    \begin{tabular}{c c c c c c}
    \toprule
    & no dropouts, no collusion  & only collusion  & only dropouts  & dropouts and collusion & LDP (for comparison) \\
      & $t=10$, $c=0$  &  $t=10$, $c=2$ & $t=8$, $c=0$  & $t=8$, $c=2$ & $t=10$, $c=0$ (best case) \\

    \midrule 
    $\sigma^2_*$  & $\to\infty$ & $\to\infty$  & $5.466$ & $6.318$ & $3.975$ \\
\midrule 
$\rho_*$  & $\to-0.111$  & $\to-0.111$  & $-0.091$ & $-0.089$ & $0$\\
\midrule 
    MSE (biased) & $0.166$ & $0.199$  & $0.554$ & $0.598$ & $0.665$\\
    \midrule 
    MSE (unbiased) & $0.199$  & $0.248$  & $1.242$ & $1.488$ & $1.988$ \\
       \bottomrule
       \\
    \end{tabular}
    \caption{Comparison of optimal parameters across varying dropout and collusion thresholds, with $t$ and $c$ denoting the minimum number of responding users, and maximum number of colluding users, respectively.}
    \label{example_tbl}
\end{table*}

\subsection{Discussion on CorDP-DME}

In this section, we illustrate the key ideas used in CorDP-DME through a simple three-user example. With the notation used in Section~\ref{problem}, user $i$ sends $M(\mathbf{x}_i)=\mathbf{x}_i+\mathbf{Z}_i$ to the server, that computes $\frac{1}{3}\sum_{i=1}^3 M(\mathbf{x}_i)=\frac{1}{3}\sum_{i=1}^3\mathbf{x}_i+\frac{1}{3}\sum_{i=1}^3\mathbf{Z}_i$  as the estimate of $\frac{1}{3}\sum_{i=1}^3\mathbf{x}_i$, considering the unbiased case. Assume that we allow for at most one user to drop out in this example. As there are four possibilities for user dropouts, the possible estimates at the server are given by $E_1=\frac{1}{2}(\mathbf{x}_1+\mathbf{x}_2+\mathbf{Z}_1+\mathbf{Z}_2)$, $E_2=\frac{1}{2}(\mathbf{x}_1+\mathbf{x}_3+\mathbf{Z}_1+\mathbf{Z}_3)$, $E_3=\frac{1}{2}(\mathbf{x}_2+\mathbf{x}_3+\mathbf{Z}_2+\mathbf{Z}_3)$, and $E_4=\frac{1}{3}(\sum_{i=1}^3\mathbf{x}_i+\sum_{i=1}^3\mathbf{Z}_i)$ for $\frac{1}{2}(\mathbf{x}_1+\mathbf{x}_2)$, $\frac{1}{2}(\mathbf{x}_1+\mathbf{x}_3)$, $\frac{1}{2}(\mathbf{x}_2+\mathbf{x}_3)$, and $\frac{1}{3}\sum_{i=1}^3\mathbf{x}_i$, respectively. The corresponding MSEs are given by, $\mathrm{MSE}_1=\frac{1}{4}\mathbb{E}[\|\mathbf{Z}_1+\mathbf{Z}_2\|^2]$, $\mathrm{MSE}_2=\frac{1}{4}\mathbb{E}[\|\mathbf{Z}_1+\mathbf{Z}_3\|^2]$, $\mathrm{MSE}_3=\frac{1}{4}\mathbb{E}[\|\mathbf{Z}_2+\mathbf{Z}_3\|^2]$, and $\mathrm{MSE}_4=\frac{1}{9}\mathbb{E}[\|\mathbf{Z}_1+\mathbf{Z}_2+\mathbf{Z}_3\|^2]$. Note that for the unbiased case, the MSE does not depend on the $\mathbf{x}_i$'s. To minimize the residual error caused by the worst case dropouts, we need to solve:
\begin{align}\label{threeuser}
    \min_{\mathcal{D}_{\mathbf{Z}_1,\mathbf{Z}_2,\mathbf{Z}_3}}\max\{\mathrm{MSE}_1,\mathrm{MSE}_2,\mathrm{MSE}_3,\mathrm{MSE}_4\}
\end{align}
under the privacy constraint in \eqref{dp_eq}, where $\mathcal{D}_{\mathbf{Z}_1,\mathbf{Z}_2,\mathbf{Z}_3}$ represents the joint distribution of $\mathbf{Z}_1,\mathbf{Z}_2,\mathbf{Z}_3$. Note that this is exactly the main optimization problem in \eqref{main_opt} for this example, with $\alpha_{\mathcal{U}}$ fixed at 1 for unbiased estimates. CorDP-DME considers the class of Gaussian distributions with arbitrary correlations among users for $\mathcal{D}_{\mathbf{Z}_1,\mathbf{Z}_2,\mathbf{Z}_3}$. Let $\mathbf{Z}_{i,j}$ denote the $j$th coordinate of $\mathbf{Z}_i$. We show that for each $j$, the optimum covariance matrix of $[\mathbf{Z}_{1,j},\mathbf{Z}_{2,j},\mathbf{Z}_{3,j}]$ has all equal diagonal and all equal off diagonal entries (Theorem~\ref{converse}). Then, minimizing the worst case MSE over $\mathcal{D}_{\mathbf{Z}_1,\mathbf{Z}_2,\mathbf{Z}_3}$ reduces to minimizing it over $\sigma^2$ and $\rho$ as shown in \eqref{cov_str}. We then show that the maximum MSE in \eqref{threeuser} corresponds to any combination of the maximum number of dropouts, i.e., $\mathrm{MSE}_i$ for $i=1,2,3$ for this example. 

Next, we incorporate the privacy constraint in \eqref{dp_eq}. Note that for this example, assuming no colluding users, the privacy constraint in \eqref{dp_eq} simplifies to,
\begin{align}\label{threeuserspriv}
    &\mathbb{P}(M(\mathbf{x}_i)\in\mathcal{A}|M(\mathbf{x}_j),M(\mathbf{x}_k))\nonumber\\
    &\quad\leq e^\epsilon\mathbb{P}(M(\mathbf{x}_i')\in\mathcal{A}|M(\mathbf{x}_j),M(\mathbf{x}_k))+\delta,
\end{align}
for all $i,j,k\in\{1,2,3\}$ where $i\neq j\neq k$. By deriving the conditional Gaussian distributions,  \eqref{threeuserspriv} simplifies to $-2\rho^2+\rho\sigma^2+\sigma^2-2\sigma^2_{\epsilon,\delta}\geq0$ (see Appendix~\ref{pfthm1} for the general proof). Analytically solving \eqref{threeuser} under this simplified privacy constraint requires finding the optimal $\rho$ for a fixed $\sigma^2$, and then analyzing the behavior of \eqref{threeuser} with respect to $\sigma^2$. (see Appendix~\ref{pfthm1} and Theorems~\ref{thm1}-\ref{unbiased} for details).

The resulting optimum $\sigma^2_*$ and $\rho_*$ in this example define the joint distribution of $\mathbf{Z}_1,\mathbf{Z}_2$, and $\mathbf{Z}_3$ that minimizes the residual error in the aggregate under the worst-case dropouts. In this setup, users directly transmit $M(\mathbf{x}_i)$ to the server, which computes the aggregate in a single step, completing the protocol in one round. Among existing single-round protocols, which are typically LDP-based, CorDP-DME achieves superior accuracy as shown in Section~\ref{results}.

\subsection{Example}\label{example_num}

In this section, we present an example with 
$n=10$ users with $d=5$, $\epsilon=2$ and $\delta=10^{-5}$, considering different dropout and collusion thresholds. We demonstrate how the optimal parameters from Theorem~\ref{thm1} vary under each scenario. Table~\ref{example_tbl} provides a detailed comparison of the optimal parameters and the corresponding minimum MSEs, as obtained from Theorems~\ref{thm1} and~\ref{unbiased}, across various conditions. Note that in Table~\ref{example_tbl}, for the case with no dropouts, the optimal noise variance added to each user's private vector approaches infinity, regardless of the number of colluding users. This generalizes the two-user scenario discussed in Section~\ref{why} under similar dropout-free conditions. Additionally, the optimal correlation coefficient converges to 
$-0.111$, the largest magnitude allowed by valid $10\times10$ covariance matrices of the form \eqref{cov_str} with negative correlation. While the optimal parameters for the case of no dropouts with $c=0$ and $c=2$ converge to the same values asymptotically, the resulting MSEs differ because the MSE is a function of $\sigma^2(1+\rho(t-1))$, and $\lim_{\sigma^2\to\infty}\sigma^2(1+\rho(t-1))$ takes different values in each case.

As shown in columns 4 and 5 of Table~\ref{example_tbl}, when dropouts are introduced, employing maximum variance and maximum (negative) correlation becomes ineffective, as even a single dropout can lead to substantial residual noise variance, similar to the behavior seen in SecAgg \cite{secagg,secagg2}. Nevertheless, negative correlation can still be applied within a single round, resulting in improved performance compared to LDP as seen by the last column (see Section~\ref{experiments} for a detailed comparison). This is reflected by the fact that the optimal correlation coefficients remain negative (rather than zero) across all cases in Table~\ref{example_tbl} (except last column which shows the LDP baseline).

As expected, columns 4 and 5 (with dropouts) show a reduction in noise variance and the magnitude of the correlation coefficient, compared to the case of no dropouts, to mitigate the impact of dropouts on the MSE. Comparing these columns, we observe that for the same dropout threshold, increasing the number of colluding users requires higher noise variance for each user and reduced noise correlation among users to maintain $(\epsilon,\delta)$-DP. This ensures that the server cannot infer too much information about honest users from colluding ones, but at the cost of increasing the MSE, as seen in columns 4-5.

The last column presents the optimal parameters and MSEs for LDP-based DME under the ideal scenario where all users respond and none collude with the server. Even in this best-case scenario, both biased and unbiased estimates perform worse than all cases examined in CorDP-DME (columns 2-5). Notably, LDP employs the lowest noise variance compared to all CorDP-DME cases. Although it may seem counterintuitive to achieve a lower MSE with higher noise variance, this is made feasible through the use of negative correlation, as indicated in the second row.  

\begin{figure*}
    \centering
    \includegraphics[scale=0.55]{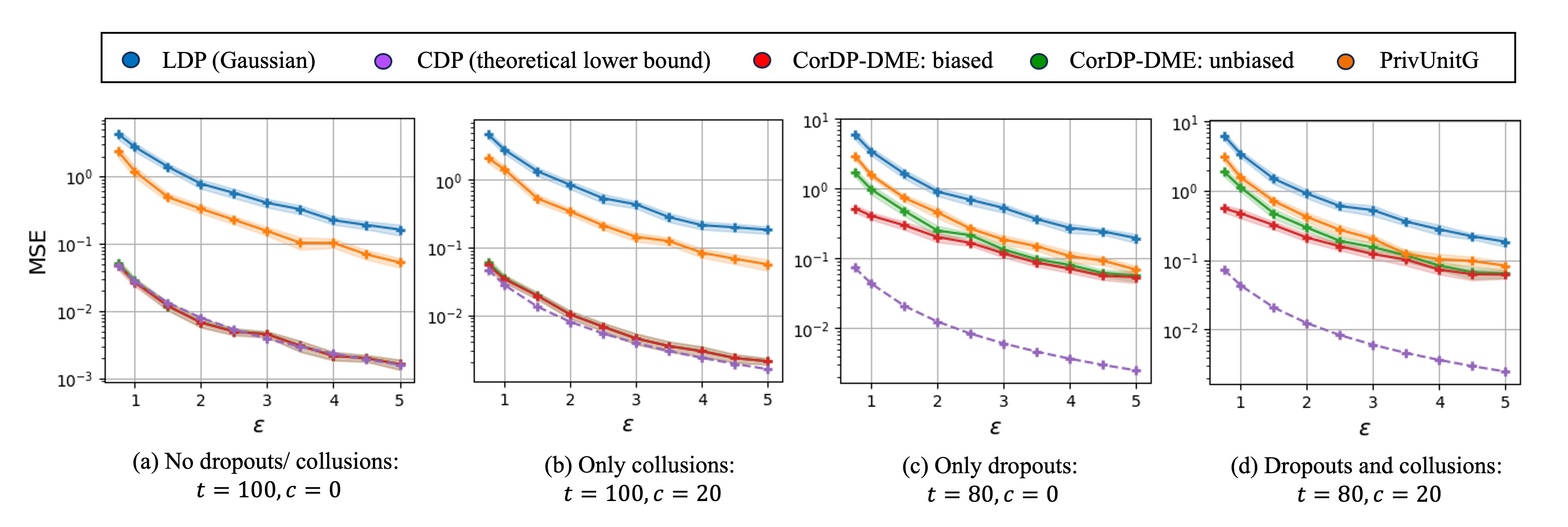}
    \caption{Privacy-utility trade-offs with (a) no dropouts and no colluding users, (b) with colluding users and no dropouts, (c) with dropouts and no colluding users (d) with both dropouts and colluding users. $t$: number of users remaining in the system after dropouts, $c$: number of colluding users. }
    \label{exp}
\end{figure*}

\subsection{Noise generation in CorDP-DME}\label{offline}

For a given DP-DME setting with $n$ users, $\epsilon,\delta$ privacy parameters and $t,c$ dropout and colluding user thresholds, CorDP-DME first calculates the corresponding optimum noise distribution and decoder parameters from Section~\ref{results}. Each user then generates their specific noise variables, $\mathbf{Z}_i$, ensuring that the joint distribution of $(\mathbf{Z}_1, \dotsc, \mathbf{Z}_n)$ follows the structure in \eqref{cov_str} with the determined optimal parameters. The following steps are followed in the noise generation protocol:
\begin{enumerate}[leftmargin=*,itemsep=0pt, topsep=0pt, partopsep=0pt]
    \item Each pair of users $(i,j)$, $i,j\in[1:n]$, $i\neq j$ generates a pairwise random seed using the Diffie-Hellman key exchange, via secure communications through the server.
    \item Using the common seed, each pair of users $(i,j)$ samples the same random vector $\mathbf{S}_{i,j}\in\mathbb{R}^d$ from $\mathcal{N}(\mathbf{0}_d,-\rho_*\sigma^2_*\mathsf{I}_d)$,\footnote{We use use negative correlation among the privacy mechanisms of different users. Hence, $-\rho_*>0$.} where $\mathbf{S}_{i,j}=\mathbf{S}_{j,i}$, and $\mathbf{S}_{i,j}$, $\mathbf{S}_{i',j'}$ are independent for any $(i,j)\neq(i',j')$.
    \item Each user $i$ independently generates another noise variable $\mathbf{N}_i\sim\mathcal{N}(\mathbf{0}_d,\sigma^2_*(1+\rho_*(n-1))\mathsf{I}_d)$.
    \item The combined noise added to $\mathbf{x}_i$, $i\in[1:n]$ is:
    \begin{align}
        \mathbf{Z}_i=\sum_{j=1,j<i}^n \mathbf{S}_{i,j}-\sum_{j=1,j>i}^n \mathbf{S}_{i,j}+\mathbf{N}_i.
    \end{align}
 Subsequently, each user $i$, $i\in[1:n]$ sends $M(\mathbf{x}_i)=\mathbf{x}_i+\mathbf{Z}_i$ to the server as per \eqref{encode}. Then, the server decodes the mean estimate using \eqref{decoder}, based on the optimum decoder calculated using the number of responding users $|\mathcal{U}|$, and the optimum noise parameters $\sigma^2_*,\rho_*$.
\end{enumerate}
Fig.~\ref{offline_fig} illustrates the overview of the noise generation process implemented during the offline phase.

\section{Experiments}\label{experiments}

We implement CorDP-DME for specific values of $\epsilon,\delta,n,t,c$, and compare the privacy-utility trade-offs against DME with LDP and CDP (with the Gaussian mechanism). As the two baselines correspond to unbiased estimates, we compare both biased (Theorem~\ref{thm1}) and unbiased (Theorem~\ref{unbiased}) versions of CorDP-DME with LDP and CDP. Fig.~\ref{exp} shows the privacy-utility trade-offs of different DP-DME cases corresponding to $n=100$, $\delta=10^{-5}$ and $d=20$. Each case was replicated 20 times and the plots show the MSEs with $95\%$ confidence intervals. Recall that $t$ and $c$ are the thresholds on the minimum responding users and maximum colluding users, respectively. CorDP-DME coincides with CDP when no dropouts or colluding users are considered ( Fig.~\ref{exp}(a)). When considering only colluding users (without any dropouts), the MSE of CorDP-DME increases slightly compared to that of CDP (FIg.~\ref{exp}(b)). This is due to the need for higher variance and reduced correlation to ensure $(\epsilon,\delta)$-DP of the honest users. The MSE of CorDP-DME with dropouts (with or without colluding users) lies in between the MSEs of (Gaussian) LDP and CDP (Figs.~\ref{exp}(c) and (d)). We also show that  for many cases (all cases considered in Fig.~\ref{exp}), CorDP-DME outperforms PrivUnitG \cite{optimal_algorithms}, which is an approximation of PrivUnit \cite{PrivUnit}, that is proven to be the optimum among all LDP mechanisms that result in unbiased estimates. However, we point out that PrivUnit is only applicable to the case where all users' private vectors have a common (fixed) $L_2$ norm, while all Gaussian mechanisms including CorDP-DME can accommodate any private vector with a bounded $L_2$ norm.



\section{Conclusions and Limitations}

In this work, we present CorDP-DME, a novel differentially private DME protocol that uses correlated Gaussian noise to achieve a favorable balance between utility, resilience to dropouts, and robustness against colluding users. CorDP-DME spans the spectrum between DME with LDP, which provides strong resilience but poor utility, and SA-based approaches, which achieve high utility but are also high in complexity and overheads. A key insight is that carefully tuned (anti) correlated noise can significantly improve utility compared to independent noise mechanisms, even in adversarial settings with dropouts and collusions.
Important directions for future work includes developing discrete or quantized variants of CorDP-DME, exploring sparse covariance structures to characterize the fundamental trade-off between communication complexity and utility (MSE), and extending the concept of noise correlation to DP-DME mechanisms beyond the class of Gaussian mechanisms considered in this work.



\newpage
\bibliographystyle{unsrt}
\bibliography{references}

\newpage
\appendices

\section{Generalized Privacy Definition}\label{privacydef}

In this section, we 1) motivate the privacy constraint in Definition~\ref{def1} by comparing with alternative DP definitions, 2) show that the privacy definitions used in DME with LDP and SA based distributed DP are special cases of the proposed privacy constraint in Definition~\ref{def1}.

\paragraph{Comparison with Alternative DP Definitions} First we restate our privacy definition here for convenience.

\textbf{Definition 1: Generalized $(\epsilon,\delta)$-DP for DME:}
Let $\mathcal{D}_i=\{\mathbf{x}\}_{j\neq i}\cup\mathbf{x}_i$ and $\mathcal{D}'_i=\{\mathbf{x}\}_{j\neq i}\cup\mathbf{x}_i'$  be two neighboring datasets that only differ in the vector of user $i$ for any $i\in\mathcal{U}_{all}\setminus\mathcal{U}_{col}$. Let $\mathcal{G}_i=\{\{M(\mathbf{x}_j)\}_{j\in\mathcal{U}_{all}\setminus\{i\}}, \{\mathbf{x}_j,\mathbf{Z}_j,\mathcal{I}_j\}_{j\in\mathcal{U}_{col}}\}$ denote all the random variables observed by the server from all users except user $i$, for any $i\in\mathcal{U}_{all}\setminus\mathcal{U}_{col}$. For a given $\epsilon\geq0$ and $\delta\in(0,1)$, a DP-DME scheme ensures $(\epsilon,\delta)$-DP if the following is satisfied.
\begin{align}
    \mathbb{P}(M(\mathbf{x}_i)\in\mathcal{A}|\mathcal{G}_i)\leq e^\epsilon\mathbb{P}(M(\mathbf{x}_i')\in\mathcal{A}|\mathcal{G}_i)+\delta,
\end{align}
$\forall \mathcal{D}_i,\mathcal{D}'_i$, $\forall i\in\mathcal{U}_{all}\setminus\mathcal{U}_{col}$ and $\forall\mathcal{A}\subset\mathbb{R}^d$ in the Borel $\sigma$-field.

To motivate this privacy definition, consider the following alternative DP definition that follows from the original DP framework. For two neighboring datasets $\mathcal{D}_i$ and $\mathcal{D}'_i$, let $\mathcal{V}_{\mathcal{D}_i}=\{M(\mathbf{x}_i),\mathcal{G}_i\}$ and $\mathcal{V}_{\mathcal{D}'_i}=\{M(\mathbf{x}'_i),\mathcal{G}_i\}$ denote the random variables that the server observes from the respective datasets. Then, the privacy mechanism $M$ is said to satisfy $(\epsilon,\delta)$-DP if the following is satisfied.
\begin{align}\label{originaldp}
    \mathbb{P}(\mathcal{V}_{\mathcal{D}_i}\in\mathcal{Y})\leq e^{\epsilon}\mathbb{P}(\mathcal{V}_{\mathcal{D}'_i}\in\mathcal{Y})+ \delta,
\end{align}
$\forall \mathcal{D}_i,\mathcal{D}'_i$, $\forall i\in\mathcal{U}_{all}\setminus\mathcal{U}_{col}$ and $\forall\mathcal{Y}\subset\mathcal{J}$, where $\mathcal{J}$ is the domain of $\mathcal{V}_\mathcal{D}$. Note that a privacy mechanism $M$ that satisfies the generalized DP constraint in Definition~\ref{def1} also satisfies \eqref{originaldp}, which can be directly observed by multiplying both sides of \eqref{dp_eq} by $\mathbb{P}(\mathcal{G}_i)$. However, the inverse argument does not hold for all cases, implying that the DP constraint in Definition~\ref{def1} is stronger than the one in \eqref{originaldp}. 

Definition~\ref{def1} generalizes the privacy constraints used in DP-DME with LDP \cite{PrivUnit,optimal_algorithms,minimax_procedures} and distributed DP with SA \cite{ddg,dpsecagg}, and offers a privacy framework to analyze and compare DP-DME mechanisms with arbitrary correlations among users. Next, we prove that LDP and Distributed DP with SA are special cases of Definition~\ref{def1}.

\paragraph{DME with LDP} Recall that each $\mathcal{G}_i$ in Definition~\ref{def1} contains information of all users except user $i$. Therefore, the random variables in $\mathcal{G}_i$ are independent of $M(\mathbf{x}_i)$, for all $i\in[1:n]\setminus\mathcal{C}$ as DME with LDP utilizes independent privacy mechanisms among users. Thus, \eqref{dp_eq} directly simplifies to,
\begin{align}
    \mathbb{P}(M(\mathbf{x}_i)\in\mathcal{A})\leq e^\epsilon\mathbb{P}(M(\mathbf{x}_i')\in\mathcal{A})+\delta,\quad \forall i\in[1:n]\setminus\mathcal{C},
\end{align}
$\forall \mathbf{x}_i,\mathbf{x}_i'\in\mathbb{B}^d$ and $\forall\mathcal{A}\subset\mathbb{R}^d$ in the Borel $\sigma$-field, which is the privacy constraint in DME with LDP.

\paragraph{SecAgg based DME with distributed DP} To prove that the privacy constraint used in SecAgg based DME with distributed DP is a special case of the privacy constraint in Definition~\ref{def1}, we first consider what is transmitted and received by the users and the server, respectively.
\begin{align}
    \text{user $i$ $\rightarrow$ \text{server}: }& \Bar{M}(\mathbf{x}_i)=M(\mathbf{x}_i)+\sum_{j<i}\mathbf{\bar{S}}_{j,i}-\sum_{j>i}\mathbf{\bar{S}}_{i,j},
\end{align}
$\forall i\in[1:n]$ where $\mathbf{\bar{S}}_{i,j}$ are uniformly distributed random variables from a finite field $\mathbb{F}_p$. Considering no dropouts and no colluding users, the privacy constraint in Definition~\ref{def1} implies,
\begin{align}\label{again1}
    &\mathbb{P}(\bar{M}(\boldsymbol{x}_i)\in\mathcal{A}|\bar{M}(\mathbf{x}_j)=y_j,\forall j\neq i)\nonumber\\
    &\quad\leq e^\epsilon\mathbb{P}(\bar{M}(\boldsymbol{x}_i')\in\mathcal{A}|\bar{M}(\mathbf{x}_j)=y_j,\forall j\neq i)\!+\!\delta,\quad \forall i
\end{align}
for any fixed $\mathbf{x}_j\in\mathbb{B}^d$, $j\neq i$,  $\forall \mathbf{x}_i,\mathbf{x}_i'\in\mathbb{B}^d$ and $\forall\mathcal{A}\subset\mathbb{R}^d$ in the Borel $\sigma$-field. Let  $\mathcal{A}'=\{a':a'=a+\sum_{j\neq i}y_j, \forall a\in\mathcal{A}\}$.
\begin{align}
    &\mathbb{P}(\bar{M}(\boldsymbol{x}_i)\in\mathcal{A}|\bar{M}(\mathbf{x}_j)=y_j,\forall j\neq i)\nonumber\\
    &=\frac{\mathbb{P}(\bar{M}(\boldsymbol{x}_i)\in\mathcal{A},\sum_{j=1}^n\bar{M}(\mathbf{x}_j)\in\mathcal{A}',\{\bar{M}(\mathbf(x)_j)=y_j\}_{j\neq i})}{\mathbb{P}(\{\bar{M}(\mathbf(x)_j)=y_j\}_{j\neq i})}\\
    &=\mathbb{P}(\bar{M}(\boldsymbol{x}_i)\in\mathcal{A}|\sum_{j=1}^n\bar{M}(\mathbf{x}_j)\in\mathcal{A}',\{\bar{M}(\mathbf(x)_j)=y_j\}_{j\neq i})\nonumber\\
    &\quad\qquad\times\mathbb{P}(\sum_{j=1}^n\bar{M}(\mathbf{x}_j)\in\mathcal{A}'|\{\bar{M}(\mathbf(x)_j)=y_j\}_{j\neq i}))\\
    &=1\times \frac{\mathbb{P}(\sum_{j=1}^n\bar{M}(\mathbf{x}_j)\in\mathcal{A}',\{\bar{M}(\mathbf(x)_j)=y_j\}_{j\neq i})}{\mathbb{P}(\{\bar{M}(\mathbf(x)_j)=y_j\}_{j\neq i})}\\
    &=\frac{\mathbb{P}(\sum_{j=1}^nM(\mathbf{x}_j)\in\mathcal{A}')\mathbb{P}(\{\bar{M}(\mathbf(x)_j)=y_j\}_{j\neq i})}{\mathbb{P}(\{\bar{M}(\mathbf(x)_j)=y_j\}_{j\neq i})}\\
    &=\mathbb{P}(\sum_{j=1}^nM(\mathbf{x}_j)\in\mathcal{A}')\label{lastsecpf}
\end{align}
where we use $\sum_{i=1}^n\bar{M}(\mathbf{x}_i)=\sum_{i=1}^nM(\mathbf{x}_i)$ and Shannon's one-time-pad theorem to derive the last two steps. Substituting \eqref{lastsecpf} in \eqref{again1} (with all $\mathbf{x}_j$, $j\neq i$ fixed and $\mathbf{x}_i$ on the LHS and $\mathbf{x}_i'$ on the RHS) gives the privacy constraint in distributed DP with SecAgg.

\section{Proof of Proposition~\ref{vec_opt_dec}}\label{pfprop1}

\noindent{\textbf{Proposition~\ref{vec_opt_dec} restated:}}
For any fixed $\mathcal{U}\subseteq\mathcal{U}_{all}$ satisfying $|\mathcal{U}|\geq t$, and for any $\sigma^2$, $r$, the optimum decoder is given by,
 \begin{align}\label{opt_decoder}
\boldsymbol{\alpha}^*_{\mathcal{U}}&=\arg\min_{\boldsymbol{\alpha}_{\mathcal{U}}}\mathrm{MSE}(\sigma^2,\rho,\mathcal{U},\boldsymbol{\alpha}_{\mathcal{U}})\\
        &=\frac{1}{1+\frac{d}{|\mathcal{U}|}(\sigma^2+r(|\mathcal{U}|-1))}\mathbf{1}_{|\mathcal{U}|}
\end{align}
The corresponding MSE is given by,
\begin{align}
    \min_{\boldsymbol{\alpha}_{\mathcal{U}}}\mathrm{MSE}(\sigma^2,\rho,\mathcal{U},\boldsymbol{\alpha}_{\mathcal{U}})=\left(1+\frac{|\mathcal{U}|/d}{\sigma^2(1+\rho(|\mathcal{U}|-1))}\right)^{-1}
\end{align}

\begin{Proof}
    Let $x_{i,k}$ and $Z_{i,k}$ denote the $k$th element of $\mathbf{x}_i$ and $\mathbf{Z}_i$, respectively. Then,
\begin{align}\label{permuted_vec}
    &\mathrm{MSE}(\sigma^2,\rho,\mathcal{U},\boldsymbol{\alpha}_{\mathcal{U}})\nonumber\\
    &=\sup_{\mathbf{x}_{j_1},\dotsc,\mathbf{x}_{j_{|\mathcal{U}|}}\in\mathbb{B}^d}\mathbb{E}\!\!\left[\left\|\frac{1}{|\mathcal{U}|}\sum_{i\in\mathcal{U}}\alpha_i(\mathbf{x}_i+\mathbf{Z}_i)\!-\!\frac{1}{|\mathcal{U}|}\sum_{i\in\mathcal{U}}\mathbf{x}_i\right\|^2\right]\\
    &=\!\!\!\!\!\!\sup_{\mathbf{x}_{j_1},\dotsc,\mathbf{x}_{j_{|\mathcal{U}|}}\in\mathbb{B}^d}\sum_{k=1}^d\mathbb{E}\!\!\left[\!\left(\!\frac{1}{|\mathcal{U}|}\!\sum_{i\in\mathcal{U}}\alpha_i({x}_{i,k}\!+\!Z_{i,k})\!-\!\frac{1}{|\mathcal{U}|}\sum_{i\in\mathcal{U}}{x}_{i,k}\!\right)^2\!\right]\\
    &=\sup_{\mathbf{x}_{j_1},\dotsc,\mathbf{x}_{j_{|\mathcal{U}|}}\in\mathbb{B}^d}\frac{1}{|\mathcal{U}|^2}\sum_{k=1}^d\left(\boldsymbol{\alpha}_{\mathcal{U}}^TA_k\boldsymbol{\alpha}_{\mathcal{U}}-2\cdot\mathbf{1}^T\mathbf{x}^{[k]}\mathbf{x}^{[k]T}\boldsymbol{\alpha}_{\mathcal{U}}\right.\nonumber\\
    &\qquad\qquad\qquad\qquad\qquad\qquad\left.+\mathbf{1}^T\mathbf{x}^{[k]}\mathbf{x}^{[k]T}\mathbf{1}\right)
\end{align}
where $\mathbf{x}^{[k]}=[x_{j_1,k},\dotsc,x_{j_{|\mathcal{U}|},k}]^T$ and $A_k=\mathbf{x}^{[k]}\mathbf{x}^{[k]T}+\Sigma$, where $\Sigma$ is the covariance matrix of $[Z_{j_1,k},\dotsc,Z_{j_{|\mathcal{U}|},k}]^T$ for all $k\in[1:d]$. Define,
\begin{align}
    &f(\boldsymbol{\alpha}_{\mathcal{U}},\mathbf{x}_{j_1},\dotsc,\mathbf{x}_{j_{|\mathcal{U}|}},\Sigma)\nonumber\\
    &=\frac{1}{|\mathcal{U}|^2}\sum_{k=1}^d\left(\boldsymbol{\alpha}_{\mathcal{U}}^TA_k\boldsymbol{\alpha}_{\mathcal{U}}-2\cdot\mathbf{1}^T\mathbf{x}^{[k]}\mathbf{x}^{[k]T}\boldsymbol{\alpha}_{\mathcal{U}}+\mathbf{1}^T\mathbf{x}^{[k]}\mathbf{x}^{[k]T}\mathbf{1}\right)
\end{align} 
For any fixed $\mathbf{x}_{j_1},\dotsc,\mathbf{x}_{j_{|\mathcal{U}|}}$ and $\Sigma$, $f(\boldsymbol{\alpha}_{\mathcal{U}},\mathbf{x}_{j_1},\dotsc,\mathbf{x}_{j_{|\mathcal{U}|}},\Sigma)$ is convex in $\boldsymbol{\alpha}_{\mathcal{U}}$ as all $A_k$, $k\in[1:d]$ are positive definite. For any given $\boldsymbol{\alpha}_{\mathcal{U}}$ in \eqref{permuted_vec}, as $\mathbf{Z}_i$, $i\in\mathcal{U}$ are i.i.d. and $\mathbf{x}_i$, $i\in\mathcal{U}$ are chosen from the same set $\mathbb{B}^d$,  we have, 
\begin{align}
    \mathrm{MSE}(\sigma^2,\rho,\mathcal{U},\boldsymbol{\alpha}_{\mathcal{U}})&=\mathrm{MSE}(n,\mathcal{U},\sigma^2,\rho,\Pi_j(\boldsymbol{\alpha}_{\mathcal{U}})),
\end{align}
for $j\in[1:|\mathcal{U}|!]$ where $\Pi_j(\boldsymbol{\alpha}_{\mathcal{U}})$ denotes the $j$th permutation of the elements of $\boldsymbol{\alpha}_{\mathcal{U}}$. Therefore, 
\begin{align}
    &\mathrm{MSE}(\sigma^2,\rho,\mathcal{U},\boldsymbol{\alpha}_{\mathcal{U}})\nonumber\\
    &=\frac{1}{|\mathcal{U}|!}\sum_{j=1}^{|\mathcal{U}|!}\mathrm{MSE}(n,\mathcal{U},\sigma^2,\rho,\Pi_j(\boldsymbol{\alpha}))\\
    &=\frac{1}{|\mathcal{U}|!}\sum_{j=1}^{|\mathcal{U}|!}\sup_{\mathbf{x}_{j_1},\dotsc,\mathbf{x}_{j_{|\mathcal{U}|}}\in\mathbb{B}^d}f\left(\Pi_j(\boldsymbol{\alpha}_{\mathcal{U}}),\mathbf{x}_{j_1},\dotsc,\mathbf{x}_{j_{|\mathcal{U}|}},\Sigma\right)\\
    &\geq  \sup_{\mathbf{x}_{j_1},\dotsc,\mathbf{x}_{j_{|\mathcal{U}|}}\in\mathbb{B}^d}\frac{1}{|\mathcal{U}|!}\sum_{j=1}^{|\mathcal{U}|!}f\left(\Pi_j(\boldsymbol{\alpha}_{\mathcal{U}}),\mathbf{x}_{j_1},\dotsc,\mathbf{x}_{j_{|\mathcal{U}|}},\Sigma\right)\\
    &\geq  \sup_{\mathbf{x}_{j_1},\dotsc,\mathbf{x}_{j_{|\mathcal{U}|}}\in\mathbb{B}^d}f\left(\frac{1}{|\tau|!}\sum_{j=1}^{|\tau|!}(\Pi_j(\boldsymbol{\alpha}_{\mathcal{U}})),\mathbf{x}_{j_1},\dotsc,\mathbf{x}_{j_{|\mathcal{U}|}},\Sigma\right)\label{convex2}\\
    &=  \sup_{\mathbf{x}_{j_1},\dotsc,\mathbf{x}_{j_{|\mathcal{U}|}}\in\mathbb{B}^d}f\left(\mathbf{\tilde{\boldsymbol{\alpha}}_{\mathcal{U}}},\mathbf{x}_{j_1},\dotsc,\mathbf{x}_{j_{|\mathcal{U}|}},\Sigma\right)
\end{align}
where \eqref{convex2} is due to the convexity of $f(\boldsymbol{\alpha}_{\mathcal{U}},\mathbf{x}_{j_1},\dotsc,\mathbf{x}_{j_{|\mathcal{U}|}},\Sigma)$ in $\boldsymbol{\alpha}_{\mathcal{U}}$, and $\mathbf{\tilde{\boldsymbol{\alpha}}_{\mathcal{U}}}=\left(\frac{1}{|\mathcal{U}|}\sum_{i\in\mathcal{U}}\alpha_i\right)\mathbf{1}_{|\mathcal{U}|}$. This implies that for any decoder $\boldsymbol{\alpha}_{\mathcal{U}}$ with $\mathcal{U}$, $\sigma^2$ and $\rho$ fixed, there exists a $\mathbf{\tilde{\boldsymbol{\alpha}}}=[\tilde{\alpha}_{j_1},\dotsc,\tilde{\alpha}_{j_{|\mathcal{U}|}}]^T$ such that $\tilde{\alpha}_k=\tilde{\alpha}_\ell$, $\forall k,\ell$ satisfying,
\begin{align}
\mathrm{MSE}(\sigma^2,\rho,\mathcal{U},\boldsymbol{\alpha}_{\mathcal{U}})\geq\mathrm{MSE}(n,\mathcal{U},\sigma^2,\rho,\mathbf{\tilde{\boldsymbol{\alpha}}_{\mathcal{U}}}).    
\end{align}
Therefore, for a given $\mathcal{U}\subseteq\mathcal{U}_{all}$, the optimum decoder is of the form $\boldsymbol{\alpha}_{\mathcal{U}}=\alpha\mathbf{1}_{|\mathcal{U}|}$, and \eqref{permuted_vec} can be written as,
\begin{align}
    &\mathrm{MSE}(n,\mathcal{U},\sigma^2,\rho,\alpha)\nonumber\\
    &=\sup_{\mathbf{x}_{j_1},\dotsc,\mathbf{x}_{j_{|\mathcal{U}|}}\in\mathbb{B}^d}\mathbb{E}\left[\left\|\frac{1}{|\mathcal{U}|}\sum_{i\in\mathcal{U}}\alpha(\mathbf{x}_i+\mathbf{Z}_i)-\frac{1}{|\mathcal{U}|}\sum_{i\in\mathcal{U}}\mathbf{x}_i\right\|^2\right]\\
    &=\sup_{\mathbf{x}_{j_1},\dotsc,\mathbf{x}_{j_{|\mathcal{U}|}}\in\mathbb{B}^d}\frac{1}{|\mathcal{U}|^2}\!\sum_{k=1}^d\!\left(\!\!(\alpha\!-\!1)^2\sum_{i\in\mathcal{U}}x_{i,k}^2\right.\nonumber\\
    &\quad\left.+(\alpha\!-\!1)^2\!\!\sum_{\substack{i,j\in\mathcal{U}\\i\neq j}}\!\!x_{i,k}x_{j,k}+\alpha^2\sum_{\substack{i,j\in\mathcal{U}\\i\neq j}}\rho\sigma^2+\alpha^2\sigma^2|\mathcal{U}|\right)\label{varian}\\
    &=\sup_{\mathbf{x}_{j_1},\dotsc,\mathbf{x}_{j_{|\mathcal{U}|}}\in\mathbb{B}^d}\frac{1}{|\mathcal{U}|^2}\left(\!\!(\alpha\!-\!1)^2\sum_{i\in\mathcal{U}}\|\mathbf{x}_i\|^2\right.\\
    &\left.+(\alpha\!-\!1)^2\!\!\sum_{\substack{i,j\in\mathcal{U}\\i\neq j}}\mathbf{x}_i^T\mathbf{x}_j+d\alpha^2\rho\sigma^2|\mathcal{U}|(|\mathcal{U}|-1)+d\alpha^2\sigma^2|\mathcal{U}|\right)
\end{align}
The worst case MSE is resulted when $\mathbf{x}_i^T\mathbf{x}_j=1$, for all $i,j\in\mathcal{U}$, that is, when $\mathbf{x}_i=\mathbf{x}_j$, $\forall i,j$, and $\mathbf{x}_i\in\mathbb{S}^{d-1}$, where $\mathbb{S}^{d-1}$ is the unit sphere in $\mathbb{R}^d$. Therefore,
\begin{align}
    &\mathrm{MSE}(n,\mathcal{U},\sigma^2,\rho,\alpha)\nonumber\\
    &\quad=\frac{1}{|\mathcal{U}|^2}\left((\alpha-1)^2|\mathcal{U}|+(\alpha-1)^2|\mathcal{U}|(|\mathcal{U}|-1)\right.\nonumber\\
    &\left.\qquad+d\alpha^2\sigma^2|\mathcal{U}|(1+\rho(|\mathcal{U}|-1))\right)\\
    &\quad=(\alpha-1)^2+\frac{d\alpha^2\sigma^2}{|\mathcal{U}|}(1+\rho(|\mathcal{U}|-1))\label{mse_vec_sph1}
\end{align}
The optimum decoder for any fixed $\mathcal{U}$, $\sigma^2$ and $r$ is computed as,
\begin{align}
    &\frac{\partial \mathrm{MSE}(n,\mathcal{U},\sigma^2,\rho,\alpha)}{\partial\alpha}\nonumber\\
    &\quad=2(\alpha-1)+\frac{2d\alpha\sigma^2}{|\mathcal{U}|}(1+\rho(|\mathcal{U}|-1))=0\\
    &\implies\alpha^*=\frac{1}{1+\frac{d\sigma^2}{|\mathcal{U}|}(1+\rho(|\mathcal{U}|-1))}.
\end{align}
The resulting MSE is obtained by substituting $\alpha^*$ in \eqref{mse_vec_sph1}.
\end{Proof}

\section{Proof of Theorem~\ref{thm1}}\label{pfthm1}

\paragraph{Theorem~\ref{thm1} restated} For any given $\epsilon>0$, $\delta\in(0,1)$, $t$ and $c$, the optimum $\mathcal{D}_Z$ that solves \eqref{main_opt} while satisfying \eqref{dp_eq} is characterized by,
\begin{align}
(\sigma^2_*,\rho_*)=\arg\min_{\sigma^2,\rho}\max_{\substack{\mathcal{U}\subseteq\mathcal{U}_{all}\\ t\leq|\mathcal{U}|\leq n}} \min_{\boldsymbol{\alpha}_{\mathcal{U}}}\mathrm{MSE}(\sigma^2,\rho,\mathcal{U},\boldsymbol{\alpha}_{\mathcal{U}})\nonumber    
\end{align}
\begin{align}
    \sigma^2_*&=\begin{cases}
        \infty, & t=n\\
        \frac{\sigma^2_{\epsilon,\delta}(n^2-2n-cn+2)}{(n-c)^2}
    \\
    +\frac{\sigma^2_{\epsilon,\delta}(n-c-1)(n+c-2nc+t(n+c-2))}{(n-c)^2\sqrt{(t-c)(n-t)(n-c-1)}}, & c<t<n.
    \end{cases}\\
    \rho_*&=\begin{cases}
    -\frac{\sigma^2_*-\sigma^2_{\epsilon,\delta}}{\sigma^2_*(n-1)-\sigma^2_{\epsilon,\delta}(n-2)}, & c=1\\
        \frac{-(n-2)\left(1-\frac{\sigma^2_{\epsilon,\delta}}{\sigma^2_*}\right)-c}{2(n-1)(c-1)}
    \\+\frac{\sqrt{\left((n-2)\left(1-\frac{\sigma^2_{\epsilon,\delta}}{\sigma^2_*}\right)-c\right)^2+4(n-c-1)\left(1-\frac{\sigma^2_{\epsilon,\delta}}{\sigma^2_*}\right)}}{2(n-1)(c-1)}, & c\neq1
    \end{cases}
    \end{align}
with $\sigma_{\epsilon,\delta}\!=\!\inf_{\hat{\sigma}>0}\{\hat{\sigma};\Phi\!\left(\frac{1}{\hat{\sigma}}\!-\!\frac{\epsilon\hat{\sigma}}{2}\right)-e^\epsilon\Phi\!\left(-\frac{1}{\hat{\sigma}}\!-\!\frac{\epsilon\hat{\sigma}}{2}\right)\leq\delta\}$, where $\Phi$ is the standard Gaussian CDF. The resulting minimum MSE is given by,
\begin{align}
    \min_{\sigma^2,\rho}&\max_{\substack{\mathcal{U}\subseteq\mathcal{U}_{all}\\ t\leq|\mathcal{U}|\leq n}} \min_{\boldsymbol{\alpha}_{\mathcal{U}}}\mathrm{MSE}(\sigma^2,\rho,\mathcal{U},\boldsymbol{\alpha}_{\mathcal{U}})\nonumber\\
    &=\left(1+\frac{t/d}{\sigma^2_*+\rho_*(t-1)}\right)^{-1}
\end{align}

\begin{Proof}
    The proof consists of three main steps, namely, 1) optimizing the decoder at the server for any fixed joint distribution of $(\boldsymbol{Z}_1,\dotsc,\boldsymbol{Z}_n)$, i.e., any fixed $\sigma^2$ and $\rho$, 2) determining the feasible regions of $\sigma^2$ and $\rho$ that satisfy the privacy constraint, 3) characterizing the overall minimum MSE by optimizing $\sigma^2$ and $\rho$. Step 1 is proved in Proposition~\ref{vec_opt_dec}. For step 2, note that the information available at the server increases with the number of responding users and the number of colluding users. Therefore, we consider the case where all $n$ users respond, and up to any $c$ users can collude with the server, to analyze the privacy constraint. Lemma ~\ref{privacylem} characterizes step 2. We analyze step 3 in two cases. In case 1, we assume no dropouts, i.e., $t=n$. In case 2, we assume that up to any $t<n$ users can dropout. The two cases are analyzed in Lemmas~\ref{optsigmarho} and~\ref{dropoutvarlem}, respectively. 

    Let $D_Z^G(\sigma^2,\rho)$ denote the multivariate Gaussian distribution of $(\mathbf{Z}_1,\dotsc,\mathbf{Z}_n)$ with following properties. $\boldsymbol{Z}_i\sim\mathcal{N}(\boldsymbol{0}_d,\sigma^2\mathsf{I}_d)$, $i\in\mathcal{U}_{all}$, with the all zeros vector of size $d\times1$ denoted by $\boldsymbol{0}_d$ and the identity matrix of size $d\times d$ denoted by $\mathsf{I}_d$. The $k$th element of $\boldsymbol{Z}_i$ is denoted by $Z_{i,k}$ for $k\in[1:d]$. $Z_{i,k}$'s are allowed to be correlated as $\mathbb{E}[Z_{i,k}Z_{j,k}]=\rho\sigma^2=r$ for $\forall i\neq j$, $k\in[1:d]$. That is, $[Z_{1,k},\dotsc,Z_{n,k}]^T\sim\mathcal{N}(\boldsymbol{0}_n,\Sigma)$ for $k\in[1:d]$ with $\Sigma_{i,i}=\sigma^2$ for $i\in[1:n]$ and $\Sigma_{i,j}=r$ for $i,j\in[1:n]$, $i\neq j$. Moreover, let $\mathbb{E}[Z_{i,k}Z_{j,k'}]=0$, $\forall i,j$, $\forall k\neq k'$.

\begin{lemma}[Privacy Condition]\label{privacylem}
    The Gaussian mechanism $\mathcal{D}_Z^G$ with given $\sigma^2$ and $r$ satisfies the privacy constraint in Definition~\ref{def1} for any $\mathcal{U}$ and $\mathcal{U}_{col}$ satisfying $|\mathcal{U}_{col}|\leq c<t\leq|\mathcal{U}|$, so long as:
    \begin{align}\label{privacy3}
        r^2(n-1)(c-1)+&r(\sigma^2(n+c-2)-\sigma^2_{\epsilon,\delta}(n-2))\nonumber\\
        &+\sigma^2(\sigma^2-\sigma^2_{\epsilon,\delta}) \geq 0,
    \end{align}
    where $\sigma^2_{\epsilon,\delta}=\inf_{\hat{\sigma}>0}\{\hat{\sigma};\Phi\left(\frac{1}{\hat{\sigma}}-\frac{\epsilon\hat{\sigma}}{2}\right)-e^\epsilon\Phi\left(-\frac{1}{\hat{\sigma}}-\frac{\epsilon\hat{\sigma}}{2}\right)\leq\delta\}$. For any fixed $\sigma^2\geq\sigma^2_{\epsilon,\delta}$, the values of $r$ satisfying \eqref{privacy3} are given by,
    \begin{align}\label{range}
        r \geq\begin{cases}
            \frac{-\sigma^2(\sigma^2-\sigma^2_{\epsilon,\delta})}{\sigma^2(n-1)-\sigma^2_{\epsilon,\delta}(n-2)}, & c=1\\
            \frac{-(n-2)(\sigma^2-\sigma^2_{\epsilon,\delta})-\sigma^2c}{2(n-1)(c-1)}\\+\frac{\sqrt{((n-2)(\sigma^2-\sigma^2_{\epsilon,\delta})-\sigma^2c)^2
            +4(n-c-1)\sigma^2(\sigma^2-\sigma^2_{\epsilon,\delta})}}{2(n-1)(c-1)}, & c>1
        \end{cases}
    \end{align}
    and for $c=0$,
    \begin{align}\label{range2}
        r \in \left[a-b,a+b\right]
    \end{align}
    where,
    \begin{align}
        a&=\frac{(n-2)(\sigma^2-\sigma^2_{\epsilon,\delta})}{2(n-1)}\\ 
        b&=\frac{\sqrt{(n-2)^2(\sigma^2-\sigma^2_{\epsilon,\delta})^2+4(n-1)\sigma^2(\sigma^2-\sigma^2_{\epsilon,\delta})}}{2(n-1)}.
    \end{align} 
\end{lemma}

\begin{Proof}[Proof of Lemma~\ref{privacylem}]
    Recall that the privacy mechanism of each user is given by $M(\mathbf{x}_i)=\mathbf{x}_i+\mathbf{Z}_i$, $i\in[1:n]$. Note that for user $i$, the added noise can be written in the following form, based on our noise generation protocol in Section~\ref{proposed}.
    \begin{align}
        \mathbf{Z}_i=\sum_{j<i}\mathbf{S}_{j,i}-\sum_{j>i}\mathbf{S}_{i,j}+N_i,\quad i\in[1:n]
    \end{align}
    where $\mathbf{S}_{i,j}\sim\mathcal{N}(\mathbf{0}_d,-r\mathsf{I}_d)$ and $\mathbf{N}_{i}\sim\mathcal{N}(\mathbf{0}_d,(\sigma^2+r(n-1))\mathsf{I}_d)$. To analyze the privacy constraint in Definition~\ref{def1}, the set of random variables observed by the server from all users except for user $i$, $i\in\mathcal{U}_{all}\setminus\mathcal{U}_{col}$ when any $\mathcal{U}_{col}\subset\mathcal{U}_{all}$, $|\mathcal{U}_{col}|\leq c$ users are allowed to collude with the server is given by,
    \begin{align}
        \mathcal{G}_i=\{\{M(\mathbf{x}_{j})\}_{j\in\mathcal{U}_{all}\setminus\{i\}},\{\mathbf{x}_k,\mathbf{N}_k,\{\mathbf{S}_{k,j}\}_{\forall j}\}_{k\in\mathcal{U}_{col}}\}.
    \end{align}
    Then, the privacy constraint in Definition~\ref{def1} stated as
    \begin{align}
        \mathbb{P}(M(\mathbf{x}_i)\in\mathcal{A}|\mathcal{G}_i)\leq e^\epsilon\mathbb{P}(M(\mathbf{x}_i')\in\mathcal{A}|\mathcal{G}_i)+\delta,
    \end{align}
    $\forall \mathcal{D}_i,\mathcal{D}_i'$, $\forall i\in\mathcal{U}_{all}\setminus\mathcal{U}_{col}$, and $\forall \mathcal{A}$, simplifies to,
    \begin{align}
        &\mathbb{P}(\hat{M}(\mathbf{x}_i)\in\mathcal{A}|\hat{M}(\mathbf{x}_j)=\mathbf{y}_j,j\in\mathcal{U}_{all}\setminus\{\mathcal{U}_{col}\cup i\})\nonumber\\
        &\leq e^\epsilon\mathbb{P}(\hat{M}(\mathbf{x}_i')\in\mathcal{A}|\hat{M}(\mathbf{x}_j)=\mathbf{y}_j,j\in\mathcal{U}_{all}\setminus\{\mathcal{U}_{col}\cup i\})+\delta,\label{simplified}
    \end{align}
    $\forall \mathcal{D}_i,\mathcal{D}_i'$, $\forall i\in\mathcal{U}_{all}\setminus\mathcal{U}_{col}$, and $\forall \mathcal{A}$ where $\hat{M}(\mathbf{x}_i)=\mathbf{x}_i+\sum_{j<i,j\notin\mathcal{U}_{col}}\mathbf{S}_{j,i}-\sum_{j>i,j\notin\mathcal{U}_{col}}\mathbf{S}_{i,j}+\mathbf{N}_i$.
    Let $\mathbf{Y}_k=\hat{M}(\mathbf{x}_k)$, $\forall k\in[1:n]$. We first derive the conditional distribution of $\mathbf{Y}_i$ given $\{\mathbf{Y}_j$, $j\in\mathcal{U}_{all}\setminus\{\mathcal{U}_{col}\cup i\}\}$. Without loss of generality assume that $i=1$ and $\mathcal{U}_{col}=\{n-c+1,\dotsc,n\}$.

\begin{Claim}\label{cond_vec}
    The conditional distribution of $\mathbf{Y}_1$, given $\{\mathbf{Y}_j$, $j\in[2:n-c]\}$ is given by,
\begin{align}
    \mathbf{Y}_1|\mathbf{Y}_2=\mathbf{y}_2,\dotsc,\mathbf{Y}_{n-c}=\mathbf{y}_{n-c}\sim N(\Tilde{\mathbf{\mu}},\Tilde{\Sigma})
\end{align}    
where,
\begin{align}\label{mean_vec}
    \Tilde{\mathbf{\mu}}&\!=\!\mathbf{x}_1\!+\!\left(\!\!r\mathbf{1}_{n-c-1}^T\!\!\begin{pmatrix}
        \sigma^2+rc \!\!&\!\! r \!\!&\!\! \dotsc \!\!&\!\! r\\
         r \!\!&\!\! \sigma^2+rc \!\!&\!\! \dotsc \!\!&\!\! r\\
         \vdots \!\!&\!\! \vdots \!\!&\!\! \vdots \!\!&\!\! \vdots \\
        r \!\!&\!\!  r \!\!&\!\! \dotsc \!\!&\!\! \sigma^2+rc 
    \end{pmatrix}_{n-c-1}^{-1}\!\!\!\!\!\!\!\!\!\!\otimes\mathsf{I}_d\right)\nonumber\\
    &\qquad\qquad\qquad\times\begin{pmatrix}
        \mathbf{y}_2-\mathbf{x}_2\\\vdots\\\mathbf{y}_{n-c}-\mathbf{x}_{n-c}
    \end{pmatrix}\\
    \Tilde{\Sigma}&\!=\!\!\left(\!\!\sigma^2\!+\!rc\!-\!r^2\mathbf{1}_{n\!-\!c\!-\!1}^T\!\!\begin{pmatrix}
        \sigma^2\!+\!rc \!\!&\!\! r \!\!&\!\! \dotsc \!\!&\!\! r\\
         r \!\!&\!\! \sigma^2\!+\!rc \!\!&\!\! \dotsc \!\!&\!\! r\\
         \vdots \!\!&\!\! \vdots \!\!&\!\! \vdots \!\!&\!\! \vdots \\
        r \!\!&\!\!  r \!\!&\!\! \dotsc \!\!&\!\! \sigma^2\!+\!rc 
    \end{pmatrix}_{\!\!\!\!(n\!-\!c\!-\!1)^2}^{-1}\!\!\!\!\!\!\!\!\!\!\!\!\!\!\!\!\mathbf{1}_{n\!-\!c\!-\!1}\!\!\right)\!\!\mathsf{I}_d\\
    &=\tilde{\sigma}^2\mathsf{I}_d\label{variance_vec}
\end{align}
\end{Claim}
\begin{Proof}[Proof of Claim~\ref{cond_vec}]
$\mathbf{Y}_\ell=\hat{M}(\mathbf{x}_\ell)=\mathbf{x}_\ell+\sum_{j<\ell,j\notin\mathcal{U}_{col}}\mathbf{S}_{j,\ell}-\sum_{j>\ell,j\notin\mathcal{U}_{col}}\mathbf{S}_{\ell,j}+\mathbf{N}_\ell$, for all $\ell\in\mathcal{U}_{all}\setminus\mathcal{U}_{col}$. Therefore, $\mathbf{Y}_\ell\sim\mathcal{N}(\mathbf{x}_\ell,(\sigma^2+rc)\mathsf{I}_d)$. The distribution of the $k$th component of $\mathbf{Y}_\ell$, across all $\ell\in\mathcal{U}_{all}\setminus\mathcal{U}_{col}=[1:n-c]$ is given by,
\begin{align}
    &\begin{pmatrix}
        Y_{1,k}\\\vdots\\\!Y_{n\!-\!c,k}\!
    \end{pmatrix}\!\!\sim\!\mathcal{N}\!\begin{pmatrix}\!\!\!
        \begin{pmatrix}
            x_{1,k}\\\vdots\\x_{n\!-\!c,k}
        \end{pmatrix}\!,\!\begin{pmatrix}
            \sigma^2\!+\!rc \!\!\!&\!\!\! r \!\!\!&\!\!\! \dotsc \!\!\!&\!\!\! r\\
            r \!\!\!&\!\!\! \sigma^2\!+\!rc  \!\!\!&\!\!\! \dotsc \!\!\!&\!\!\! r\\
            \vdots \!\!\!&\!\!\! \vdots \!\!\!&\!\!\! \vdots \!\!\!&\!\!\! \vdots \!\!\!&\!\!\!\\
            r \!\!\!&\!\!\!  r \!\!\!&\!\!\! \dotsc \!\!\!&\!\!\! \sigma^2\!+\!rc
        \!\!\!\!\!\end{pmatrix}_{\!\!\!(n\!-\!c)^2}\!\!
    \end{pmatrix}\label{y_cov}
\end{align}
for $k\in[1:d]$, as $\text{cov}(Y_{i,k},Y_{j,k})=-\mathbb{E}[S_{i,j}^2]=r$ for $i\neq j$ and $\text{var}(Y_{i,k})=-r(n-c-1)+\sigma^2+r(n-1)=\sigma^2+rc$, $\forall i$. From the above distributions and from the fact that $\text{cov}(Y_{i,k},Y_{j,k'})=0$, $ i,j\in\mathcal{U}_{all}\setminus\mathcal{U}_{col}$ and $\forall k\neq k'$, we derive,
\begin{align}
    &\begin{pmatrix}
        \mathbf{Y}_{1}\\\vdots\\\!\mathbf{Y}_{n\!-\!c}\!
    \end{pmatrix}\!\!\sim\!\mathcal{N}\!\begin{pmatrix}\!\!\!
        \begin{pmatrix}
            \mathbf{x}_{1}\\\vdots\\\mathbf{x}_{n\!-\!c}
        \end{pmatrix}\!,\!\begin{pmatrix}
            (\sigma^2\!+\!rc)\mathsf{I}_d \!\!\!\!&\!\!\!\! r\mathsf{I}_d \!\!\!\!&\!\!\!\! \dotsc \!\!\!\!&\!\!\!\! r\mathsf{I}_d\\
            r\mathsf{I}_d \!\!\!\!&\!\!\!\! (\sigma^2\!+\!rc)\mathsf{I}_d  \!\!\!\!&\!\!\!\! \dotsc \!\!\!\!&\!\!\!\! r\mathsf{I}_d\\
            \vdots \!\!\!\!&\!\!\!\! \vdots \!\!\!\!&\!\!\!\! \vdots \!\!\!\!&\!\!\!\! \vdots \!\!\!\!&\!\!\!\!\\
            r\mathsf{I}_d \!\!\!\!&\!\!\!\!  r\mathsf{I}_d \!\!\!\!&\!\!\!\! \dotsc \!\!\!\!&\!\!\!\! (\sigma^2\!+\!rc)\mathsf{I}_d
        \!\!\!\!\!\end{pmatrix}\!\!
    \end{pmatrix}
\end{align}
We use the following standard result on the conditional distributions of multivariate Gaussian distributions \cite{eaton1983multivariate}. Let $\mathbf{V}\in\mathbb{R}^d$ be $\mathbf{V}\sim\mathcal{N}(\mathbf{\hat{\mu}},\hat{\Sigma})$. Consider the partition $\mathbf{V}=[\mathbf{V}_1,\mathbf{V}_2]^T$ with $\mathbf{V}_1\in\mathbb{R}^p$ and $\mathbf{V}_2\in\mathbb{R}^{d-p}$, and let the corresponding partitions of $\mathbf{\hat{\mu}}$ and $\hat{\Sigma}$ be $\mathbf{\hat{\mu}}=[\mathbf{\hat{\mu}}_1,\mathbf{\hat{\mu}}_2]^T$ and $\hat{\Sigma}=\begin{pmatrix}
            \hat{\Sigma}_{1,1} & \hat{\Sigma}_{1,2}\\
            \hat{\Sigma}_{2,1} & \hat{\Sigma}_{2,2}  
        \end{pmatrix}$. Then, the conditional distribution of $\mathbf{V}_1|\mathbf{V}_2=\mathbf{v}_2$ is given by $\mathbf{V}_1|\mathbf{V}_2=\mathbf{v}_2\sim\mathcal{N}(\mathbf{\mu}_*,\Sigma_*)$ where,
        \begin{align}
            \mathbf{\mu}_*&=\mathbf{\hat{\mu}}_1+\hat{\Sigma}_{1,2}\hat{\Sigma}_{2,2}^{-1}(\mathbf{v}_2-\mathbf{\hat{\mu}}_2)\\
            \Sigma_*&=\hat{\Sigma}_{1,1}-\hat{\Sigma}_{1,2}\hat{\Sigma}_{2,2}^{-1}\hat{\Sigma}_{2,1}
        \end{align}
Based on this, we have, $\mathbf{Y}_1|\mathbf{Y}_2\!=\!\mathbf{y}_2,\dotsc,\mathbf{Y}_{n\!-\!c}\!=\!\mathbf{y}_{n\!-\!c}\sim N(\Tilde{\mathbf{\mu}},\Tilde{\Sigma})$ where,
\begin{align}
    \tilde{\mathbf{\mu}}&=\mathbf{x}_1+(r\mathbf{1}_{n-c-1}^T\otimes\mathsf{I}_d)\nonumber\\
    &\times\left(\!\!\!\begin{pmatrix}
            (\sigma^2+rc) \!\!\!&\!\!\! r \!\!\!&\!\!\! \dotsc \!\!\!&\!\!\! r\\
            r \!\!\!&\!\!\! (\sigma^2+rc)  \!\!\!&\!\!\! \dotsc \!\!\!&\!\!\! r\\
            \vdots \!\!\!&\!\!\! \vdots \!\!\!&\!\!\! \vdots \!\!\!&\!\!\! \vdots \\
            r \!\!\!&\!\!\!  r \!\!\!&\!\!\! \dotsc \!\!\!&\!\!\! (\sigma^2+rc)
        \end{pmatrix}\!\!\otimes\mathsf{I}_d\!\right)^{\!\!-1}\!\!\!\!\!\!\!\!\begin{pmatrix}
            \mathbf{y}_2-\mathbf{x}_2\\\vdots\\\mathbf{y}_{n\!-\!c}\!-\!\mathbf{x}_{n\!-\!c}
        \end{pmatrix}\\
        &=\!\mathbf{x}_1\!+\!\!\!\left(\!r\mathbf{1}^T_{n\!-\!c\!-\!1}\begin{pmatrix}
            (\sigma^2+rc) \!\!\!&\!\!\! r \!\!\!&\!\!\! \dotsc \!\!\!&\!\!\! r\\
            r \!\!\!&\!\!\! (\sigma^2+rc)  \!\!\!&\!\!\! \dotsc \!\!\!&\!\!\! r\\
            \vdots \!\!\!&\!\!\! \vdots \!\!\!&\!\!\! \vdots \!\!\!&\!\!\! \vdots \!\!\!&\!\!\!\\
            r \!\!\!&\!\!\!  r \!\!\!&\!\!\! \dotsc \!\!\!&\!\!\! (\sigma^2+rc)
        \end{pmatrix}^{-1}\!\!\!\!\!\otimes\mathsf{I}_d\!\!\right)\nonumber\\
        &\qquad\qquad\qquad\qquad\qquad\qquad\times\begin{pmatrix}
            \mathbf{y}_2-\mathbf{x}_2\\\vdots\\\mathbf{y}_{n-c}-\mathbf{x}_{n-c}
        \end{pmatrix}
\end{align}
using the properties $(A\otimes B)^{-1}=A^{-1}\otimes B^{-1}$ and $(A\otimes B)(C\otimes D)=(AC)\otimes(BD)$ of Kronecker products. Moreover, 
\begin{align}
    \tilde{\Sigma}&=(\sigma^2+rc)\mathsf{I}_d-(r\mathbf{1}_{n-c-1}^T\otimes\mathsf{I}_d)\nonumber\\
    &\times\left(\!\!\!\begin{pmatrix}
            \sigma^2+rc & r & \dotsc & r\\
            r & \sigma^2+rc  & \dotsc & r\\
            \vdots & \vdots & \vdots & \vdots &\\
            r &  r & \dotsc & \sigma^2+rc
        \end{pmatrix}_{(n-c-1)^2}\!\!\!\!\!\!\!\!\!\!\!\otimes\mathsf{I}_d\right)^{-1}\nonumber\\
        &\qquad\qquad\qquad\qquad\times(r\mathbf{1}_{n-c-1}\otimes\mathsf{I}_d)\\
        &=(\sigma^2+rc)\mathsf{I}_d\nonumber\\
        &\quad-r^2\mathbf{1}^T_{n\!-\!c\!-\!1}\!\!\begin{pmatrix}
            \sigma^2\!+\!rc \!\!\!&\!\!\! r \!\!\!&\!\!\! \dotsc \!\!\!&\!\!\! r\\
            r \!\!\!&\!\!\! \sigma^2\!+\!rc  \!\!\!&\!\!\! \dotsc \!\!\!&\!\!\! r\\
            \vdots \!\!\!&\!\!\! \vdots \!\!\!&\!\!\! \vdots \!\!\!&\!\!\! \vdots \\
            r \!\!\!&\!\!\!  r \!\!\!&\!\!\! \dotsc \!\!\!&\!\!\! \sigma^2\!+\!rc
        \end{pmatrix}^{\!-1}\!\!\!\!\!\mathbf{1}_{n\!-\!c\!-\!1}\otimes\mathsf{I}_d\\
        &\!=\!\!\left(\!\!(\sigma^2\!+\!rc)\!-\!r^2\mathbf{1}^T_{n\!-\!c\!-\!1}\!\!\begin{pmatrix}
            \sigma^2+rc \!\!\!\!&\!\!\!\! r \!\!\!\!&\!\!\!\! \dotsc \!\!\!\!&\!\!\!\! r\\
            r \!\!\!\!&\!\!\!\! \sigma^2+rc  \!\!\!\!&\!\!\!\! \dotsc \!\!\!\!&\!\!\!\! r\\
            \vdots \!\!\!\!&\!\!\!\! \vdots \!\!\!\!&\!\!\!\! \vdots \!\!\!\!&\!\!\!\! \vdots \!\!\!\!&\!\!\!\!\\
            r \!\!\!\!&\!\!\!\!  r \!\!\!\!&\!\!\!\! \dotsc \!\!\!\!&\!\!\!\! \sigma^2+rc\!\!\!\!\!
        \end{pmatrix}^{-1}\!\!\!\!\!\!\!\!\!\mathbf{1}_{n\!-\!c\!-\!1}\!\!\!\right)\!\!\mathsf{I}_d\\
        &=\tilde{\sigma}^2\mathsf{I}_d
\end{align}
\end{Proof}
Next, we apply Claim~\ref{cond_vec} in the privacy constraint in \eqref{simplified} to obtain the condition on $\sigma^2$ and $r$ to ensure $(\epsilon,\delta)$-DP.
\begin{Claim}\label{privacy_vec_lem}
    The DP-DME system in Theorem~\ref{thm1} satisfies $(\epsilon,\delta)$-DP when,
\begin{align}\label{sig_lb_}
    \Tilde{\sigma}^2\geq\sigma^2_{\epsilon,\delta}.
\end{align}
\end{Claim}

\begin{Proof}[Proof of Claim~\ref{privacy_vec_lem}]
    For given values of $\mathbf{Y}_j=\mathbf{y}_j$, $\forall j\in[2:n-c]$, consider the variation of $\mathbf{x}_1$ by fixing all $\mathbf{x}_j$, $\forall j\in[2:n-c]$, and define $f(\mathbf{x}_1)=\tilde{\mathbf{\mu}}$. Then, define a new random variable $\mathbf{W}=f(\mathbf{x}_1)+\mathbf{N}$, where $\mathbf{N}\sim\mathcal{N}\left(\mathbf{0}_{d},\Tilde{\sigma}^2\mathsf{I}_d\right)$. Note that for any given values of $\mathbf{y}_j$, $\mathbf{x}_j$, $\forall j\in[2:n-c]$, $\mathbf{W}\sim \mathbf{Y}_1|\mathbf{Y}_j=\mathbf{y}_j,j\in[2:n-c]$, (statistically equivalent). Now, consider the $(\epsilon,\delta)$-DP constraint in \eqref{simplified}.
\begin{align}
    &\mathbb{P}(\hat{M}(\mathbf{x}_1)\in\mathcal{A}|\hat{M}(\mathbf{x}_j)=\mathbf{y}_j,j\in[2:n-c])\nonumber\\
    &\quad\leq e^\epsilon\mathbb{P}(\hat{M}(\mathbf{x}_1')\in\mathcal{A}|\hat{M}(\mathbf{x}_j)=\mathbf{y}_j,j\in[2:n-c])+\delta
\end{align}
which is equivalent to,
\begin{align}\label{gauss2}
    &\mathbb{P}(f(\mathbf{x}_1)+\mathbf{N}\in\mathcal{A})\leq e^\epsilon\mathbb{P}(f(\mathbf{x}'_1)+\mathbf{N}\in\mathcal{A})+\delta.
\end{align}
As \eqref{gauss2} represent the standard Gaussian mechanism in DP for the query $f(\mathbf{x}_1)$, we use the results from \cite{gauss,reviewGauss} to obtain the values of $\sigma^2$ and $r$ that satisfy \eqref{gauss2}. We restate Theorem 8 of \cite{gauss} here for completeness.

\noindent\textbf{Theorem 8 of \cite{gauss}} Let $f:\mathbb{X}\to\mathbb{R}^d$ be a function with global $L_2$ sensitivity $\Delta$. For any $\epsilon>0$ and $\delta\in[0,1]$ the Gaussian output perturbation mechanism $M(x)=f(x)+Z$ with $Z\sim\mathcal{N}(\mathbf{0}_d,\hat{\sigma}^2\mathsf{I}_d)$ is $(\epsilon,\delta)$-DP if and only if,
\begin{align}\label{thm8}
    \Phi\left(\frac{\Delta}{2\hat{\sigma}}-\frac{\epsilon\hat{\sigma}}{\Delta}\right)-e^\epsilon\Phi\left(-\frac{\Delta}{2\hat{\sigma}}-\frac{\epsilon\hat{\sigma}}{\Delta}\right)\leq\delta,
\end{align}
where $\Phi(\cdot)$ is the CDF of the standard Gaussian distribution. 

Applying Theorem 8 of \cite{gauss} directly on \eqref{gauss2} gives,
\begin{align}\label{lb_siggam2}
    \Tilde{\sigma}^2\geq\sigma^2_{\epsilon,\delta},
\end{align}
where $\sigma^2_{\epsilon,\delta}=\inf_{\hat{\sigma}>0}\{\hat{\sigma};\Phi\left(\frac{\Delta}{2\hat{\sigma}}-\frac{\epsilon\hat{\sigma}}{\Delta}\right)-e^\epsilon\Phi\left(-\frac{\Delta}{2\hat{\sigma}}-\frac{\epsilon\hat{\sigma}}{\Delta}\right)\leq\delta\}$ with $\Delta=2$. The value of $\Delta$ is calculated as,
\begin{align}\label{sensitivity}
    \Delta&=\sup_{\mathbf{x}_1\in\mathbb{B}^d}\|f(\mathbf{x}_1)-f(\mathbf{x'}_1)\|=\|\mathbf{x}_1-\mathbf{x'}_1\|=2
\end{align}
The lower bound in \eqref{lb_siggam2} is due to the fact that $\Phi\left(\frac{\Delta}{2\hat{\sigma}}-\frac{\epsilon\hat{\sigma}}{\Delta}\right)-e^\epsilon\Phi\left(-\frac{\Delta}{2\hat{\sigma}}-\frac{\epsilon\hat{\sigma}}{\Delta}\right)$ for any fixed $\epsilon$ and $\Delta=2$ is a decreasing function in $\hat{\sigma}$, which is proved next. 

\begin{Claim}\label{Phidecrease}
    $\Phi\left(\frac{1}{\hat{\sigma}}-\frac{\epsilon\hat{\sigma}}{2}\right)-e^\epsilon\Phi\left(-\frac{1}{\hat{\sigma}}-\frac{\epsilon\hat{\sigma}}{2}\right)$ is a decreasing function in $\hat{\sigma}$.
\end{Claim}

\begin{Proof}
    Using the definition of the standard Gaussian CDF, for a fixed $\epsilon$ we have,
    \begin{align}
        f(\hat{\sigma})&=\Phi\left(\frac{1}{\hat{\sigma}}-\frac{\epsilon\hat{\sigma}}{2}\right)-e^\epsilon\Phi\left(-\frac{1}{\hat{\sigma}}-\frac{\epsilon\hat{\sigma}}{2}\right)\\
        &=\frac{1}{2\pi}\int_{-\infty}^{\frac{1}{\hat{\sigma}}-\frac{\epsilon\hat{\sigma}}{2}}e^{-\frac{t^2}{2}}dt-e^{\epsilon}\int_{-\infty}^{-\frac{1}{\hat{\sigma}}-\frac{\epsilon\hat{\sigma}}{2}}e^{-\frac{t^2}{2}}dt.
    \end{align}
    As $f(\hat{\sigma})$ is a smooth and continuous function, its derivative is given by,
    \begin{align}
        f'(\hat{\sigma})&=\frac{1}{2\pi}\left(-\frac{1}{\sigma^2}-\frac{\epsilon}{2}\right)e^{-\frac{1}{2}\left(\frac{1}{\hat{\sigma}^2}+\frac{\epsilon^2\hat{\sigma}^2}{4}-\epsilon\right)}\nonumber\\
        &\quad-\frac{e^\epsilon}{2\pi}\left(\frac{1}{\sigma^2}-\frac{\epsilon}{2}\right)e^{-\frac{1}{2}\left(\frac{1}{\hat{\sigma}^2}+\frac{\epsilon^2\hat{\sigma}^2}{4}+\epsilon\right)}\\
        &=-\frac{1}{\sigma^2\pi}e^{-\frac{1}{2}\left(\frac{1}{\hat{\sigma}^2}-\frac{\epsilon\hat{\sigma}}{2}\right)^2}<0.
    \end{align}
\end{Proof}
This concludes the proof of Claim~\ref{privacy_vec_lem}.
\end{Proof}

Substituting for $\tilde{\sigma}^2$ in \eqref{sig_lb_} from \eqref{variance_vec} gives,
\begin{align}
    &\left(\!\!\sigma^2\!+\!rc\!-\!r^2\mathbf{1}_{n\!-\!c\!-\!1}^T\!\!\begin{pmatrix}
        \sigma^2+rc \!\!\!&\!\!\! r \!\!\!&\!\!\! \dotsc \!\!\!&\!\!\! r\\
         r \!\!\!&\!\!\! \sigma^2+rc \!\!\!&\!\!\! \dotsc \!\!\!&\!\!\! r\\
         \vdots \!\!\!&\!\!\! \vdots \!\!\!&\!\!\! \vdots \!\!\!&\!\!\! \vdots \\
        r \!\!\!&\!\!\!  r \!\!\!&\!\!\! \dotsc \!\!\!&\!\!\! \sigma^2+rc \!\!
    \end{pmatrix}^{-1}\!\!\!\!\!\!\!\!\mathbf{1}_{n-c-1}\!\!\right)\!\geq\!\sigma^2_{\epsilon,\delta}\\
    &\sigma^2+rc-r^2\mathbf{1}_{n-c-1}^T\left((\sigma^2+r(c-1))\mathsf{I}_{n-c-1}\nonumber\right.\\
    &\left.\qquad\qquad+r\mathbf{1}_{n-c-1}\mathbf{1}_{n-c-1}^T\right)^{-1}\mathbf{1}_{n-c-1}\geq\sigma^2_{\epsilon,\delta}\label{privacycondition}
\end{align}
Using the Sherman-Morrison formula, \eqref{privacycondition} simplifies to,
    \begin{align}
        &\sigma^2+rc-r^2\mathbf{1}^T_{n-c-1}\left(\frac{1}{\sigma^2+r(c-1)}I_{n-c-1}\right.\nonumber\\
        &\left.\quad-\frac{\frac{r}{(\sigma^2+r(c-1))^2}\mathbf{1}_{n-c-1}\mathbf{1}_{n-c-1}^T}{1+\frac{r}{\sigma^2+r(c-1)}\mathbf{1}_{n-c-1}^T\mathbf{1}_{n-c-1}}\right)\mathbf{1}_{n-1}\geq\sigma^2_{\epsilon,\delta}\\
        &\frac{(\sigma^2+r(c-1))(\sigma^2+r(n-1))}{\sigma^2+r(n-2)}\geq\sigma^2_{\epsilon,\delta}
    \end{align}
which is equivalent to the expression in \eqref{privacy3}. The analysis of the roots of \eqref{privacy3} along with the constraint $\sigma^2+r(n-1)>0$ (for a positive variance of $\mathbf{N}_i$s) result in the values of $r$ given in \eqref{range}, that makes the system satisfy $(\epsilon,\delta)$-DP for a fixed $\sigma^2$. It can be shown that for any fixed $\sigma^2<\sigma^2_{\epsilon,\delta}$, $\sigma^2+r(n-1)\leq0$ for any $r$ satisfying \eqref{privacy3}. Matrices of the form $\Sigma=(\sigma^2+r(c-1))\mathsf{I}_{n}+r\mathbf{1}_n\mathbf{1}_n^T$ with such $\sigma$ and $r$ are not valid covariance matrices of $(Y_{1,k},\dotsc,Y_{n-c,k})$, $k\in[1:d]$, as they are not positive definite, i.e., $\mathbf{1}_{n-c}^T\Sigma\mathbf{1}_{n-c}=(n-c)(\sigma^2+r(n-1))\leq0$. This verifies the requirement $\sigma^2\geq\sigma^2_{\epsilon,\delta}$.

Next, we prove that the covariance matrix of $[\mathbf{Y}_{1,k},\dotsc,\mathbf{Y}_{n-c,k}]^T$ for any $0\leq c<t$, $\forall k$, i.e., matrix $\Sigma=(\sigma^2+r(c-1))\mathsf{I}_{n-c}+r\mathbf{1}_{n-c}\mathbf{1}_{n-c}^T$ with any fixed $\sigma^2\geq\sigma^2_{\epsilon,\delta}$ and any corresponding $r$ in the range \eqref{range} is positive definite.

\begin{Claim}\label{cl5}
    The matrix $\Sigma$ with $\Sigma_{i,i}=\sigma^2+rc$, $\forall i\in[1:n-c]$ and $\Sigma_{i,j}=r$, $\forall i,\in[1:n-c]$, $i\neq j$, is positive definite for any fixed $\sigma^2\geq\sigma^2_{\epsilon,\delta}$ and for any $r$ in the range \eqref{range}.
\end{Claim}

\begin{Proof}[Proof of Claim~\ref{cl5}]
    From the properties of circulant matrices \cite{circulant}, the  eigenvalues of $\Sigma$ are given by,
    \begin{align}
        \lambda_j&=\sigma^2+r\sum_{k=1}^{n-1}\omega^{kj},\quad j=0,\dotsc,n-1\\
        &=\begin{cases}
            \sigma^2+r(n-1), & j=0\\
            \sigma^2+r(c-1), & j=1,\dotsc,n-1
        \end{cases}
    \end{align}
where $\omega$ is the $n$th root of unity. It remains to prove that $\lambda_j>0$, $\forall j$ to prove the positive definiteness of $\Sigma$.

\noindent\textbf{case 1: $c=0$:} From direct substitution of the upper and lower bounds  in \eqref{range2} on $r$, one can observe that $\sigma^2-r>0$ and $\sigma^2+r(n-1)>0$, respectively. The explicit calculations are given below. This proves $\lambda_j>0$, $\forall j$.

\noindent\emph{Proof of $\sigma^2+r(n-1)>0$}: Considering the lower bound on $r$ in \eqref{range2}, we have,
\begin{align}
        &\sigma^2+r(n-1)\nonumber\\
        &\geq\sigma^2+\left(\frac{(n-2)(\sigma^2-\sigma^2_{\epsilon,\delta})}{2(n-1)}\right.\nonumber\\
        &\left.-\frac{\sqrt{(n-2)^2(\sigma^2-\sigma^2_{\epsilon,\delta})^2+4(n-1)\sigma^2(\sigma^2-\sigma^2_{\epsilon,\delta})}}{2(n-1)}\right)(n-1)\\
        &=\frac{1}{2}\left(2\sigma^2+(n-2)(\sigma^2-\sigma^2_{\epsilon,\delta})\right.\nonumber\\
        &\left.\quad-\sqrt{(n-2)^2(\sigma^2-\sigma^2_{\epsilon,\delta})^2+4(n-1)\sigma^2(\sigma^2-\sigma^2_{\epsilon,\delta})}\right)\\
        &=\frac{1}{2}(p-q),
    \end{align}
    where $p=2\sigma^2+(n-2)(\sigma^2-\sigma^2_{\epsilon,\delta})>0$ and $q=\sqrt{(n-2)^2(\sigma^2-\sigma^2_{\epsilon,\delta})^2+4(n-1)\sigma^2(\sigma^2-\sigma^2_{\epsilon,\delta})}\geq0$. Note that,
    \begin{align}
        p^2\!-\!q^2 &\!=\!4\sigma^4\!+\!4\sigma^2(n-2)(\sigma^2-\sigma^2_{\epsilon,\delta})\!-\!4(n-1)\sigma^2(\sigma^2\!-\!\sigma^2_{\epsilon,\delta})\\
        &=4\sigma^2(\sigma^2-\sigma^2+\sigma^2_{\epsilon,\delta})>0,
    \end{align}
which implies that $p>q$.

\noindent\emph{Proof of $\sigma^2-r>0$:} Considering the upper bound on $r$ in \eqref{range2},
\begin{align}
    \sigma^2-r&\geq\sigma^2-\frac{(n-2)(\sigma^2-\sigma^2_{\epsilon,\delta})}{2(n-1)}\nonumber\\
    &-\frac{\sqrt{(n-2)^2(\sigma^2-\sigma^2_{\epsilon,\delta})^2+4(n-1)\sigma^2(\sigma^2-\sigma^2_{\epsilon,\delta})}}{2(n-1)}\\
    &=\tilde{p}-\tilde{q},
\end{align}
where $\tilde{p}=\sigma^2-\frac{(n-2)(\sigma^2-\sigma^2_{\epsilon,\delta})}{2(n-1)}>0$ and $\tilde{q}=\frac{\sqrt{(n-2)^2(\sigma^2-\sigma^2_{\epsilon,\delta})^2+4(n-1)\sigma^2(\sigma^2-\sigma^2_{\epsilon,\delta})}}{2(n-1)}\geq0$. Note that $\tilde{p}^2-\tilde{q}^2>0$, which implies that $\Tilde{p}>\Tilde{q}$.

\noindent\textbf{case 2: $c=1$: } If $r\geq0$, $\lambda_j>0$, $\forall j$, and $\Sigma$ is positive definite. For $r<0$ satisfying \eqref{range}, $\sigma^2+r(c-1)\geq\sigma^2+r(n-1)$. Therefore, it remains to prove that $\sigma^2+r(n-1)>0$ for $r$ satisfying \eqref{range} for any $\sigma^2\geq\sigma^2_{\epsilon,\delta}$. Using the lower bound in \eqref{range},
\begin{align}
    \sigma^2+r(n-1)&\geq\sigma^2-\frac{\sigma^2(\sigma^2-\sigma^2_{\epsilon,\delta})(n-1)}{\sigma^2(n-1)-\sigma^2_{\epsilon,\delta}(n-2)}\\
    &=\frac{\sigma^2\sigma^2_{\epsilon,\delta}}{\sigma^2(n-1)-\sigma^2_{\epsilon,\delta}(n-2)}>0.
\end{align}

\noindent\textbf{case 3: $c>1$: } Similar to case 2, if $r\geq0$, $\lambda_j>0$, $\forall j$, and $\Sigma$ is positive definite. For $r<0$ satisfying \eqref{range}, $\sigma^2+r(n-1)\leq\sigma^2+r(c-1)$. Therefore, it remains to prove that $\sigma^2+r(n-1)>0$ for $r$ satisfying \eqref{range} for any $\sigma^2\geq\sigma^2_{\epsilon,\delta}$. From the lower bound on $r$ in \eqref{range}, we have,
\begin{align}
    &\sigma^2+r(n-1)\nonumber\\
    &\geq\sigma^2+\frac{-(n-2)(\sigma^2-\sigma^2_{\epsilon,\delta})-\sigma^2c}{2(c-1)}\nonumber\\
    &+\frac{\sqrt{((n-2)(\sigma^2-\sigma^2_{\epsilon,\delta})-\sigma^2c)^2+4(n-c-1)\sigma^2(\sigma^2-\sigma^2_{\epsilon,\delta})}}{2(c-1)}\\
    &\geq\frac{-((n-2)(\sigma^2-\sigma^2_{\epsilon,\delta})-\sigma^2c)-2\sigma^2}{2(c-1)}\nonumber\\
    &+\frac{\sqrt{((n-2)(\sigma^2-\sigma^2_{\epsilon,\delta})-\sigma^2c)^2+4(n-c-1)\sigma^2(\sigma^2-\sigma^2_{\epsilon,\delta})}}{2(c-1)}
\end{align}
If $-((n-2)(\sigma^2-\sigma^2_{\epsilon,\delta})-\sigma^2c)-2\sigma^2\geq0$, $\sigma^2+r(n-1)>0$ and the claim is proved. Consider the case where $-((n-2)(\sigma^2-\sigma^2_{\epsilon,\delta})-\sigma^2c)-2\sigma^2<0$. To prove that $\sigma^2+r(n-1)>0$, we prove that $p>|q|$, where,
\begin{align}
    p&=\sqrt{((n-2)(\sigma^2-\sigma^2_{\epsilon,\delta})-\sigma^2c)^2+4(n-c-1)\sigma^2(\sigma^2-\sigma^2_{\epsilon,\delta})}\\
    q&=-((n-2)(\sigma^2-\sigma^2_{\epsilon,\delta})-\sigma^2c)-2\sigma^2
\end{align}
As $p>0$ and $|q|>0$, consider,
\begin{align}
    &p^2-|q|^2\nonumber\\
    &=((n-2)(\sigma^2-\sigma^2_{\epsilon,\delta})-\sigma^2c)^2+4(n-c-1)\sigma^2(\sigma^2-\sigma^2_{\epsilon,\delta})\nonumber\\
    &\quad-((n-2)(\sigma^2-\sigma^2_{\epsilon,\delta})-\sigma^2c)^2\nonumber\\
    &\quad-4\sigma^2((n-2)(\sigma^2-\sigma^2_{\epsilon,\delta})-\sigma^2c)-4\sigma^4\\
    &=-4\sigma^2(\sigma^2-\sigma^2_{\epsilon,\delta})(c-1)+4\sigma^4(c-1)\\
    &=4\sigma^2\sigma^2_{\epsilon,\delta}(c-1)>0
\end{align}
as $c>1$. This proves $p>|q|$.
\end{Proof}

Claims~\ref{cond_vec} and~\ref{privacy_vec_lem} collectively provide the feasible range of $r$ for a fixed $\sigma^2$ to ensure privacy and Claim~\ref{cl5} proves that these values of $\sigma^2$ and (corresponding) $r$ result in valid covariance matrices, which proves Lemma~\ref{privacylem}.
\end{Proof}

Next, we find the optimum values of $\sigma^2$ and $r$ that minimizes the MSE while satisfying the privacy constraint in Lemma~\ref{privacylem}.

\begin{lemma}[Optimum noise parameters - No dropouts]\label{optsigmarho}
For any fixed $\sigma^2\geq\sigma^2_{\epsilon,\delta}$, the optimum $r=\rho\sigma^2$ that satisfies $(\epsilon,\delta)$-DP is given by,
\begin{align}
    r_*&=\arg\min_{r} \min_{\boldsymbol{\alpha}_{\mathcal{U}_{all}}} \mathrm{ MSE}(n,\mathcal{U}_{all},\sigma^2,r,\boldsymbol{\alpha}_{\mathcal{U}_{all}})\\
    &=\begin{cases}
        a-b, & c=0\\
        \frac{-\sigma^2(\sigma^2-\sigma^2_{\epsilon,\delta})}{\sigma^2(n-1)-\sigma^2_{\epsilon,\delta}(n-2)}, & c=1\\
            \frac{-(n-2)(\sigma^2-\sigma^2_{\epsilon,\delta})-\sigma^2c}{2(n-1)(c-1)}\\
            +\frac{\sqrt{((n-2)(\sigma^2-\sigma^2_{\epsilon,\delta})-\sigma^2c)^2
            +4(n-c-1)\sigma^2(\sigma^2-\sigma^2_{\epsilon,\delta})}}{2(n-1)(c-1)}, & c>1
    \end{cases} \label{optrho}
\end{align}
 where,
\begin{align}
        a&=\frac{(n-2)(\sigma^2-\sigma^2_{\epsilon,\delta})}{2(n-1)}\\ 
        b&=\frac{\sqrt{(n-2)^2(\sigma^2-\sigma^2_{\epsilon,\delta})^2+4(n-1)\sigma^2(\sigma^2-\sigma^2_{\epsilon,\delta})}}{2(n-1)}.
    \end{align} 
For the case where $t=n$, $\min_{r} \min_{\boldsymbol{\alpha}_{\mathcal{U}_{all}}}\mathrm{MSE}(n,\mathcal{U}_{all},\sigma^2,r,\boldsymbol{\alpha}_{\mathcal{U}_{all}})$ is a decreasing function in $\sigma^2$ and,  
\begin{align}
    \min_{\sigma^2\geq\sigma^2_{\epsilon,\delta}}\min_{r} &\min_{\boldsymbol{\alpha}_{\mathcal{U}_{all}}}\mathrm{MSE}(n,\mathcal{U}_{all},\sigma^2,r,\boldsymbol{\alpha}_{\mathcal{U}_{all}})\nonumber\\
    &=\lim_{\sigma^2\to\infty}\min_{r} \min_{\boldsymbol{\alpha}}\mathrm{MSE}(n,\mathcal{U}_{all},\sigma^2,r,\boldsymbol{\alpha}_{\mathcal{U}_{all}})\\
    &=\begin{cases}
        O\left(\frac{d\sigma^2_{\epsilon,\delta}}{n(n-c)}\right), & \text{if } n(n-c)\gg d\\
        O(1), & otherwise.
    \end{cases}
\end{align}
\end{lemma}

\begin{Proof}[Proof of Lemma~\ref{optsigmarho}] The proof consists of two steps, namely, 1) optimize $r$ for fixed $\sigma^2$, 2) optimize $\sigma^2$ to minimize the MSE. The first step is characterized in Claim~\ref{cl6}. 

\begin{Claim}\label{cl6}
    For any fixed $\sigma^2\geq\sigma^2_{\epsilon,\delta}$ the optimum $r$ is given in \eqref{optrho}.
\end{Claim}

\begin{Proof}
   For any given $\sigma^2$, $r$, $\mathcal{U}$, from Proposition~\ref{vec_opt_dec}, 
    \begin{align}
        \min_{\boldsymbol{\alpha}_{\mathcal{U}}}\mathrm{MSE}(n,\mathcal{U},\sigma^2,r,\boldsymbol{\alpha}_{\mathcal{U}})&=\left(1+\frac{|\mathcal{U}|/d}{\sigma^2+r(|\mathcal{U}|-1)}\right)^{-1}
    \end{align}
    is an increasing function in $r$. For a given $\sigma^2\geq\sigma^2_{\epsilon,\delta}$, any $r$ in the range specified in Lemma~\ref{privacylem} satisfies $(\epsilon,\delta)$-DP, and results in a valid covariance matrix (Claim~\ref{cl5}). Therefore, the optimum $r$ is given by the minimum value in the respective ranges \eqref{range}-\eqref{range2}.
\end{Proof}
To find the optimum $\sigma^2$, we begin with the following claim.

\begin{Claim}\label{cl7}
For the case of no dropouts, i.e., $t=n$, $\min_{r}\min_{\boldsymbol{\alpha}_{\mathcal{U}_{all}}}\mathrm{MSE}(n,\mathcal{U}_{all},\sigma^2,r,\boldsymbol{\alpha}_{\mathcal{U}_{all}})$ is a decreasing function in $\sigma^2$ for any $\mathcal{U}_{col}$ satisfying $0\leq|\mathcal{U}_{col}|<n$.
\end{Claim}

\begin{Proof}
From Proposition~\ref{vec_opt_dec} and Lemma~\ref{privacylem},
    \begin{align}\label{minrhoalp}
        \min_{r}\min_{\boldsymbol{\alpha}_{\mathcal{U}_{all}}}\mathrm{MSE}(n,\mathcal{U}_{all},\sigma^2,r,\boldsymbol{\alpha})=\left(1+\frac{n/d}{\sigma^2+r_*(n-1)}\right)^{-1}
    \end{align}
as $|\mathcal{U}|=n$. Let $f(\sigma^2)=\sigma^2+r_*(n-1)$. We prove that \eqref{minrhoalp} decreases with $\sigma^2$ by showing that $f(\sigma^2)$ is a decreasing function in $\sigma^2$ for each of the three cases in \eqref{optrho}. 

\noindent\textbf{case 1: $c=0$:}
\begin{align}
        \frac{df(\sigma^2)}{d\sigma^2}&=1+(n-1)\frac{dr_*}{d\sigma^2}\\
        &=1+\frac{n-2}{2}\nonumber\\
        &-\frac{(\sigma^2-\sigma^2_{\epsilon,\delta})(n-2)^2+2(n-1)(2\sigma^2-\sigma^2_{\epsilon,\delta})}{2\sqrt{(n-2)^2(\sigma^2-\sigma^2_{\epsilon,\delta})^2+4(n-1)\sigma^2(\sigma^2-\sigma^2_{\epsilon,\delta})}}\\
        &=\frac{1}{2}(\hat{p}-\hat{q})
    \end{align}
where $\hat{p}=n>0$ and $\hat{q}=\frac{(\sigma^2-\sigma^2_{\epsilon,\delta})(n-2)^2+2(n-1)(2\sigma^2-\sigma^2_{\epsilon,\delta})}{\sqrt{(n-2)^2(\sigma^2-\sigma^2_{\epsilon,\delta})^2+4(n-1)\sigma^2(\sigma^2-\sigma^2_{\epsilon,\delta})}}>0$. As $\hat{p}^2-\hat{q}^2<0$ implies $\frac{df(\sigma^2)}{d\sigma^2}<0$, consider:
\begin{align}
    &(\hat{p}^2-\hat{q}^2)A^2\nonumber\\
    &=n^2(n-2)^2(\sigma^2-\sigma^2_{\epsilon,\delta})^2+4n^2(n-1)\sigma^2(\sigma^2-\sigma^2_{\epsilon,\delta})\nonumber\\
    &\quad-(\sigma^2-\sigma^2_{\epsilon,\delta})^2(n-2)^4-4(n-1)^2(2\sigma^2-\sigma^2_{\epsilon,\delta})^2\nonumber\\
    &\quad-4(n-1)(n-2)^2(\sigma^2-\sigma^2_{\epsilon,\delta})(2\sigma^2-\sigma^2_{\epsilon,\delta})\\
    &=4(n-1)(\sigma^2-\sigma^2_{\epsilon,\delta})^2(n-2)^2\nonumber\\
    &\quad+4(n-1)(\sigma^2-\sigma^2_{\epsilon,\delta})(n^2\sigma^2-(n-2)^2(2\sigma^2-\sigma^2_{\epsilon,\delta}))\nonumber\\
    &\quad-4(n-1)^2(2\sigma^2-\sigma^2_{\epsilon,\delta})^2\\
    &=4(n-1)(\sigma^2-\sigma^2_{\epsilon,\delta})^2(n-2)^2+16(n-1)^2(\sigma^2-\sigma^2_{\epsilon,\delta})\sigma^2\nonumber\\
    &\quad-4(n-1)(n-2)^2(\sigma^2-\sigma^2_{\epsilon,\delta})^2-4(n-1)^2(2\sigma^2-\sigma^2_{\epsilon,\delta})^2\\
    &=4(n-1)^2(4\sigma^4-4\sigma^2\sigma^2_{\epsilon,\delta}-4\sigma^4+4\sigma^2\sigma^2_{\epsilon,\delta}-\sigma^4_{\epsilon,\delta})\\
    &=-4\sigma^4_{\epsilon,\delta}(n-1)^2<0
\end{align}
where $A=\sqrt{(n-2)^2(\sigma^2-\sigma^2_{\epsilon,\delta})^2+4(n-1)\sigma^2(\sigma^2-\sigma^2_{\epsilon,\delta})}$.

\noindent\textbf{case 2: $c=1$:}
\begin{align}
    r_*&=\frac{-\sigma^2(\sigma^2-\sigma^2_{\epsilon,\delta})}{\sigma^2(n-1)-\sigma^2_{\epsilon,\delta}(n-2)}\\
    \frac{df(\sigma^2)}{d\sigma^2}&=1+(n-1)\frac{dr_*}{d\sigma^2}\\
    &\!=\!1\!+\!(n\!-\!1)\left(\frac{(\sigma^2_{\epsilon,\delta}-2\sigma^2)(\sigma^2(n-1)-\sigma^2_{\epsilon,\delta}(n-2))}{(\sigma^2(n-1)-\sigma^2_{\epsilon,\delta}(n-2))^2}\right.\nonumber\\
    &\left.\qquad+\frac{(n-1)\sigma^2(\sigma^2-\sigma^2_{\epsilon,\delta})}{(\sigma^2(n-1)-\sigma^2_{\epsilon,\delta}(n-2))^2}\right)\\
    &=-\sigma^4_{\epsilon,\delta}(n-2)<0
\end{align}

\noindent\textbf{case 3: $c>1$:}
\begin{align}
    &r*=\frac{-(n\!-\!2)(\sigma^2\!-\!\sigma^2_{\epsilon,\delta})\!-\!\sigma^2c}{2(n\!-\!1)(c\!-\!1)}\nonumber\\
    &+\frac{\sqrt{((n\!-\!2)(\sigma^2\!-\!\sigma^2_{\epsilon,\delta})\!-\!\sigma^2c)^2
            +4(n\!-\!c\!-\!1)\sigma^2(\sigma^2\!-\!\sigma^2_{\epsilon,\delta})}}{2(n\!-\!1)(c\!-\!1)}\\
    &f(\sigma^2)=\sigma^2+r_*(n-1)\\
    &=\frac{\sigma^2(c\!-\!n)\!+\!\sigma^2_{\epsilon,\delta}(n\!-\!2)}{2(c\!-\!1)}\nonumber\\
    &+\frac{\sqrt{((n\!-\!2)(\sigma^2\!-\!\sigma^2_{\epsilon,\delta})-c\sigma^2)^2\!+\!4(n-c-1)\sigma^2(\sigma^2-\sigma^2_{\epsilon,\delta})}}{2(c-1)}\\
    &\frac{df(\sigma^2)}{d\sigma^2}=\frac{c-n}{2(c-1)}\nonumber\\
    &+\frac{\sigma^2(n-c)^2-\sigma^2_{\epsilon,\delta}(n^2-nc-2n+2)}{2\!(c\!-\!1)\!\sqrt{((n\!-\!2)(\sigma^2-\sigma^2_{\epsilon,\delta})\!-\!c\sigma^2)^2\!+\!4(n\!-\!c\!\!-\!1)\sigma^2(\sigma^2\!-\!\!\sigma^2_{\epsilon,\delta})}}
\end{align}
If $\sigma^2(n-c)^2-\sigma^2_{\epsilon,\delta}(n^2-nc-2n+2)\leq0$, $\frac{df(\sigma^2)}{d\sigma^2}<0$. Consider the case where $\sigma^2(n-c)^2-\sigma^2_{\epsilon,\delta}(n^2-nc-2n+2)>0$. Let $p=\frac{\sigma^2(n-c)^2-\sigma^2_{\epsilon,\delta}(n^2-nc-2n+2)}{2(c-1)\sqrt{((n-2)(\sigma^2-\sigma^2_{\epsilon,\delta})-c\sigma^2)^2+4(n-c-1)\sigma^2(\sigma^2-\sigma^2_{\epsilon,\delta})}}>0$ and $q=\frac{n-c}{2(c-1)}>0$, where $\frac{df(\sigma^2)}{d\sigma^2}=p-q$. To prove that $\frac{df(\sigma^2)}{d\sigma^2}<0$, it remains to prove $p^2-q^2<0$.
\begin{align}
    p^2-q^2&=\frac{1}{4(c-1)^2}(B-(n-c)^2)\\
    &=\frac{\sigma^2_{\epsilon,\delta}(n-1)}{B(c-1)}(c^2-nc+n-1)
\end{align}
where $B=\frac{(\sigma^2(n-c)^2-\sigma^2_{\epsilon,\delta}(n^2-nc-2n+2))^2}{((n-2)(\sigma^2-\sigma^2_{\epsilon,\delta})-c\sigma^2)^2+4(n-c-1)\sigma^2(\sigma^2-\sigma^2_{\epsilon,\delta})}>0$. Note that
$c^2-nc+n-1<0$ for $2\leq c\leq n-2$. For $c=n-1$, $c^2-nc+n-1=0$. Therefore, $\frac{df(\sigma^2)}{d\sigma^2}<0$ for $2\leq c\leq n-2$ and $\frac{df(\sigma^2)}{d\sigma^2}=0$ when $c=n-1$, which corresponds to LDP.
\end{Proof}

Therefore, whenever $0\leq c<n-1$ $\min_{\sigma^2\geq\sigma^2_{\epsilon,\delta}}\min_{r}\min_{\boldsymbol{\alpha}_{\mathcal{U}_{all}}}\mathrm{MSE}(n,\mathcal{U}_{all},\sigma^2,r,\boldsymbol{\alpha}_{\mathcal{U}_{all}})$ is achieved when $\sigma^2\to\infty$. For $c=n-1$ which corresponds to the case of LDP, any $\sigma^2\geq\sigma^2_{\epsilon,\delta}$ (with the corresponding $r_*$) gives the same MSE. Hence, the solution $\sigma^2=\sigma^2_{\epsilon,\delta}$ and $r_*=0$ suffices to reach the MMSE when $c=n-1$. 

\begin{Claim}\label{limit} For any $0\leq c< n-1$, 
\begin{align}
    &\lim_{\sigma^2\to\infty} \min_{r}\min_{\boldsymbol{\alpha}_{\mathcal{U}_{all}}}\mathrm{MSE}(n,\mathcal{U}_{all},\sigma^2,r,\boldsymbol{\alpha}_{\mathcal{U}_{all}})\nonumber\\
    &\qquad=\begin{cases}
        O\left(\frac{d\sigma^2_{\epsilon,\delta}}{n(n-c)}\right), & \text{if } n(n-c)\gg d\\
        O(1), & otherwise.
    \end{cases}
\end{align}
\end{Claim}
\begin{Proof}
    Here we prove the claim for the most general case of $c>1$. The proof of other two cases $c=0$ and $c=1$ follow similar steps. Recall that for the case where $|\mathcal{U}|=n$,
    \begin{align}\label{copied}
        \min_{r}\min_{\boldsymbol{\alpha}_{\mathcal{U}_{all}}}\mathrm{MSE}(n,\mathcal{U}_{all},\sigma^2,r,\boldsymbol{\alpha}_{\mathcal{U}_{all}})=\left(1+\frac{n/d}{\sigma^2+r_*(n-1)}\right)^{-1}
    \end{align}
    where,
    \begin{align}
        &r_*=\frac{-(n-2)(\sigma^2-\sigma^2_{\epsilon,\delta})-\sigma^2c}{2(n-1)(c-1)}\nonumber\\
        &\quad+\frac{\sqrt{((n-2)(\sigma^2-\sigma^2_{\epsilon,\delta})-\sigma^2c)^2
            +4(n-c-1)\sigma^2(\sigma^2-\sigma^2_{\epsilon,\delta})}}{2(n-1)(c-1)}
    \end{align}
    To find the limit of \eqref{copied} as $\sigma^2\to\infty$, we first consider the following.
    \begin{align}
        &\sqrt{((n-2)(\sigma^2-\sigma^2_{\epsilon,\delta})-\sigma^2c)^2
            +4(n-c-1)\sigma^2(\sigma^2-\sigma^2_{\epsilon,\delta})}\nonumber\\
        &\quad=(\sigma^2-\sigma^2_{\epsilon,\delta})^{\frac{1}{2}}\left(\sigma^2(n-c)^2+\frac{c^2\sigma^2\sigma^2_{\epsilon,\delta}}{\sigma^2-\sigma^2_{\epsilon,\delta}}-\sigma^2_{\epsilon,\delta}(n-2)^2\right)^\frac{1}{2}\label{sqrt}\\
        &\quad=\sigma\left(1-\frac{1}{2}\frac{\sigma^2_{\epsilon,\delta}}{\sigma^2}-\frac{1}{4}\frac{1}{2!}\frac{\sigma^4_{\epsilon,\delta}}{\sigma^4}+\dotsc\right)\times\sigma(n-c)\nonumber\\
        &\quad\times\left(1+\frac{1}{2}\frac{\frac{c^2\sigma^2\sigma^2_{\epsilon,\delta}}{\sigma^2-\sigma^2_{\epsilon,\delta}}-\sigma^2_{\epsilon,\delta}(n-2)^2}{\sigma^2(n-c)^2}\right.\nonumber\\
        &\left.\quad-\frac{1}{4}\frac{1}{2!}\left(\frac{\frac{c^2\sigma^2\sigma^2_{\epsilon,\delta}}{\sigma^2-\sigma^2_{\epsilon,\delta}}-\sigma^2_{\epsilon,\delta}(n-2)^2}{\sigma^2(n-c)^2}\right)^2+\dotsc\right)
    \end{align}
where \eqref{sqrt} is obtained by applying the binomial expansion.
\begin{align}
    &\sigma^2+r_*(n-1)\nonumber\\
    &=\frac{1}{2(c-1)}\left(\sigma^2(c-n)+\sigma^2_{\epsilon,\delta}(n-2)\right.\nonumber\\
    &\left.+\sigma^2(n\!-\!c)\!\!\left(\!\!1\!-\!\frac{1}{2}\frac{\sigma^2_{\epsilon,\delta}}{\sigma^2}+\frac{1}{2}\frac{\frac{c^2\sigma^2\sigma^2_{\epsilon,\delta}}{\sigma^2-\sigma^2_{\epsilon,\delta}}\!-\!\sigma^2_{\epsilon,\delta}(n\!-\!2)^2}{\sigma^2(n-c)^2}
    \!+\!O(\frac{1}{\sigma^4})\!\!\right)\!\!\right)\\
    &=\frac{1}{2(c-1)}\left(\sigma^2_{\epsilon,\delta}(n-2)-\frac{1}{2}\sigma^2_{\epsilon,\delta}(n-c)\right.\nonumber\\
    &\quad\left.+\frac{c^2\sigma^2\sigma^2_{\epsilon,\delta}}{2(n-c)(\sigma^2-\sigma^2_{\epsilon,\delta})}-\frac{\sigma^2_{\epsilon,\delta}(n-2)^2}{2(n-c)}+O(\frac{1}{\sigma^2})\right)\\
    &=\frac{1}{2(c-1)}\left(\frac{-\sigma^2_{\epsilon,\delta}(c-2)^2}{2(n-c)}+\frac{c^2\sigma^2\sigma^2_{\epsilon,\delta}}{2(n-c)(\sigma^2-\sigma^2_{\epsilon,\delta})}\right.\nonumber\\
    &\left.\quad+O\left(\frac{1}{\sigma^2}\right)\right)
\end{align}
Next, consider the limit,
    \begin{align}
        &\lim_{\sigma^2\to\infty}\min_{r}\min_{\boldsymbol{\alpha}_{\mathcal{U}_{all}}}\mathrm{MSE}(n,\mathcal{U}_{all},\sigma^2,r,\boldsymbol{\alpha}_{\mathcal{U}_{all}})\nonumber\\
        &=\lim_{\sigma^2\to\infty}\left(1+\frac{n/d}{\sigma^2+r_*(n-1)}\right)^{-1}\\
        &=\lim_{\sigma^2\to\infty}\!\!\!\left(\!\!1\!+\!\frac{n/d}{\frac{1}{2(c-1)}\!\!\!\left(\!\frac{-\sigma^2_{\epsilon,\delta}(c-2)^2}{2(n-c)}\!+\!\frac{c^2\sigma^2\sigma^2_{\epsilon,\delta}}{2(n-c)(\sigma^2-\sigma^2_{\epsilon,\delta})}\!+\!O\left(\frac{1}{\sigma^2}\right)\right)}\!\!\right)^{\!-1}\\
        &=\lim_{\sigma^2\to\infty}\left(1+\frac{n(n-c)}{d\sigma^2_{\epsilon,\delta}}\right)^{-1}\\
        &=\begin{cases}
            O\left(\frac{d\sigma^2_{\epsilon,\delta}}{n(n-c)}\right), & n(n-c)>>d\\
            O(1), & otherwise.
        \end{cases}
    \end{align}
Claims~\ref{cl6}-~\ref{limit} collectively prove Lemma~\ref{optsigmarho}.
\end{Proof}

\end{Proof}

Next, we derive the optimum noise parameters for the case with dropouts, i.e., any $\mathcal{U}_{col}$ and $\mathcal{U}$ with $|\mathcal{U}|<n$. We first note that the DME process is the same as what is considered in the case with no dropouts with a reduced number of users. As the noise parameters of the privacy mechanism ($\sigma^2$ and $r$) are defined prior to the aggregation stage at the server, they are optimized for the worst case with the largest number of dropouts, resulting in an increase in the MSE compared to the case with no dropouts. However, the decoder can be optimized for the surviving number of users at the time of aggregation, as aggregation is performed after obtaining all responses from the remaining users. Therefore, the proof of the optimum noise parameters and the decoder consists of two steps, namely, 1) determining the optimum decoder for any given number of surviving users, 2) calculating the optimum noise parameters of the privacy mechanism, considering the worst case dropouts and colluding users.

Step 1 is direct from Proposition~\ref{vec_opt_dec}. Proposition~\ref{vec_opt_dec} provides the optimum decoder for any fixed $\sigma^2$ and $r=\rho\sigma^2$ with any set of responding users $\mathcal{U}\subseteq\mathcal{U}_{all}$, $|\mathcal{U}|\geq t$. Even though all users may not respond when allowing for dropouts, the same conditions in Lemma~\ref{privacylem} must be satisfied by the noise parameters to satisfy the strongest form of $(\epsilon,\delta)$-DP, considering the maximum information leakage that occurs when all $n$ users respond. Therefore, the problem to be solved is given by,
\begin{align}\label{prob}
&\min_{r,\sigma^2}\max_{\substack{\mathcal{U}\subset\mathcal{U}_{all}\\ t\leq|\mathcal{U}|<n}} \min_{\boldsymbol{\alpha}_{\mathcal{U}}}\mathrm{MSE}(n,\mathcal{U},\sigma^2,r,\boldsymbol{\alpha}_{\mathcal{U}})\nonumber\\
&=\min_{\sigma^2\geq\sigma^2_{\epsilon,\delta}}\!\!\min_{r}\max_{\substack{\mathcal{U}\subset\mathcal{U}_{all}\\ t\leq|\mathcal{U}|<n}}\left(1+\frac{|\mathcal{U}|/d}{\sigma^2+r(|\mathcal{U}|-1)}\right)^{-1}\\
    &=\min_{\sigma^2\geq\sigma^2_{\epsilon,\delta}}\min_{r}\left(1+\frac{t/d}{\sigma^2+r(t-1)}\right)^{-1},
\end{align}
as $\left(1+\frac{|\mathcal{U}|/d}{\sigma^2+r(|\mathcal{U}|-1)}\right)^{-1}$ is decreasing in $|\mathcal{U}|$ for any fixed $\sigma^2$ and $r$ that result in a valid covariance matrix, i.e., satisfies $\sigma^2+r(n-1)>0$. For any fixed $\sigma^2\geq\sigma^2_{\epsilon,\delta}$, $\left(1+\frac{t/d}{\sigma^2+r(t-1)}\right)^{-1}$ increases with $r$, and the optimal $r$ is given by the $r_*$ in \eqref{optrho}. The next step is to solve,
\begin{align}\label{sig_opt_d}
    \sigma^{2}_{opt}&=\arg\min_{\sigma^2\geq\sigma^2_{\epsilon,\delta}}\min_{r}\max_{\substack{\mathcal{U}\subset\mathcal{U}_{all}\\ t\leq|\mathcal{U}|<n}}\min_{\boldsymbol{\alpha}_{\mathcal{U}}}\mathrm{MSE}(n,\mathcal{U},\sigma^2,r,\boldsymbol{\alpha}_{\mathcal{U}})\\
    &=\arg\min_{\sigma^2\geq\sigma^2_{\epsilon,\delta}}\left(1+\frac{t/d}{\sigma^2+r_*(t-1)}\right)^{-1}.
\end{align}

\begin{lemma}[Optimum noise variance - with dropouts]\label{dropoutvarlem} For the case with $n$ users out of which up to any $n-t$ are allowed to dropout, the optimum noise variance $\sigma^2_{opt}$ that solves \eqref{sig_opt_d} is given by,
\begin{align}\label{optsigmaa}
    \sigma^2_{opt}&=\frac{\sigma^2_{\epsilon,\delta}(n^2-2n-cn+2)}{(n-c)^2}\nonumber\\
    &+\frac{\sigma^2_{\epsilon,\delta}(n-c-1)(n+c-2nc+t(n+c-2))}{(n-c)^2\sqrt{(t-c)(n-t)(n-c-1)}}.
\end{align}
\end{lemma}

\begin{Proof}
To find $\sigma^2_{opt}$ consider the following notation for any given $\sigma^2\geq\sigma^2_{\epsilon,\delta}$ with $\epsilon,\delta,n,c$ and $t$ fixed.
\begin{align}
    G(\sigma^2)&=\min_{r}\max_{\substack{\mathcal{U}\subset\mathcal{U}_{all}\\ t\leq|\mathcal{U}|<n}}\min_{\boldsymbol{\alpha}_{\mathcal{U}}}\mathrm{MSE}(n,\mathcal{U},\sigma^2,r,\boldsymbol{\alpha}_{\mathcal{U}})\\
    &=\left(1+\frac{t/d}{\sigma^2+r_*(t-1)}\right)^{-1}    
\end{align}

\begin{Claim}\label{cl8}
    The critical point of $G(\sigma^2)$ is given by,
    \begin{align}\label{critical_p}
        \sigma^2_{cr}&=\frac{\sigma^2_{\epsilon,\delta}(n^2-2n-cn+2)}{(n-c)^2}\nonumber\\
        &\quad+\frac{\sigma^2_{\epsilon,\delta}\lambda\sqrt{(n-c-1)(n-t)(t-c)}}{(n-c)^2(n-t)(t-c)}
    \end{align}
    where $\lambda=(n+c)(t+1)-2(nc+t)$.
\end{Claim}

\begin{Proof}
    $G(\sigma^2)=\left(1+\frac{t/d}{\sigma^2+r_*(t-1)}\right)^{-1}$ for any $\sigma^2\geq\sigma^2_{\epsilon,\delta}$ is a smooth and continuous function in $\sigma^2$ as $\sigma^2+r_*(t-1)$ is smooth and $\sigma^2+r_*(t-1)>0$ (proof in Section~\ref{pfthm1}, Claim~\ref{cl5}). Consider the derivative of $G(\sigma^2)$ with respect to $\sigma^2$ denoted by $G'(\sigma^2)$ to find its critical points.
\begin{align}\label{der}
    G'(\sigma^2)\!=\!-\!\left(\!\frac{-\frac{t}{d}\left(1\!+\!(t\!-\!1)\frac{dr_*}{d\sigma^2}\!\right)}{(\sigma^2+r_*(t-1))^2}\right)\!\!\left(\!1\!+\!\frac{t/d}{\sigma^2+r_*(t\!-\!1)}\right)^{-2}\!\!\!=0\\
    1+(t-1)\frac{dr_*}{d\sigma^2}=0\label{83dup}
\end{align}
which simplifies to \eqref{critical_p}.
\end{Proof}
Next, we show that the critical point in Claim~\ref{cl8} is the global minimum of $G(\sigma^2)$. Let $y=\frac{\sigma^2}{\sigma^2_{\epsilon,\delta}}$ and $y_{cr}=\frac{\sigma^2_{cr}}{\sigma^2_{\epsilon,\delta}}$. 
\paragraph{case 1: $c>1$} Recall that the $G'(\sigma^2)$ in \eqref{der} is of the form,
\begin{align}\label{K}
    &G'(\sigma^2)\nonumber\\
    &=K(t,c,\sigma^2,r_*)\left(1+(t-1)\frac{dr_*}{d\sigma^2}\right)\\
    &=K(t,c,\sigma^2,r_*)\nonumber\\
    &\times\!\!\left(\!\!1\!+
    \!\frac{t-1}{2(n-1)(c-1)}\!\left(\!-(n+c-2)+\frac{1}{2}\frac{1}{\sqrt{A(y)}}\frac{dA(y)}{dy}\!\right)\!\right)
\end{align}
where $K(t,c,\sigma^2,r_*)=\frac{t/d}{(\sigma^2+r_*(t-1))^2}\left(1+\frac{t/d}{\sigma^2+r_*(t-1)}\right)^{-2}>0$ and
\begin{align}
    A(y)=((n-2)(y-1)-yc)^2
            +4(n-c-1)y(y-1)
\end{align}
Let $\phi(y)=\sqrt{A(y)}$, and Define,
\begin{align}
    g(y)&=1+\frac{t-1}{2(n-1)(c-1)}\left(-(n+c-2)+h(y)\right)\\ 
    h(y)&=\frac{1}{2}\frac{1}{\sqrt{A(y)}}\frac{dA(y)}{dy}\\
    &=\frac{(y(n-c-2)-(n-2))(n-c-2)}{\phi(y)}\nonumber\\
    &\qquad\frac{-2(-n+c+1)(2y-1)}{\phi(y)}
\end{align}
where $g(y_{cr})=0$. Note that $h(y)=\phi'(y)$. Consider the derivative of $h(y)$.
\begin{align}
    h'(y)&=\frac{((n-c-2)^2+4(n-c-1))\phi(y)-\phi'(y)h(y)\phi(y)}{\phi^2(y)}\\
    &=\frac{(n-c-2)^2+4(n-c-1)-h^2(y)}{\phi(y)}\\
    &=\frac{4(c-1)(n^2-n(c+2)+(c+1))}{\phi(y)}\\
    &=\frac{4(c-1)(n-c-1)(n-1)}{\phi(y)}\geq 0\label{ineqeq}
\end{align}
where the equality in \eqref{ineqeq} holds when $c=n-1$, which corresponds to the case of LDP for which $r_*=0$ and $\sigma^2_*=\sigma^2_{\epsilon,\delta}$. Thus, for $1<c<n-1$, $h'(y)>0$. Then, for any arbitrarily small $\tilde{\delta}>0$,
\begin{align}
    &G'(\sigma^2_{cr}+\tilde{\delta})\nonumber\\
    &=K(t,c,\sigma^2_{cr}+\tilde{\delta},r_*)\nonumber\\
    &\times\left(1+\frac{t-1}{2(n-1)(c-1)}\left(-(n-c-2)+h(\frac{\sigma^2_{cr}+\tilde{\delta}}{\sigma^2_{\epsilon,\delta}})\right)\right)\\
    &=K(t,c,\sigma^2_{cr}+\tilde{\delta},r_*)\nonumber\\
    &\times\left(1+\frac{t-1}{2(n-1)(c-1)}\left(-(n-c-2)+h(y_{cr})+\delta\right)\right)\\
    &=K(t,c,\sigma^2_{cr}+\tilde{\delta},r_*)\frac{\delta(t-1)}{2(n-1)(c-1)}>0\label{g_func1}
\end{align}
Similarly,
\begin{align}
    &G'(\sigma^2_{cr}-\tilde{\delta})\nonumber\\
    &=K(t,c,\sigma^2_{cr}-\tilde{\delta},r_*)\nonumber\\
    &\times\left(1+\frac{t-1}{2(n-1)(c-1)}\left(-(n-c-2)+h(\frac{\sigma^2_{cr}-\tilde{\delta}}{\sigma^2_{\epsilon,\delta}})\right)\right)\\
    &=K(t,c,\sigma^2_{cr}-\tilde{\delta},r_*)\nonumber\\
    &\times\left(1+\frac{t-1}{2(n-1)(c-1)}\left(-(n-c-2)+h(y_{cr})-\delta\right)\right)\\
    &=K(t,c,\sigma^2_{cr}-\tilde{\delta},r_*)\frac{-\delta(t-1)}{2(n-1)(c-1)}<0\label{g_func2}
\end{align}
This proves that $\sigma^2_{cr}$ is a minimum of $G(\sigma^2)$.  

\paragraph{case 2: $c=1$} When $c=1$, $y_{cr}$ corresponding to \eqref{cl8} is given by,
\begin{align}
    y_{cr}=\frac{n-2}{n-1}+\frac{1}{n-1}\sqrt{\frac{(n-2)(t-1)}{n-t}}
\end{align}

Recall that the $G'(\sigma^2)$ in \eqref{der} is of the form,
\begin{align}\label{K}
    G'(\sigma^2)&=K(t,c,\sigma^2,r_*)\left(1+(t-1)\frac{dr_*}{d\sigma^2}\right)\\
    &=K(t,c,\sigma^2,r_*)\nonumber\\
    &\quad\left(1-\frac{t-1}{n-1}\left(1+\frac{n-2}{(y(n-1)-(n-2))^2}\right)\right)
\end{align}
Let $\tilde{g}(y)=1-\frac{t-1}{n-1}(1+\tilde{h}(y))$ and $\tilde{h}(y)=\frac{n-2}{(y(n-1)-(n-2))^2}$. Note that, $h'(y)=-\frac{2(n-1)(n-2)}{(y(n-1)-(n-2))^3}$, and $h'(y)<0$ for $y>y_{cr}-\frac{1}{n-1}\sqrt{\frac{(n-2)(t-1)}{n-t}}$. Therefore, for any small $\delta\in(0,\frac{\sigma^2_{\epsilon,\delta}}{n-1}\sqrt{\frac{(n-2)(t-1)}{n-t}})$,
\begin{align}
    G'(\sigma^2_{cr}+\delta)&\!=\!K(t,c,\sigma^2,r_*)\!\!\left(\!1\!-\!\frac{t-1}{n-1}\left(1+h(\frac{\sigma^2_{cr}+\delta}{\sigma^2_{\epsilon,\delta}})\!\right)\!\right)\\
    &=K(t,c,\sigma^2,r_*)\left(1-\frac{t-1}{n-1}\left(1+h(y_{cr})-\tilde{\delta}\right)\right)\\
    &=K(t,c,\sigma^2,r_*)\left(\frac{\tilde{\delta}(t-1)}{n-1}\right)>0
\end{align}
Similarly,    
\begin{align}
    G'(\sigma^2_{cr}-\delta)&=K(t,c,\sigma^2,r_*)\left(1-\frac{t-1}{n-1}\left(1+h(\frac{\sigma^2_{cr}-\delta}{\sigma^2_{\epsilon,\delta}})\right)\right)\\
    &=K(t,c,\sigma^2,r_*)\left(1-\frac{t-1}{n-1}\left(1+h(y_{cr})+\tilde{\delta}\right)\right)\\
    &=K(t,c,\sigma^2,r_*)\left(\frac{-\tilde{\delta}(t-1)}{n-1}\right)<0,
\end{align}
which proves that $\sigma^2_{cr}$ is a minimum of $G(\sigma^2)$. 

\paragraph{case 3: $c=0$} As the $r_*$ is the same for both $c>1$ and $c=0$, proving that $\sigma^2_{cr}$ is the minimum of $G(\sigma^2)$ follows the same steps as case 1.
\end{Proof}
As there are no other critical points satisfying $\sigma^2\geq\sigma^2_{\epsilon,\delta}$, the optimum $\sigma^2$ is given by $\sigma^2_{cr}$ and is presented in its complete form in \eqref{optsigmaa}.
\end{Proof}

\section{Proof of Proposition~\ref{ubsnodrop}}\label{pfcoro1}

\paragraph{Proposition~\ref{ubsnodrop} restated:} The following upper bounds hold for $\sigma^2_{\epsilon,\delta}$.
    \begin{align}\label{ubs}
        \sigma^2_{\epsilon,\delta}&\leq\begin{cases}
            \frac{8\ln(1.25/\delta)}{\epsilon^2}, & \epsilon,\delta\in(0,1)\\
            \frac{2\eta^2}{\epsilon}, & \epsilon\geq1, \delta\in(0,1)
        \end{cases}
    \end{align}
    where
    \begin{align}\label{constant}
       \eta=\begin{cases}
        1+2\sqrt{\ln\frac{1}{2\delta}}, & 0<\delta\leq0.05\\
        1+2\sqrt{\ln10}, & 0.05<\delta<1.
    \end{cases} 
    \end{align}    
\begin{Proof} We use the upper bounds on $\sigma_{\epsilon,\delta}$ derived in \cite{reviewGauss} and \cite{DP1}.
    \begin{align}\label{ub_rem}
        \sigma_{\epsilon,\delta}&\leq\begin{cases}
            \frac{2}{\epsilon}\sqrt{2\ln\frac{1.25}{\delta}}, & \epsilon,\delta\in(0,1)\\
            \frac{2}{\epsilon}\sqrt{2\ln\frac{1}{2\delta}}+\sqrt{\frac{2}{\epsilon}}, & \epsilon\geq1, \delta\in(0,0.05].
        \end{cases}
    \end{align}
Note that in the second case of \eqref{ub_rem}, $\frac{2}{\epsilon}\sqrt{2\ln\frac{1}{2\delta}}+\sqrt{\frac{2}{\epsilon}}\leq\sqrt{\frac{2}{\epsilon}}\left(2\sqrt{\ln\frac{1}{2\delta}}+1\right)$ as $\epsilon\geq1$. Moreover, it is clear that any $\hat{\sigma}$ satisfying $\Phi\left(\frac{1}{\hat{\sigma}}-\frac{\epsilon\hat{\sigma}}{2}\right)-e^\epsilon\Phi\left(-\frac{1}{\hat{\sigma}}-\frac{\epsilon\hat{\sigma}}{2}\right)\leq\delta_1$ also satisfies $\Phi\left(\frac{1}{\hat{\sigma}}-\frac{\epsilon\hat{\sigma}}{2}\right)-e^\epsilon\Phi\left(-\frac{1}{\hat{\sigma}}-\frac{\epsilon\hat{\sigma}}{2}\right)\leq\delta_2$, whenever $\delta_1\leq\delta_2$. Therefore, for any $\epsilon\geq1$ and $\delta\in(0.05,1]$ we have, 
\begin{align}
    \sigma_{\epsilon,\delta}\leq\sigma_{\epsilon,0.05}\leq\sqrt{\frac{2}{\epsilon}}\left(2\sqrt{\ln10}+1\right),
\end{align}
where the first inequality comes from the definition of $\sigma_{\epsilon,\delta}$, given by $\sigma_{\epsilon,\delta}=\inf_{\hat{\sigma}>0}\{\hat{\sigma};\Phi\left(\frac{1}{\hat{\sigma}}-\frac{\epsilon\hat{\sigma}}{2}\right)-e^\epsilon\Phi\left(-\frac{1}{\hat{\sigma}}-\frac{\epsilon\hat{\sigma}}{2}\right)\leq\delta\}$.
This determines the values of $\eta$ in \eqref{constant}.
    
\end{Proof}

\section{Proof of Proposition~\ref{unbiased}}\label{pfunbiased}

The proof of Proposition~\ref{unbiased} is direct from the proof of Theorem 1, as in Theorem~\ref{thm1}, we essentially solved
\begin{align}\label{previous}
(\sigma^2_*,\rho_*)=\arg\min_{\sigma^2,\rho}\max_{\substack{\mathcal{U}\subseteq\mathcal{U}_{all}\\ t\leq|\mathcal{U}|\leq n}}\sup_{\mathbf{x}_{j_1},\dotsc,\mathbf{x}_{j_{|\mathcal{U}|}}\in\mathbb{B}^d}\sigma^2(1+\rho(|\mathcal{U}|-1),
\end{align}
while satisfying \eqref{dp_eq}, and then applied the decoder in Proposition~\ref{vec_opt_dec}. Note that \eqref{previous} is equivalent to \eqref{unbiased_eq}, which shows that the optimum noise parameters do not change in the biased and unbiased cases. However, the resulting MSE changes as the decoder further normalizes the direct MSE resulted by the parameters in \eqref{previous}. 

\section{Additional Details on Section~\ref{why}}

In this section, we go over the rigorous proofs of the geometric interpretation provided for the two-user example in Section~\ref{why}.

\subsection{Privacy Constraint}\label{addwhy} For this example, \eqref{dp_eq} simplifies to,
\begin{align}\label{simplify}
    &\mathbb{P}(\mathbf{x}_i+\mathbf{Z}_i\in\mathcal{A}|\mathbf{x}_j+\mathbf{Z}_j=\mathbf{y}_j)\nonumber\\
    &\quad\leq e^\epsilon\mathbb{P}(\mathbf{x}_i'+\mathbf{Z}_i\in\mathcal{A}|\mathbf{x}_j+\mathbf{Z}_j=\mathbf{y}_j)+\delta
\end{align}
for each $i\neq j$, $\forall\mathbf{x}_i,\mathbf{x}_i'\in\mathbb{B}^d$, and $\forall\mathcal{A}\subset\mathbb{R}^d$ in the Borel sigma algebra. Let $\mathbf{Y}_i=\mathbf{x}_i+\mathbf{Z}_i$ for $i=1,2$. The first step is to quantify the conditional distribution $\mathbf{Y}_i|\mathbf{Y}_j=\mathbf{y}_j$ for $i\neq j$. WLOG assume that $i=1$. 
\begin{align}
    \begin{pmatrix}
        \mathbf{Y}_1\\\mathbf{Y}_2
    \end{pmatrix}\sim\mathcal{N}\left(\begin{pmatrix}
        \mathbf{x}_1\\\mathbf{x}_2
    \end{pmatrix},\begin{pmatrix}
        \sigma^2\mathsf{I}_d & \rho\sigma^2\mathsf{I}_d\\
         \rho\sigma^2\mathsf{I}_d & \sigma^2\mathsf{I}_d 
    \end{pmatrix}\right)
\end{align}
Then, from the theorems of conditional Gaussian distributions, we have,  $\mathbf{Y}_1|\mathbf{Y}_2=\mathbf{y}_2\sim\mathcal{N}(\mathbf{\mu}_*,\Sigma_*)$ where,
        \begin{align}
            \mathbf{\mu}_*&=
        \mathbf{x}_1+\rho\mathsf{I}_d(\mathbf{y}-\mathbf{x}_2)\\
            \Sigma_*&=\sigma^2\mathsf{I}_d-\rho^2\sigma^2\mathsf{I}_d=\sigma^2(1-\rho^2)\mathsf{I}_d=\Tilde{\sigma}^2\mathbf{I}_d
        \end{align}
Define $f(\mathbf{x}_i)=\mathbf{x}_1+\rho\mathsf{I}_d(\mathbf{y}-\mathbf{x}_2)$. Then, define a new random variable $\mathbf{W}=f(\mathbf{x}_1)+\mathbf{N}$, where $\mathbf{N}\sim\mathcal{N}\left(\mathbf{0}_{d},\Tilde{\sigma}^2\mathsf{I}_d\right)$. Note that for any given values of $\mathbf{y}_2$, $\mathbf{x}_2$, $\mathbf{W}\sim \mathbf{Y}_1|\mathbf{Y}_2=\mathbf{y}_2$, (statistically equivalent). Now, consider the $(\epsilon,\delta)$-DP constraint in \eqref{simplify}, which is equivalent to,
\begin{align}\label{stdgauss2}
    &\mathbb{P}(f(\mathbf{x}_1)+\mathbf{N}\in\mathcal{A})\leq e^\epsilon\mathbb{P}(f(\mathbf{x}'_1)+\mathbf{N}\in\mathcal{A})+\delta.
\end{align}
As \eqref{stdgauss2} represents the standard Gaussian mechanism on $f(\mathbf{x}_i)$ (with a sensitivity of $\sup_{\mathbf{x},\mathbf{x}'}\|f(\mathbf{x}_i)-f(\mathbf{x}_i')\|_2=\|\mathbf{x}_i-\mathbf{x}_i'\|=2$), the variance of $N$ must satisfy,
\begin{align}
    \tilde{\sigma}^2\geq\sigma^2_{\epsilon,\delta}
\end{align}
to ensure $(\epsilon,\delta)$-DP, where $\sigma^2_{\epsilon,\delta}=\inf_{\hat{\sigma}>0}\{\hat{\sigma};\Phi\left(\frac{\Delta}{2\hat{\sigma}}-\frac{\epsilon\hat{\sigma}}{\Delta}\right)-e^\epsilon\Phi\left(-\frac{\Delta}{2\hat{\sigma}}-\frac{\epsilon\hat{\sigma}}{\Delta}\right)\leq\delta\}$ with $\Delta=2$. This simplifies to,
\begin{align}
    \sigma^2(1-\rho^2)&\geq\sigma^2_{\epsilon,\delta}\\
    \implies\quad\sigma\sin\theta&\geq\sigma_{\epsilon,\delta}
\end{align}
Recall that $\mathbb{E}[\|\mathbf{Z}_i\|^2]=\sigma^2d=\|\mathbf{Z}_i\|_{\mathcal{H}}^2$ from the vector representation explained in Section~\ref{why}. Therefore, the privacy constraint in \eqref{simplify} simplifies to,
\begin{align}
    \|\mathbf{Z}_i\|_{\mathcal{H}}\sin\theta\geq\sigma_{\epsilon,\delta}\sqrt{d}=\gamma_{\epsilon,\delta}
\end{align}
which can be interpreted as the component of $\mathbf{Z}_i$ that is orthogonal to $\mathbf{Z}_j$ for 
$i\neq j$ having a variance that is lower bounded by a constant to ensure $(\epsilon,\delta)$-DP.

\subsection{Optimizing the noise distribution}\label{opt_para}

To obtain the optimum distribution of $(\mathbf{Z}_1,\mathbf{Z}_2)$ that minimizes $\|\mathbf{Z}_1+\mathbf{Z}_2\|_{\mathcal{H}}$ (MSE) while satisfying $\|\boldsymbol{Z}_i^\bot\|_{\mathcal{H}}=\|\boldsymbol{Z}_i\|_{\mathcal{H}}\sin\theta \geq\gamma_{\epsilon,\delta}$ (privacy), we optimize $\|\mathbf{Z}_i\|_{\mathcal{H}}$ and $\theta$, which corresponds to optimizing $\sigma^2$ and $\rho$. The MSE is minimized when $\sigma^2\to\infty$ and $\rho\to-1$.\footnote{The proof of $\sigma^2_*\to\infty$ and $\rho_*\to-1$ is given in the general proof of Lemma~\ref{optsigmarho} in Appendix~\ref{pfthm1}, with $c=0$.} This is explained as follows. As illustrated in Fig.~\ref{mag}, $\|\mathbf{Z}_1+\mathbf{Z}_2\|_{\mathcal{H}}$ decreases as $\|\mathbf{Z}_i\|_{\mathcal{H}}$ and $\theta$ increase while satisfying $\|\mathbf{Z}_i\|_{\mathcal{H}}\sin\theta=\gamma_{\epsilon,\delta}$ for privacy. In the limit, when $\|\mathbf{Z}_i\|_{\mathcal{H}}\to\infty$ and $\theta\to\pi$,\footnote{Note that increasing $\theta$ beyond $\pi$ is not optimal as $\theta=\pi+\eta$ and $\theta=\pi-\eta$ correspond to the same setting.} i.e., when $\sigma^2\to\infty$ and $\rho\to-1$, $\mathbf{Z}_1+\mathbf{Z}_2$ aligns perpendicular to $\mathbf{Z}_2$, and $\|\mathbf{Z}_1+\mathbf{Z}_2\|_{\mathcal{H}}\to\gamma_{\epsilon,\delta}$, which is the minimum achievable $\|\mathbf{Z}_1+\mathbf{Z}_2\|_{\mathcal{H}}$ while ensuring $\|\mathbf{Z}_i\|_{\mathcal{H}}\sin\theta\geq\gamma_{\epsilon,\delta}$. The resulting minimum MSE is given by,
\begin{align}
    \mathsf{MMSE}&=\lim_{\substack{\sigma^2\to\infty\\\rho\to\pi}}\frac{1}{4}\mathbb{E}[\|\mathbf{Z}_1+\mathbf{Z}_2\|^2]=\lim_{\substack{\sigma^2\to\infty\\\rho\to\pi}}\frac{1}{4}\|\mathbf{Z}_1+\mathbf{Z}_2\|_{\mathcal{H}}^2\nonumber\\
    &=\frac{\gamma^2_{\epsilon,\delta}}{4}.\label{msecor}
\end{align}

\subsection{Deriving $\gamma_{\epsilon,\delta}$ and comparison with CDP}\label{gammaderive}

We will outline how $\gamma_{\epsilon,\delta}$ is derived for a specified $\epsilon$ and $\delta$. For this, we use known results on the Gaussian mechanism in DP \cite{gauss,reviewGauss} as follows. The standard Gaussian mechanism in DP states that for any given $\epsilon$ and $\delta$, $\mathbf{Z}\sim\mathcal{N}(\mathbf{0}_d,\sigma^2\mathsf{I}_d)$ with $\sigma^2\geq\sigma^2_{\epsilon,\delta,\Delta}$ ensures:
\begin{align}\label{stddp}
    \mathbb{P}(\mathbf{v}+\mathbf{Z}\in\mathcal{A})\leq e^\epsilon\mathbb{P}(\mathbf{v}'+\mathbf{Z}\in\mathcal{A})+\delta
\end{align}
for any $\mathbf{v},\mathbf{v}'\in\mathbb{R}^d$ with $\|\mathbf{v}-\mathbf{v}'\|\leq\Delta$, and $\forall\mathcal{A}\subset\mathbb{R}^d$, where $\sigma^2_{\epsilon,\delta,\Delta}$ is given by,
\begin{align}\label{std_gau}
    \sigma_{\epsilon,\delta,\Delta}=\inf_{\tilde{\sigma}>0}\left\{\tilde{\sigma}:\Phi\left(\frac{\delta}{2\tilde{\sigma}}-\frac{\epsilon\tilde{\sigma}}{\Delta}\right)-e^\epsilon\Phi\left(-\frac{\Delta}{2\tilde{\sigma}}-\frac{\epsilon\tilde{\sigma}}{\Delta}\right)\leq\delta\right\} 
\end{align}
 where $\Phi$ denotes the standard Gaussian CDF. The corresponding MSE of the encoded version of $\mathbf{v}$, i.e., $\mathbf{v}+\mathbf{Z}$ is lower bounded as:
 \begin{align}\label{singleuser}
     \mathbb{E}[\|\mathbf{Z}\|^2]=\sigma^2d\geq\sigma^2_{\epsilon,\delta,\Delta}d.
 \end{align}

Recall that when considering the privacy of $\mathbf{x}_i$ in the two user case, the \emph{effective} noise added to $\mathbf{x}_i$ is quantified by $\mathbf{Z}_i^\bot$ as the rest of $\mathbf{Z}_i$ can be inferred by $\mathbf{Z}_j$, for  $i,j\in\{1,2\}$, $j\neq i$. The privacy constraint in \eqref{dp_eq} reduces to $\mathbb{P}(\mathbf{x}_i+\mathbf{Z}_i^\bot\in\mathcal{A}')\leq e^\epsilon \mathbb{P}(\mathbf{x}'_i+\mathbf{Z}_i^\bot\in\mathcal{A}')+\delta$ for $i=\{1,2\}$, $\forall\mathbf{x},\mathbf{x}'\in\mathbb{B}^d$, and $\forall\mathcal{A}'\subset\mathbb{R}^d$ (see Appendix~\ref{addwhy} for a rigorous proof). This imposes a lower bound on the variance of $\mathbf{Z}_i^\bot$ and hence on the MSE resulted by the \emph{effective} noise: $\mathbb{E}[\|\mathbf{Z}_i^\bot\|^2]=\|\mathbf{Z}_i^\bot\|_{\mathcal{H}}^2\geq\sigma^2_{\epsilon,\delta}d$ to ensure $(\epsilon,\delta)$-DP based on the standard Gaussian mechanism, similar to \eqref{singleuser}. Here we denote $\sigma^2_{\epsilon,\delta}=\sigma^2_{\epsilon,\delta,\Delta_x=2}$, as $\Delta_x=\sup_{\mathbf{x},\mathbf{x}'}\|\mathbf{x}_i-\mathbf{x}'_i\|=2$ for $i=\{1,2\}$. This, together with \eqref{geopriv} characterizes the constant $\gamma_{\epsilon,\delta}$ as:
\begin{align}\label{done}
\gamma_{\epsilon,\delta}=\sigma_{\epsilon,\delta}\sqrt{d}
\end{align}

For the same example with two users with vectors from $\mathbb{B}^d$, consider the corresponding CDP setting with respect to the standard Gaussian mechanism in \eqref{stddp}. Then, $\mathbf{v}=\frac{1}{2}(\mathbf{x}_1+\mathbf{x}_2)$ and $\mathbf{v'}=\frac{1}{2}(\mathbf{x}'_1+\mathbf{x}_2)$ with $\mathbf{x}_2$ fixed (or vice-versa). From \eqref{singleuser}, we have, $\mathbb{E}[\|Z\|^2]\geq\sigma^2_{\epsilon,\delta,\Delta_v}d$, with $\Delta_v=\sup_{\mathbf{v},\mathbf{v}'}\|\mathbf{v}-\mathbf{v}'\|=1$. Using the bounds on \eqref{std_gau} from \cite{reviewGauss,gauss} which shows that $\sigma^2_{\epsilon,\delta,\Delta}\approx\Delta^2\kappa_{\epsilon,\delta}$ where $\kappa_{\epsilon,\delta}$ is fixed for a given $\epsilon$ and $\delta$, we have,
\begin{align}
    \mathrm{MSE}_{\mathrm{CDP}}=\mathbb{E}[\|Z\|^2]\geq\kappa_{\epsilon,\delta}d.
\end{align}
Similarly, for the same $\epsilon,\delta$, we bound the MSE of the correlated Gaussian mechanism, using \eqref{msecor} and \eqref{done} for this example as:
\begin{align}
    \mathrm{MSE}_{\mathrm{Corr-Gaussian}}=\frac{1}{4}\mathbb{E}[\|\mathbf{Z}_1+\mathbf{Z}_2\|^2]\geq\frac{\Delta_x^2\kappa_{\epsilon,\delta}d}{4}=\kappa_{\epsilon,\delta}d.
\end{align}
This two-user example shows that the same MSE can be achieved without the requirement of a trusted server by carefully choosing the parameters of the correlated privacy mechanism. The insights of Fig. \ref{privacy_fig} generalize for more than two users. Specifically, in Theorem~\ref{thm1} and Corollary~\ref{cor1}, we show that an MSE of $\frac{\sigma^2_{\epsilon,\delta}d}{n^2}$ (same as CDP) is achieved for the general case of $n \geq 2$ users by the correlated Gaussian mechanism, even with no trusted server. 

Next, we consider the case of LDP. The privacy constraint in LDP-based DME is the same as \eqref{stddp}, with $\mathbf{v}=\mathbf{x}_i$ and $\mathbf{v}'=\mathbf{x}_i'$ representing any two possible vectors generated by the same user $i$. In LDP, the privacy constraint is defined for each user independently as the privacy mechanisms are independent, i.e., $\mathbf{Z}_i\sim\mathcal{N}(\mathbf{0}_{d},\sigma^2\mathsf{I}_{d})$, $\forall i$ with no correlation among each other. To satisfy $(\epsilon,\delta)$-DP, $\sigma^2\geq\sigma^2_{\epsilon,\delta}$ must hold, as the sensitivity is $\Delta_x=\sup_{\mathbf{x},\mathbf{x}'}\|\mathbf{x}-\mathbf{x}'\|=2$ (recall that $\sigma^2_{\epsilon,\delta}=\sigma^2_{\epsilon,\delta,\Delta_x=2}$). The estimation error is $\frac{1}{n}\sum_{i=1}^n\mathbf{Z}_i$ for the general case of $n$ users, which results in a minimum MSE of $\mathbb{E}[\|\frac{1}{n}\sum_{i=1}^n\mathbf{Z}_i\|^2]=\frac{\sigma^2_{\epsilon,\delta}d}{n}$, which is $n$ times larger than the MSE achieved by CorDP-DME for the same level of privacy.

\section{Proof of Theorem~\ref{converse}}\label{pfconverse}

\textbf{Theorem~\ref{converse} restated:}  Let $\tilde{\mathbf{S}}_{\mathcal{U}}=\frac{1}{|\mathcal{U}|}\sum_{i\in\mathcal{U}}(\mathbf{x}+\tilde{\mathbf{Z}}_i)$ be an unbiased estimate of $\mathbf{S}_\mathcal{U}$, where $\tilde{\mathbf{Z}}_i\sim\mathcal{N}(\mathbf{0}_d,\sigma^2_i\mathsf{I}_d)$ for $i\in[1:n]$. Let $[\tilde{Z}_{1,k},\dotsc,\tilde{Z}_{n,k}]^T\sim\mathcal{N}(\mathbf{0}_n,\Sigma)$ for $k\in[1:d]$, where $\tilde{Z}_{i,k}$ is the $k$th coordinate of $\tilde{\mathbf{Z}}_i$ and $\Sigma$ is symmetric positive definite. Define the corresponding MSE of $\tilde{\mathbf{S}}_{\mathcal{U}}$ as, 
\begin{align}\label{unbiased_def}
\mathrm{MSE}(\Sigma)&=\max_{\substack{\mathcal{U}\subseteq\mathcal{U}_{all}\\ t\leq|\mathcal{U}|\leq n}}\quad\sup_{\mathbf{x}_{j_1},\dotsc,\mathbf{x}_{j_{|\mathcal{U}|}}\in\mathbb{B}^d}\nonumber\\
&\qquad\mathbb{E}\left[\left\|\frac{1}{|\mathcal{U}|}\sum_{i\in\mathcal{U}}(\mathbf{x}_i+\tilde{\mathbf{Z}}_i)-\frac{1}{|\mathcal{U}|}\sum_{i\in\mathcal{U}}\mathbf{x}_i\right\|^2\right]
\end{align} 
$\forall\Sigma=\Sigma^T,\quad\Sigma\succ0$ satisfying $(\epsilon,\delta)$-DP in Definition~\ref{def1} with $\mathcal{U}_{col}=\emptyset$, 
\begin{align}\label{conv_eq}
    \mathrm{MSE}\left(\frac{1}{n!}\sum_{i=1}^{n!}\Pi_i(\Sigma)\right)\leq\mathrm{MSE}(\Sigma)
\end{align}
where $\Pi_i(\Sigma)=P_i\Sigma P_i^T$ is the $i$-th permutation of $\Sigma$ defined by the permutation matrix $P_i$. Moreover, $\tilde{\Sigma}=\frac{1}{n!}\sum_{i=1}^{n!}\Pi_i(\Sigma)$ satisfies $(\epsilon,\delta)$-DP in Definition~\ref{def1} with $\mathcal{U}_{col}=\emptyset$.

\begin{Proof}    
For the case of non-colluding users, the privacy constraint in Definition~\ref{def1} stated as
    \begin{align}
        \mathbb{P}(M(\mathbf{x}_i)\in\mathcal{A}|\mathcal{G}_i)\leq e^\epsilon\mathbb{P}(M(\mathbf{x}_i')\in\mathcal{A}|\mathcal{G}_i)+\delta,
    \end{align}
    $\forall \mathcal{D}_i,\mathcal{D}_i'$, $\forall i\in\mathcal{U}_{all}\setminus\mathcal{U}_{col}$, and $\forall \mathcal{A}$, simplifies to,
    \begin{align}
        &\mathbb{P}(M(\mathbf{x}_i)\in\mathcal{A}|M(\mathbf{x}_j)=\mathbf{y}_j,j\in\mathcal{U}_{all},j\neq i)\nonumber\\
        &\leq e^\epsilon\mathbb{P}(M(\mathbf{x}_i')\in\mathcal{A}|M(\mathbf{x}_j)=\mathbf{y}_j,j\in\mathcal{U}_{all},j\neq i)+\delta,\label{simplified}
    \end{align}
    $\forall \mathcal{D}_i,\mathcal{D}_i'$, $\forall i\in\mathcal{U}_{all}$, and $\forall \mathcal{A}$.
    Let $\mathbf{Y}_k=M(\mathbf{x}_k)$, $\forall k\in[1:n]$. We first derive the conditional distribution of $\mathbf{Y}_i$ given $\{\mathbf{Y}_j$, $j\in\mathcal{U}_{all},j\neq i\}\}$. WLOG assume that $i=1$. Then, following the same steps as the proof of lemma~\ref{privacylem} gives,
    $\mathbf{Y}_1|\{\mathbf{Y}_j,j\in[2:n]\sim N(\Tilde{\mathbf{\mu}},\Tilde{\Sigma})$ where,
\begin{align}\label{mean_vec}
    \Tilde{\mathbf{\mu}}&=\mathbf{x}_1+\left([r_{1,2},\dotsc,r_{1,n}]\otimes\mathsf{I}_d\right)\nonumber\\
    &\times\left(\begin{pmatrix}
        \sigma^2_2 & r_{2,3} & \dotsc & r_{2,n}\\
         r_{3,2} & \sigma^2_3 & \dotsc & r_{3,n}\\
         \vdots & \vdots & \vdots & \vdots \\
        r_{n,2} &  r_{n,3} & \dotsc & \sigma^2_{n}
    \end{pmatrix}^{-1}\otimes\mathsf{I}_d\right)\begin{pmatrix}
        \mathbf{y}_2-\mathbf{x}_2\\\vdots\\\mathbf{y}_{n}-\mathbf{x}_{n}
    \end{pmatrix}\\
    \Tilde{\Sigma}&=\left(\sigma^2_1\otimes\mathsf{I}_d\right)-\left([r_{1,2},\dotsc,r_{1,n-c}]\otimes\mathsf{I}_d\right)\nonumber\\
    &\times\left(\begin{pmatrix}
        \sigma^2_2 & r_{2,3} & \dotsc & r_{2,n}\\
         r_{3,2} & \sigma^2_3 & \dotsc & r_{3,n}\\
         \vdots & \vdots & \vdots & \vdots \\
        r_{n,2} &  r_{n,3} & \dotsc & \sigma^2_{n}r_{n,j} 
    \end{pmatrix}^{-1}\otimes\mathsf{I}_d\right)\nonumber\\
    &\qquad\qquad \times\left([r_{1,2},\dotsc,r_{1,n}]\otimes\mathsf{I}_d)\right)^T
\end{align}
Following the same arguments as in the proof of Lemma~\ref{privacylem} gives the privacy constraint on user 1 as:
\begin{align}\label{priv_new}
    &\begin{bmatrix}
        r_{1,2}\\\vdots\\r_{1,n}
    \end{bmatrix}^T\begin{pmatrix}
        \sigma^2_2 & r_{2,3} & \dotsc & r_{2,n}\\
         r_{3,2} & \sigma^2_3 & \dotsc & r_{3,n}\\
         \vdots & \vdots & \vdots & \vdots \\
        r_{n,2} &  r_{n,3} & \dotsc & \sigma^2_{n}
    \end{pmatrix}^{-1}\!\!\!\!\begin{bmatrix}
        r_{1,2}\\\vdots\\r_{1,n}
    \end{bmatrix}\leq\sigma^2_1-\sigma^2_{\epsilon,\delta}
\end{align}
Define $\mathbf{r}_1=[r_{1,2},\dotsc,r_{1,n}]^T$, and
\begin{align}
    &\Sigma_1=\begin{pmatrix}
        \sigma^2_2 & r_{2,3} & \dotsc & r_{2,n}\\
         r_{3,2} & \sigma^2_3 & \dotsc & r_{3,n}\\
         \vdots & \vdots & \vdots & \vdots \\
        r_{n,2} &  r_{n,3} & \dotsc & \sigma^2_{n}
    \end{pmatrix}
\end{align}
Note that,
\begin{align}
(\mathbf{Y}_1,\dotsc,\mathbf{Y}_{n})\sim\mathcal{N}\left([\mathbf{x}_1,\dotsc,\mathbf{x}_{n}]^T,\Sigma\otimes\mathsf{I}_d\right)    
\end{align}
where,
\begin{align}
    \Sigma=\begin{pmatrix}
        \sigma^2_1 & r_{1,3} & \dotsc & r_{1,n}\\
         r_{2,1} & \sigma^2_2 & \dotsc & r_{2,n}\\
         \vdots & \vdots & \vdots & \vdots \\
        r_{n,2} &  r_{n,3} & \dotsc & \sigma^2_{n}
    \end{pmatrix}
\end{align}
which gives,
\begin{align}
    \Sigma=\begin{bmatrix}
        \sigma_1^2 & \mathbf{r}_1^T\\
        \mathbf{r}_1 & \Sigma_1
    \end{bmatrix}
\end{align}
Let $\Sigma^{-1}=\begin{bmatrix}
    z & \mathbf{Y}^T\\
    \mathbf{Y} & X
\end{bmatrix}$
where $z$ is a scalar, $\mathbf{Y}$ is a vector of size $(n-1)\times1$ and $X$ is a matrix of size $(n-1)\times(n-1)$. As $\Sigma\Sigma^{-1}=\mathsf{I}_{n}$, we have,
\begin{align}
    &\sigma_1^2z+\mathbf{r}^TY=1\label{a}\\
    &\sigma_1^2\mathbf{Y}^T+\mathbf{r}_1^TX=\mathbf{0}_{n-1}^T\label{b}\\
    &z\mathbf{r}_1+\Sigma_1Y=\mathbf{0}_{n-1}\label{c}\\
    &\mathbf{r}_1\mathbf{Y}^T+\Sigma_1X=\mathsf{I}_{n-1}\label{d}
\end{align}
From the above equations, we obtain,
\begin{align}
    \Sigma_1^{-1}&=X-\frac{1}{z}\mathbf{Y}\mathbf{Y}^T\\
    \mathbf{r}_1^TX&=-\frac{(1-\mathbf{r}_1^T\mathbf{Y})}{z}\mathbf{Y}^T\label{b2}
\end{align}
The privacy constraint on user 1 in \eqref{priv_new} is given by,
\begin{align}
    \mathbf{r}_1^T\Sigma_1^{-1}\mathbf{r}_1&\leq\sigma^2_1-\sigma^2_{\epsilon,\delta}\\
    \mathbf{r}_1^T\left(X-\frac{1}{z}\mathbf{Y}\mathbf{Y}^T\right)\mathbf{r}_1&\leq\sigma^2_1-\sigma^2_{\epsilon,\delta}\\
    \left(-\sigma_1^2-\frac{1}{z}\mathbf{r}_1^T\mathbf{Y}\right)\mathbf{Y}^T\mathbf{r}_1&\leq\sigma^2_1-\sigma^2_{\epsilon,\delta}\label{a2}\\
    -\frac{1}{z}\mathbf{Y}^T\mathbf{r}_1&\leq\sigma^2_1-\sigma^2_{\epsilon,\delta}\label{c2}\\
    z&\leq\frac{1}{\sigma^2_{\epsilon,\delta}}\label{d2}
\end{align}
where \eqref{a2} follows from \eqref{a} and \eqref{b2}, and \eqref{c2},\eqref{d2} follow from \eqref{a}. Therefore, the privacy constraint on user 1 simplifies to $\Sigma^{-1}_{1,1}\leq\frac{1}{\sigma^2_{\epsilon,\delta}}$, where $\Sigma_{1,1}^{-1}$ denotes the first diagonal element of $\Sigma^{-1}$. Similarly, the general privacy constraint must be satisfied for all users $i\in\mathcal{U}_{all}$, i.e.,
\begin{align}\label{simpl}
    \Sigma^{-1}_{i,i}\leq\frac{1}{\sigma^2_{\epsilon,\delta}}, \quad i\in\mathcal{U}_{all}
\end{align}
The privacy constraint in \eqref{simpl} can be written as,
\begin{align}\label{pi1}
    &\mathbf{e}_i^T\Sigma^{-1}\mathbf{e}_i\leq\frac{1}{\sigma^2_{\epsilon,\delta}},\quad i\in\mathcal{U}_{all}
\end{align}
where $\mathbf{e}_i$ denotes the $i$th column of $\mathsf{I}_{n}$. 

Next, we simplify the MSE expression to formalize the optimization problem to be solved. Define:
\begin{align}
\mathrm{MSE}(\Sigma)&=\max_{\substack{\mathcal{U}\subseteq\mathcal{U}_{all}\\ t\leq|\mathcal{U}|\leq n}}\quad\sup_{\mathbf{x}_{j_1},\dotsc,\mathbf{x}_{j_{|\mathcal{U}|}}\in\mathbb{B}^d}\nonumber\\
&\qquad\mathbb{E}\left[\left\|\frac{1}{|\mathcal{U}|}\sum_{i\in\mathcal{U}}(\mathbf{x}_i+\tilde{\mathbf{Z}}_i)-\frac{1}{|\mathcal{U}|}\sum_{i\in\mathcal{U}}\mathbf{x}_i\right\|^2\right]\\
&=\max_{\substack{\mathcal{U}\subseteq\mathcal{U}_{all}\\ t\leq|\mathcal{U}|\leq n}}\left(\sum_{i\in\mathcal{U}}\mathbf{e}_i\right)^T\Sigma\left(\sum_{i\in\mathcal{U}}\mathbf{e}_i\right)
\end{align} 
The optimization problem to be solved is:
\begin{align}
    &\min_{\Sigma} \max_{\substack{\mathcal{U}\subseteq\mathcal{U}_{all}\\ t\leq|\mathcal{U}|\leq n}}\left(\sum_{i\in\mathcal{U}}\mathbf{e}_i\right)^T\Sigma\left(\sum_{i\in\mathcal{U}}\mathbf{e}_i\right)\nonumber\\
    &s.t. \quad \mathbf{e}_i^T\Sigma^{-1}\mathbf{e}_i\leq\frac{1}{\sigma^2_{\epsilon,\delta}},\quad i\in\mathcal{U}_{all}\label{opt}
\end{align}

\begin{Claim}\label{claimm9}
    Assume that $\Sigma_*$ is a solution to \eqref{opt}. Then, any permutation of $\Sigma_*$ denoted by $\Pi_j(\Sigma_*)=P_j\Sigma_*P_j^T$, $j\in[1:n!]$ is also a solution to \eqref{opt}.
\end{Claim}

\begin{Proof}
\begin{align}
    \mathrm{MSE}(\Pi_j(\Sigma_*)) &=\max_{\substack{\mathcal{U}\subseteq\mathcal{U}_{all}\\ t\leq|\mathcal{U}|\leq n}}\left(\sum_{i\in\mathcal{U}}\mathbf{e}_i\right)^TP_j\Sigma_*P_j^T\left(\sum_{i\in\mathcal{U}}\mathbf{e}_i\right)\\
    &=\max_{\substack{\mathcal{U}\subseteq\mathcal{U}_{all}\\ t\leq|\mathcal{U}|\leq n}}\left(\sum_{i\in\mathcal{U}}\mathbf{e}_i\right)^T\Sigma_*\left(\sum_{i\in\mathcal{U}}\mathbf{e}_i\right)\\
    &=\mathrm{MSE}(\Sigma_*)
\end{align}
as the maximum over all $\mathcal{U}$ is chosen. If $\Sigma_*$ satisfies the privacy constraint in \eqref{opt}, any permutation of $\Sigma_*$ also satisfies it by definition.
    
\end{Proof}

\begin{Claim}\label{part1} $\mathrm{MSE}\left(\frac{1}{n!}\sum_{j=1}^{n!}\Pi_j(\Sigma_*)\right)\leq \mathrm{MSE}(\Sigma_*)$.
\end{Claim}

\begin{Proof}
    \begin{align}
        &\mathrm{MSE}\left(\frac{1}{n!}\sum_{j=1}^{n!}\Pi_j(\Sigma_*)\right)\nonumber\\
        &=\max_{\substack{\mathcal{U}\subseteq\mathcal{U}_{all}\\ t\leq|\mathcal{U}|\leq n}}\left(\sum_{i\in\mathcal{U}}\mathbf{e}_i\right)^T\left(\frac{1}{n!}\sum_{j=1}^{n!}P_j\Sigma_*P_j^T\right)\left(\sum_{i\in\mathcal{U}}\mathbf{e}_i\right)\\
        &\leq\frac{1}{n!}\sum_{j=1}^{n!}\max_{\substack{\mathcal{U}\subseteq\mathcal{U}_{all}\\ t\leq|\mathcal{U}|\leq n}}\left(\sum_{i\in\mathcal{U}}\mathbf{e}_i\right)^T\left(P_j\Sigma_*P_j^T\right)\left(\sum_{i\in\mathcal{U}}\mathbf{e}_i\right)\\
        &=\frac{1}{n!}\sum_{j=1}^{n!}\mathrm{MSE}(\Sigma_*)\\
        &=\mathrm{MSE}(\Sigma_*)
    \end{align}
\end{Proof}

\begin{Claim}\label{claimm11}
$\tilde{\Sigma}=\frac{1}{n!}\sum_{j=1}^{n!}\Pi_j(\Sigma_*)$ satisfies the privacy constraint in \eqref{opt}.
\end{Claim}

\begin{Proof}
    The matrix inverse operation is a convex function \cite{inv}. Therefore, 
    \begin{align}
        \mathbf{e}_i^T\tilde{\Sigma}^{-1}\mathbf{e}_i&=\mathbf{e}_i^T\left(\frac{1}{n!}\sum_{j=1}^{n!}\Pi_j(\Sigma_*)\right)^{-1}\mathbf{e}_i^T\\
        &\leq\frac{1}{n!}\sum_{j=1}^{n!}\mathbf{e}_i^TP_j^T\Sigma_*^{-1}P_j\mathbf{e}_i\\
        &=\frac{1}{n!}\sum_{j=1}^{n!}\mathbf{e}_{i_j}^T\Sigma_*^{-1}\mathbf{e}_{i_j}\\
        &\leq\frac{1}{\sigma^2_{\epsilon,\delta}}
    \end{align}
    where $i_j=\Pi_j(i)$, and the last inequality is obtained from the constraint in \eqref{opt} on $\Sigma_*$.
\end{Proof}
Claims~\ref{claimm9}-\ref{claimm11} collectively prove Theorem~\ref{converse}.
\end{Proof}

\section{Details on SecAgg \cite{secagg} }\label{secaggdet}

In this Section, we outline the basic steps of SecAgg \cite{secagg}. Note that this is not the exact protocol, and we only provide the conceptual steps to give an overview for comparison.
\begin{enumerate}
    \item Basic noise generation: each pair of users $i,j\in[1:N]$, $i\neq j$ samples a common random noise variable from a finite field, i.e., $S_{i,j}=S_{j,i}\sim\text{unif}(\mathbb{F}_q)$ $\implies$ $O(n)$ communications.
    \item Additional shared randomness for dropout handling: each user $i$ distributes $(n,t)$-secret shares of $S_{i,j}$, $\forall j$ with all other users $\implies$ $O(n)$ communications
    \item Precautions for delayed user-responses: each user $i$ samples an additional random variable $b_i$, and sends its $(n,t)$-secret shares to all other users $\implies$ $O(n)$ communications.
    \item User $i$ $\to$ server: $Y_i=x_i+\sum_{j>i}S_{i,j}-\sum_{j<i}S_{i,j}+b_i$, where $x_i$ is the private value of user $i$ $\implies$ $O(1)$ (or $O(d)$ for vectors) communications.
    \item Server computes: $A_1=\sum_{i\in\mathcal{U}}Y_i$, where $\mathcal{U}$ are the set of responding users.
    \item server broadcasts $[1:n]\setminus\mathcal{U}$ to inform the dropouts to the remaining users.
    \item Server collects secret shares of each $S_{v,i}$, $i\in\mathcal{U}$ from $t$ users for each $v\in[1:n]\setminus\mathcal{U}$, and reconstructs each $S_{v,i}$.
    \item Server collects $t$ secret shares of $b_i$, $i\in\mathcal{U}$, and reconstructs $b_i$ of all responding users.
    \item Server computes:
    \begin{align}
        A_2&=A_1-\sum_{i\in\mathcal{U}}b_i\nonumber\\
        &\quad+\sum_{v\in[1:n]\setminus\mathcal{U}}\left(-\sum_{v<j,j\in\mathcal{U}}S_{v,j}+\sum_{v>j,j\in\mathcal{U}}S_{v,j}\right)
    \end{align}
    $\implies$ $A_2=\sum_{i\in\mathcal{U}}x_i$.
\end{enumerate}
In DP-DME with SecAgg, additionally, users add discrete Gaussian noise with variance $O\left(\frac{1}{n}\right)$ to $Y_i$ in step 4, to get the same performance as CDP with $(\epsilon,\delta)$-DP.

\subsection{Example of SecAgg \cite{secagg}}

Consider a four-user example with $X_i\in\mathbb{F}_q$ denoting the private value of user $i$ for $i=1,2,3,4$, where $\mathbb{F}_q$ is a  finite field. The users send the following vectors to the server:
\begin{align}
    \text{user 1: } Y_1&=X_1+A+B+C\\
    \text{user 2: } Y_2&=X_2-A+D+F\label{y2}\\
    \text{user 3: } Y_3&=X_3-B-D+G\\
    \text{user 4: } Y_4&=X_4-C-F-G
\end{align}
where $A,B,C$ are random noise uniformly distributed over $\mathbb{F}_q$. When all users respond, the server computes:
\begin{align}
    Y_1+Y_2+Y_3+Y_4=X_1+X_2+X_3+X_4
\end{align}
to obtain the sum of the users' private values without learning any information on the individual values (conditioned on the sum), due to one-time-padding. 

Now, assume that user 2 is unresponsive. Then, the server only receives:
\begin{align}
    \text{user 1: } Y_1&=X_1+A+B+C\\
    \text{user 3: } Y_3&=X_3-B-D+G\\
    \text{user 4: } Y_4&=X_4-C-F-G
\end{align}
which provides no useful information, as $Y_1,Y_3,Y_4$ and any function of them are uniformly distributed over $\mathbb{F}_q$. The server computes:
\begin{align}\label{stop1}
    Y_1+Y_3+Y_4=X_1+X_3+X_4+A-D-F
\end{align}
and requires additional information from users 1, 3, and 4 regarding $A,D$ and $F$ in subsequent rounds, to recover $X_1+X_3+X_4$. If it is guaranteed that users 1,3,4 do not dropout in the next round, the server can simply request for $A,D,F$ from users $1,3,4$ respectively, and recover $X_1+X_3+X_4$. However, the users remaining after round 1 can drop in round 2. For example, assume user 3 dropped out in round 2. Now, there is no way that the server can recover $D$ to obtain $X_1+X_3+X_4$, as the only two users with access to $D$ have dropped out. To avoid this, SecAgg requires each user to share components of their pair-wise noise terms with all other users in the initialization stage itself. This is done by each user computing $t$-out-of-$n$ secret shares of the pair-wise noise terms and distributing them over all users. Here, $t$ is the minimum number of users required in the system after all rounds and $n$ is the number of users at the beginning. For this example, assume $t=2$ and $n=4$. Therefore, at the initialization stage user 1 distributes secret-shares of $A,B$ and $C$ with users 2,3, and 4, user 2 distributes secret-shares of $B,D$ and $G$ with users 1,3, and 4, and so on. 

Now, let's go back to the discussion on recovering $A,D,F$ in \eqref{stop1}, where user 3 dropped out in round 2. To recover $A,D,F$, the server simply requests $A$ and $F$ from users 1 and 4, and also requests for the two secret shares of $D$ from them. The server then reconstructs $D$ (recall that 2 secret shares are sufficient to reconstruct the original message in 2-out-of-4 secret sharing), and recovers $X_1+X_3+X_4$, by removing $A,D,F$ from $Y_1+Y_2+Y_3$. This recovery process is successful as long as at least $t=2$ users are remaining in the system.

Now consider the situation where user 2 sends a delayed response. In this case, user 2 simply sends $Y_2$ in \eqref{y2}, without knowing that it was already treated as a dropout. Then, the server can decode $X_2$ from $Y_2$, as it already has access to $A,D,F$, violating the privacy of user 2. To avoid this, each user adds another random noise variable $b_i$ to their original uploads in round 1 as follows.
\begin{align}
    \text{user 1: } Y_1&=X_1+A+B+C+b_1\\
    \text{user 2: } Y_2&=X_2-A+D+F+b_2\label{del}\\
    \text{user 3: } Y_3&=X_3-B-D+G+b_3\\
    \text{user 4: } Y_4&=X_4-C-F-G+b_4
\end{align}
The $t$-out-of-$n$ secret shares of $b_i$ are also shared among the users just like for the pair-wise random noise terms. Now, when the server thought that user 2 dropped out, it calculates: 
\begin{align}
    Y_1+Y_3+Y_4=X_1+X_3+X_4+A-D-F+b_1+b_3+b_4.
\end{align}
Next, it recovers $A,D,F$ as before, and reaches $X_1+X_3+X_4+b_1+b_3+b_4$. Now, even if user 2 sends the delayed response in \eqref{del}, the server can not decode $X_2$ due to the $b_2$ term. Finally, to recover $X_1+X_3+X_4$, the server contacts users 1 and 4 again, requesting $b_1,b_4$ and the two secret shares of $b_3$ to remove them from the aggregate (recall that user 3 dropped out in round 2 of this example, and therefore requires the help of users 1 and 4 to recover $b_3$). This completes the basic steps of SecAgg.

\end{document}